\documentclass[aps,nofootinbib,twocolumn,superscriptaddress,toc,showkeys]{revtex4-2}
\pdfoutput=1

\usepackage[usenames,dvipsnames]{xcolor}
\usepackage{hyperref}
\hypersetup{
	colorlinks=true,
	urlcolor=Maroon,
	linkcolor=RoyalBlue,
	citecolor=Maroon,
	bookmarks=true,
	pdftitle={Cutting-Edge Tools for Cutting Edges},
	pdfauthor={Ruth Britto, Claude Duhr, Holmfridur S. Hannesdottir, Sebastian Mizera},
	pdfdisplaydoctitle=true,
	pdfstartview=FitH
}

\usepackage{amsmath}
\usepackage{amsfonts}
\usepackage{amssymb}
\usepackage{mathtools}
\usepackage{microtype}
\usepackage{bbm}
\usepackage[utf8]{inputenc}
\usepackage{blkarray}
\usepackage{bigstrut}
\usepackage{nccmath}
\usepackage{enumitem}
\usepackage{braket}
\usepackage{dsfont}
\usepackage[export]{adjustbox}
\usepackage[english]{babel}
\usepackage{stmaryrd}
\usepackage{orcidlink}
\usepackage{mathrsfs}
\usepackage{comment}

\makeatletter
\newcommand{\xMapsto}[2][]{\ext@arrow 0599{\Mapstofill@}{#1}{#2}}
\def\Mapstofill@{\arrowfill@{\Mapstochar\Relbar}\Relbar\Rightarrow}
\makeatother

\allowdisplaybreaks
\graphicspath{{figures/}}

\DeclareMathOperator*{\SumInt}{%
\mathchoice%
  {\ooalign{$\displaystyle\sum$\cr\hidewidth$\displaystyle\int$\hidewidth\cr}}
  {\ooalign{\raisebox{.14\height}{\scalebox{.7}{$\textstyle\sum$}}\cr\hidewidth$\textstyle\int$\hidewidth\cr}}
  {\ooalign{\raisebox{.2\height}{\scalebox{.6}{$\scriptstyle\sum$}}\cr$\scriptstyle\int$\cr}}
  {\ooalign{\raisebox{.2\height}{\scalebox{.6}{$\scriptstyle\sum$}}\cr$\scriptstyle\int$\cr}}
}

\renewcommand{\Re}{\mathrm{Re}}
\renewcommand{\Im}{\mathrm{Im}}

\renewcommand{\geq}{\geqslant}

\def \be {\begin{equation}}
\def \ee {\end{equation}}

\def \C  {\mathbb{C}}

\def \R  {\mathbb{R}}

\def \e  {\mathrm{e}}
\def \d  {\mathrm{d}}
\def \N  {\mathcal{N}}

\def \I  {\mathcal{I}}

\def \O  {\mathcal{O}}
\def \T  {\mathcal{T}}
\def \L  {\mathrm{L}}

\def \D {\mathrm{D}}
\def \E {\mathrm{E}}
\def \Li{\mathrm{Li}}
\def \disc{\Delta}
\def \Disc{\mathrm{Disc}}
\def \Cut{\mathrm{Cut}}
\def \eps {\varepsilon}

\def \ddelta {\hat{\delta}}
\def \UCut {\mathrm{Cut}^U}

\def \M {\mathcal{M}}

\def \zb {\bar{z}}

\definecolor{yellowish}{rgb}{0.880722,0.611041,0.142051}

\usepackage{ntheorem}
\usepackage[framemethod=tikz]{mdframed}
\theorembodyfont{\upshape}

\newmdtheoremenv[hidealllines=true,topline=true,bottomline=true,linecolor=black!10!white,linewidth=1pt,innerleftmargin=0.5em,innerrightmargin=0.5em,frametitlerule=true]{mdexample}{Example}

\newmdtheoremenv[hidealllines=true,topline=true,bottomline=true,linecolor=black!10!white,linewidth=1pt,innerleftmargin=0.5em,innerrightmargin=0.5em]{mddefinition}{Definition}

\usepackage{tikz}
\usetikzlibrary{snakes}
\usetikzlibrary{calc}
\usetikzlibrary{decorations.pathmorphing,decorations.markings,arrows.meta}
\tikzset{
    arrow style at position/.style 2 args={
        decoration={
            markings,
            mark=at position #1 with {\arrow{#2}}
        },
        postaction={decorate}
    }
}

\tikzset{
    quark/.style={
        decoration={},
        decorate
    },
    lepton/.style={
        decoration={},
        decorate
    },
    gluon/.style={
        decoration={coil, aspect=0.75, mirror, segment length=1.5mm},
        decorate
    },
    gluon_small/.style={
        decoration={coil, aspect=0.75, mirror, segment length=1mm, amplitude =0.6mm},
        decorate
    },
    vector/.style={
        decoration={snake, aspect=0.75, mirror, segment length=2mm,amplitude=0.6mm},
        decorate
    },
    Higgs/.style={
        decoration={},
        densely dashed,
        decorate
    },
}

\theoremstyle{definition}

\begin{document}

\begin{flushright}
    BONN-TH-2024-05
\end{flushright}

\title{Cutting-Edge Tools for Cutting Edges}

\author{Ruth Britto \!\orcidlink{0000-0003-2462-6481}}
\affiliation{School of Mathematics and Hamilton Mathematics Institute, Trinity College, Dublin 2, Ireland}

\author{Claude Duhr \!\orcidlink{0000-0001-5820-3570}}
\affiliation{Bethe Center for Theoretical Physics, Universit\"at Bonn, D-53115, Germany}

\author{Holmfridur S. Hannesdottir \!\orcidlink{0000-0002-5440-2086}}
\affiliation{Institute for Advanced Study, Einstein Drive, Princeton, NJ 08540, USA}

\author{Sebastian Mizera \!\orcidlink{0000-0002-8066-5891}}
\affiliation{Institute for Advanced Study, Einstein Drive, Princeton, NJ 08540, USA}

\begin{abstract}
We review different notions of cuts appearing throughout the literature on scattering amplitudes. Despite similar names, such as \emph{unitarity cuts} or \emph{generalized cuts}, they often represent distinct computations and distinct physics. We consolidate this knowledge, summarize how cuts are used in various computational strategies, and explain their relations to other quantities including imaginary parts, discontinuities, and monodromies. Differences and nuances are illustrated on explicit examples.
\end{abstract}


\maketitle

\twocolumngrid

\setcounter{tocdepth}{2}
\tableofcontents

\section{Introduction}

At the heart of modern research into scattering amplitudes are {\it cuts}.
Diagrammatically, cutting a diagram means tagging a subset of its internal edges, here denoted with dashed lines, e.g.
\be
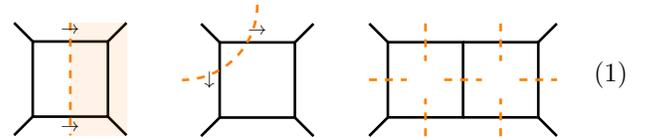
\label{eq:intro}
\begin{gathered}
\begin{tikzpicture}[line width=1,scale=0.5]
	\coordinate (v1) at (-1,-1);
    \coordinate (v2) at (-1,1);
    \coordinate (v3) at (1,1);
    \coordinate (v4) at (1,-1);
    \draw[] (v1) -- (v2) -- (v3) -- (v4) -- (v1);
    \draw[] (v1) -- ++(-135:0.7);
    \draw[] (v2) -- ++(135:0.7);
    \draw[] (v3) -- ++(45:0.7);
    \draw[] (v4) -- ++(-45:0.7);
    \draw[dashed, orange] (0,1.5) -- (0,-1.5);
    \fill[fill=orange, opacity=0.1] (0,1.5) -- (0,-1.5) -- (1.5,-1.5) -- (1.5,1.5);
    \node[scale=0.7,rotate=0] at (0,1.3) {$\rightarrow$};
    \node[scale=0.7,rotate=0] at (0,-1.3) {$\rightarrow$};
\end{tikzpicture}
\qquad
\begin{tikzpicture}[line width=1,scale=0.5]
	\coordinate (v1) at (-1,-1);
    \coordinate (v2) at (-1,1);
    \coordinate (v3) at (1,1);
    \coordinate (v4) at (1,-1);
    \draw[] (v1) -- (v2) -- (v3) -- (v4) -- (v1);
    \draw[] (v1) -- ++(-135:0.7);
    \draw[] (v2) -- ++(135:0.7);
    \draw[] (v3) -- ++(45:0.7);
    \draw[] (v4) -- ++(-45:0.7);
    \draw[dashed, orange] (0,2) arc (0:-90:2);
    \node[scale=0.7,rotate=0] at (0,1.3) {$\rightarrow$};
    \node[scale=0.7,rotate=-90] at (-1.3,0) {$\rightarrow$};
\end{tikzpicture}
\qquad
\begin{tikzpicture}[line width=1,scale=0.5]
	\coordinate (v1) at (-1,-1);
    \coordinate (v2) at (-1,1);
    \coordinate (v3) at (1,1);
    \coordinate (v4) at (1,-1);
    \coordinate (v5) at (3,-1);
    \coordinate (v6) at (3,1);
    \draw[] (v1) -- (v2) -- (v3) -- (v4) -- (v1);
    \draw[] (v3) -- (v6) -- (v5) -- (v4);
    \draw[] (v1) -- ++(-135:0.7);
    \draw[] (v2) -- ++(135:0.7);
    \draw[] (v6) -- ++(45:0.7);
    \draw[] (v5) -- ++(-45:0.7);
    \draw[dashed, orange] (0,1.5) -- (0,0.5);
    \draw[dashed, orange] (2,1.5) -- (2,0.5);
    \draw[dashed, orange] (0,-1.5) -- (0,-0.5);
    \draw[dashed, orange] (2,-1.5) -- (2,-0.5);
    \draw[dashed, orange] (-1.5,0) -- (-0.5,0);
    \draw[dashed, orange] (2.5,0) -- (3.5,0);
    \draw[dashed, orange] (0.5,0) -- (1.5,0);
\end{tikzpicture}
\end{gathered}
\ee
However, this operation can have distinct meanings depending on the physical or mathematical context in which it is employed. The purpose of this article is to review different flavors of cuts appearing in the literature, as well as explain how they are used in physical applications and as a computational scheme.

While details differ depending on the application, cuts always come with a notion of placing particles with momentum $p$ and mass $m$ on their mass shell, i.e., $p^2 = m^2$, which can be implemented as a Dirac delta function or a residue prescription. As a result, in perturbation theory the loop integrand simplifies and the number of loop integrations drops, which is ultimately why cuts become a useful tool.
Here, we will categorize cuts by their purpose. Broadly, they are employed to compute:
\begin{itemize}[leftmargin=*]
\item \textbf{Imaginary parts and total discontinuities.} Historically the oldest version, which is codified by the \emph{Cutkosky rules} and directly descends from unitarity. Unitarity cuts appear as a consequence of integrating over the phase-space of on-shell states and become useful for computing cross sections through the optical theorem, as well as sum rules via dispersion relations, see Sec.~\ref{sec:imaginary}. For unitarity cuts, we use the convention (\ref{eq:intro}, left) where the energy flows from the unshaded region to the shaded one.

\item \textbf{Individual discontinuities and monodromies.} A finer version, which relates cuts to discontinuities around individual branch points of scattering amplitudes. They give constraints on the analytic structure, which can be used in perturbative bootstrap approaches to scattering amplitudes, see Sec.~\ref{sec:individual}. For this type of cut, we use the convention (\ref{eq:intro}, middle) with arrows indicating the energy flow.

\item \textbf{Integrands and bases of functions.} Generalized cuts work at the level of loop integrands and are useful for recursively constructing them by gluing lower-point amplitudes. They can be successfully employed in a variety of applications, including generalized unitarity, differential equations, and double-copy relations, see Sec.~\ref{sec:generalized}. For generalized cuts, we use the convention (\ref{eq:intro}, right) where energies of cut particles are not constrained at all.
\end{itemize}

\section{Preliminaries}

Throughout this review we will use the following standard notation. We stick to all-incoming conventions and mostly-minus signature, i.e., momentum conservation reads $\sum_{i=1}^{n} p_i = 0$ with $p_i^0 > 0$ for incoming particles and $p_i^0 < 0$ for outgoing. We use the short-hand notation $p_{ij\cdots k} = p_i + p_j + \ldots + p_k$. For $n=4$, we use the Mandelstam invariants
\be
s = (p_1 + p_2)^2, \quad t = (p_2 + p_3)^2,\quad u = (p_1 + p_3)^2\, ,
\ee
which satisfy $s+t+u = \sum_{i=1}^{4} p_i^2$ due to momentum conservation.

\begin{mddefinition}
A Feynman integral $\I$ of a connected diagram with $\L$ loops and $\E$ internal edges in $\D = \D_0 - 2\epsilon$ space-time dimensions, where $\D_0$ is a positive integer, is given by
\be
\I := \int_{\ell_1, \ldots, \ell_\L} \frac{\N}{\prod_{e=1}^{\E} (q_e^2 - m_e^2 + i\varepsilon)}\, ,
\label{eq:FeynInt-def}
\ee
where $q_e$ is the momentum flowing through the $e^{\textrm{th}}$ edge, $\ell_1, \ldots, \ell_{\L}$ form a basis of loop momenta, and $\N$ is any numerator that is polynomial in the loop momenta. Note that $q_e$ and $\N$ can depend on both the loop momenta $\ell_i$, as well as the external momenta labelled with $p_i$. We use the short-hand notation
\be
\int_{\ell_1, \ldots, \ell_\L} := \lim_{\varepsilon \to 0^+} \int \prod_{a=1}^{\L} \frac{\e^{\gamma_E \epsilon} \mu^{2\epsilon}}{i\pi^{\D/2}}  \d^{\D} \ell_a\, ,
\label{eq:phasespace}
\ee
where $\gamma_E$ is the Euler--Mascheroni constant and $\mu$ is a mass scale. We henceforth work in units in which $\mu=1$.
\end{mddefinition}

We refer to \cite{SZABO200628} for an introduction to perturbative scattering amplitudes. For reviews of nonperturbative aspects, see \cite{BUCHHOLZ2006456,BROS2006465,IAGOLNITZER2006475}.

In all examples, we take the square root, logarithms and dilogarithms to be on their principal branches.

\begin{mdexample}\label{ex:bubble}
    We label the bubble diagram with both internal masses equal to $m$ as
    \be\label{eq:bub-diagram}
    \begin{gathered}
    \begin{tikzpicture}[line width=1,scale=0.5]
	\coordinate (v1) at (-1,0);
    \coordinate (v2) at (1,0);
    \draw[] (v1) [out=55,in=125] to (v2);
    \draw[] (v1) [out=-55,in=-125] to (v2);
    \draw[] (v1) -- ++(-135:0.7) node[left] {$p_1$};
    \draw[] (v1) -- ++(135:0.7) node[left] {$p_2$};
    \draw[] (v2) -- ++(45:0.7) node[right] {$p_3$};
    \draw[] (v2) -- ++(-45:0.7) node[right] {$p_4$};
    \node[scale=0.85] at (0,1.5) {$p_{12}{-}\ell$};
    \node[scale=0.85,yshift=-5] at (0,1.2) {$\rightarrow$};
    \node[scale=0.85,yshift=-2] at (-0.1,-1.3) {$\ell$};
    \node[scale=0.85,yshift=5] at (0,-1.3) {$\rightarrow$};
    \end{tikzpicture}
    \end{gathered}
    \ee
    In two dimensions, the corresponding Feynman integral is given by
    \begin{align}
    \I_{\text{bub}}^{\D=2} :=& \int_{\ell} \frac{1}{(\ell^2 - m^2) ((\ell - p_{12})^2 - m^2)}\nonumber\\
    =& \frac{2i}{\sqrt{ s (4m^2 {-} s)}} 
    \log \left( \frac{\sqrt{4m^2 {-}s } {-} i \sqrt{s}}{\sqrt{4m^2 {-}s } {+} i \sqrt{s}} \right).\label{eq:I_bub_int}
    \end{align}
This expression is valid for any complex $s$, except on the branch cut $s>4m^2$. 
\end{mdexample}

\begin{mdexample}\label{ex:I-0m}
The massless box diagram is labelled with
\be\label{eq:box-diagram}
\begin{gathered}
\begin{tikzpicture}[line width=1,scale=0.5]
	\coordinate (v1) at (-1,-1);
    \coordinate (v2) at (-1,1);
    \coordinate (v3) at (1,1);
    \coordinate (v4) at (1,-1);
    \draw[] (v1) -- (v2) -- (v3) -- (v4) -- (v1);
    \draw[] (v1) -- ++(-135:0.7) node[left] {$p_1$};
    \draw[] (v2) -- ++(135:0.7) node[left] {$p_2$};
    \draw[] (v3) -- ++(45:0.7) node[right] {$p_3$};
    \draw[] (v4) -- ++(-45:0.7) node[right] {$p_4$};
    \node[scale=0.85] at (0,1.8) {$p_{12}{-}\ell$};
    \node[scale=0.85,yshift=-5] at (0,1.6) {$\rightarrow$};
    \node[scale=0.85,yshift=-2] at (-0.1,-1.6) {$\ell$};
    \node[scale=0.85,yshift=5] at (0,-1.6) {$\rightarrow$};
\end{tikzpicture}
\end{gathered}
\ee
with all $p_i^2 = 0$. In ``Euclidean kinematics'' $s,t < 0$, it is given by
\begin{gather}
\I^{0m}(s,t) := \int_{\ell} \frac{1}{\ell^2 (\ell-p_1)^2 (\ell-p_{12})^2 (\ell - p_{123})^2}
\nonumber
\\
= \frac{1}{s t} \Big[
\frac{2}{\epsilon^2} \!\!\left( ({-}s)^{-\epsilon} {+} ({-}t)^{-\epsilon} \right)
{-} \left( \log(-s) - \log(-t) \right)^2 
\nonumber
\\ \hspace{2cm} {-} 4\pi^2/3
 {+} \mathcal{O}(\epsilon)\Big] \label{eq:I-0m}
\end{gather}
where we have expanded up to $\mathcal{O}(\epsilon^0)$ around $\D=4-2\epsilon$ spacetime dimensions.
The answer is written in such a way that the analytic continuation to other kinematic channels is given by replacing $s \to s+i\varepsilon$ and $t \to t + i\varepsilon$.
Note that analytic continuation in general does not amount to such a simple replacement.
\end{mdexample}

\begin{mdexample}\label{ex:I-2m}
The ``two-mass easy box'' diagram corresponds to \eqref{eq:box-diagram} with $p_1^2 \neq 0$ and $p_3^2 \neq 0$. It can be interpreted as either a contribution to a $4$-point amplitude with two massive legs, or to a $6$-point massless amplitude:
\be
\begin{gathered}
\begin{tikzpicture}[line width=1,scale=0.5]
	\coordinate (v1) at (-1,-1);
    \coordinate (v2) at (-1,1);
    \coordinate (v3) at (1,1);
    \coordinate (v4) at (1,-1);
    \draw[] (v1) -- (v2) -- (v3) -- (v4) -- (v1);
    \draw[] (v1) -- ++(-150:0.7);
    \draw[] (v1) -- ++(-120:0.7) node[left,yshift=-2] {$p_1$};
    \draw[] (v2) -- ++(135:0.7) node[left] {$p_2$};
    \draw[] (v3) -- ++(60:0.7) node[right,yshift=2] {$p_3$};
    \draw[] (v3) -- ++(30:0.7);
    \draw[] (v4) -- ++(-45:0.7) node[right] {$p_4$};
    \node[yshift=-6,RoyalBlue] at (0,-1) {\footnotesize $1$};
    \node[xshift=-6,RoyalBlue] at (-1,0) {\footnotesize $2$};
    \node[yshift=6,RoyalBlue] at (0,1) {\footnotesize $3$};
    \node[xshift=6,RoyalBlue] at (1,0) {\footnotesize $4$};
\end{tikzpicture}
\end{gathered}
\ee
Either way, in an expansion around $\D=4-2\epsilon$ spacetime dimensions, it evaluates to
\begin{gather}
    \I^{2m} (s,t,p_1^2,p_3^2) = \frac{1}{s t-p_1^2 p_3^2}
    \label{eq:I2m_full}
    \\ \times
    \Bigg\{ \left(\frac{2}{\epsilon^2}-\frac{\pi^2}{6}\right) \left[ ({-}s)^{-\epsilon} {+} ({-}t)^{-\epsilon} {-} ({-}p_1^2)^{-\epsilon} {-} ({-}p_3^2)^{-\epsilon} \right]
    \nonumber
    \\ -
    2 \Li_2\Big(1{-}\frac{p_1^2}{s} \Big)-2 \Li_2 \Big(1{-}\frac{p_1^2}{t} \Big) -
    2 \Li_2\Big(1{-}\frac{p_3^2}{s} \Big)
    \nonumber
    \\
    -2 \Li_2 \Big(1{-}\frac{p_3^2}{t} \Big)
    +2 \Li_2 \Big(1- \frac{p_1^2 p_3^2}{s t} \Big)
    - \log^2\Big(\frac{s}{t} \Big) 
    \Bigg\}
    +
    \mathcal{O} (\epsilon) \,.
    \nonumber
\end{gather}
when $s,t,p_1^2,p_3^2 < 0$.
The analytic continuation of this expression to other kinematic regions is nontrivial, and must be done carefully so that it matches onto the $i \varepsilon$ prescription of the Feynman integral. We can, for example, rotate the invariants from negative to positive in the upper half-plane and analytically continue the expression in~\eqref{eq:I2m_full}. Note that this does not correspond to numerically substituting $s \to s+i\varepsilon$ etc. for the kinematic invariants, since we might need to change the branch of the dilogarithms during the analytic continuation; recall  that~\eqref{eq:I2m_full} assumes they are written on their principal branch. 

We define the ``one-mass box'' as
\begin{equation}
    \I^{1m} (s,t,p_1^2) = \lim_{p_3^2 \to 0} \I^{2m} (s,t,p_1^2,p_3^2).
    \label{eq:I-1m}
\end{equation}
This equation holds for a fixed dimension $\D$. Note that the limit in $p_3^2 \to 0$ and the expansion in $\epsilon$ do not commute in general, but~\eqref{eq:I2m_full} is written in such a way that substituting $p_3^2 = 0$ on the right-hand side gives $\I^{1m} (s,t,p_1^2)$ up to terms of $\mathcal{O(\epsilon)}$. 
\label{eq:Ex-1mbox}
\end{mdexample}

Cuts of $\I$ will be defined by replacing propagators with delta functions of the form
\begin{subequations}
\begin{align}
\frac{-1}{2\pi i}\ddelta^{\pm}(q_e^2-m_e^2) &=
\delta^{\pm}(q_e^2-m_e^2) \\
&= \delta(q_e^2-m_e^2) \theta(\pm q_e^0)\, .
\end{align}
\label{eq:delta-hat}
\end{subequations}
The $\pm$ sign selects a definite flow of the energy component $q_e^0$ across the cut. The specific choice of which signs to use will depend on the application.

\section{\label{sec:imaginary}Unitarity cuts and total discontinuities}

\subsection{Nonperturbative unitarity}
\label{sec:non-pert-unitarity}

Historically, the first incarnation of cut integrals were unitarity cuts. They are a direct consequence of the unitarity of the $S$-matrix, $S^\dagger S=\mathbbm{1}$, which embodies the physical principle of probability conservation. If we define the transfer matrix $T$ in the usual way by $S = \mathbbm{1} + i T$, then the unitarity of the $S$-matrix translates into the statement that
\be\label{eq:unitarity}
T-T^\dagger = i T^\dagger T\,.
\ee
By definition, the scattering amplitude $\mathcal{M}(\textrm{in}\to \textrm{out})$ corresponds to the matrix elements of the transfer matrix with asymptotic external states $\langle \textrm{out}|$ and $|\textrm{in}\rangle$,
\begin{subequations}
\begin{align}
\bm{\hat{\delta}}_{\text{in,out}} \, \mathcal{M}(\textrm{in}\to \textrm{out})  &= \langle\textrm{out}|T|\textrm{in}\rangle\,,\\
\bm{\hat{\delta}}_{\text{in,out}} \mathcal{M}(\textrm{out}\to \textrm{in})^\ast  &= \langle\textrm{out}|T^\dag|\textrm{in}\rangle\,,
\end{align}
\end{subequations}
where $\bm{\hat{\delta}}_{\text{in,out}} = (2\pi)^\D \delta^\D (p_{\text{out}} + p_{\text{in}})$ is the overall momentum-conserving delta function.
Note that $\mathcal{M}$ contains all connected and disconnected terms apart from the fully disconnected term $\bm{\hat{\delta}}_{\text{in,out}} \langle \textrm{out}| \mathbbm{1} |\textrm{in} \rangle$. 
Hence, if we contract \eqref{eq:unitarity} with asymptotic sates, and we insert a complete set of asymptotic states into the right-hand side, we can cast the unitarity of the $S$-matrix a nonlinear equation for the scattering amplitudes:
\begin{align}
\label{eq:unitarity_amps}
\mathcal{M}(\textrm{in}\to \textrm{out}) & -  \mathcal{M}(\textrm{out}\to \textrm{in})^\ast =
\\
&\,i \, \SumInt_X \bm{\hat{\delta}}_{\text{in,X}} \mathcal{M}(\textrm{in}\to X)\,
\mathcal{M}(\textrm{out}\to X)^*\,.
\nonumber
\end{align}
The term on the right-hand side is called the \emph{unitarity cut}.\footnote{
Recall that in even space-time dimensions $\D$, CPT invariance gives
\be
\mathcal{M}(\text{in} \to \text{out}) = \mathcal{M}(\overline{\text{out}} \to \overline{\text{in}})\, ,
\ee
where the bar denotes applying charge and parity conjugation to the state. Likewise, CP invariance (or equivalently, T invariance) means
\be
\quad\mathcal{M}(\text{in} \to \text{out}) = \mathcal{M}(\overline{\text{in}} \to \overline{\text{out}}) = \mathcal{M}(\text{out} \to \text{in}) = \mathcal{M}(\overline{\text{out}} \to \overline{\text{in}})\, ,
\ee
so the left-hand side of \eqref{eq:unitarity_amps} can be replaced with $2i \, \Im\, \mathcal{M}(\text{in} \to \text{out})$ for CP-invariant theories.
} It involves summing over all possible intermediate states $X$ in the theory and integrating over their on-shell phase-space. More concretely,
\be
\SumInt_X := \sum_{X_j} \prod_{i \in X_j} \int \frac{\d^\D \ell_i}{(2\pi)^{\D-1}} \delta^+(\ell_i^2 - m_i^2)\, ,
\ee
where $m_i$ is the mass of particle $i$, and each $X_j$ consists of a possible set of particles that can be exchanged.
Graphically, we depict \eqref{eq:unitarity_amps} as
\be\label{eq:unitarity-diagram}
    \begin{tikzpicture}[line width=1,scale=0.25,baseline={([yshift=1.2ex]current bounding box.center)}]
    	\coordinate (v0) at (0,0);
        \draw[] (v0) -- ++(-150:2.5);
        \draw[] (v0) -- ++(-180:2.5);
        \draw[] (v0) -- ++(-210:2.5);
        \draw[] (v0) -- ++(30:2.5);
        \draw[] (v0) -- ++(0:2.5);
        \draw[] (v0) -- ++(-30:2.5);
        \filldraw[color=black, fill=black!5] (v0) circle[radius=1.5] node[xshift=-1] {$\mathcal{M}$};
        \draw[] (-2,-3) node {\footnotesize$\text{in}$};
        \draw[] (2,-3) node {\footnotesize$\text{out}$};
        \fill[fill=orange, opacity=0.1] (1.8,2) -- (1.8,-2) -- (2.8,-2) -- (2.8,2);
        \draw[dashed, orange] (1.8,2) -- (1.8,-2);
    \end{tikzpicture}
    -
    \begin{tikzpicture}[line width=1,scale=0.25,baseline={([yshift=1.2ex]current bounding box.center)}]
    	\coordinate (v0) at (0,0);
        \draw[] (v0) -- ++(-150:2.5);
        \draw[] (v0) -- ++(-180:2.5);
        \draw[] (v0) -- ++(-210:2.5);
        \draw[] (v0) -- ++(30:2.5);
        \draw[] (v0) -- ++(0:2.5);
        \draw[] (v0) -- ++(-30:2.5);
        \filldraw[color=black, fill=black!5] (v0) circle[radius=1.5] node[xshift=1] {$\mathcal{M}^\ast$};
        \draw[] (-2,-3) node {\footnotesize$\text{in}$};
        \draw[] (2,-3) node {\footnotesize$\text{out}$};
        \fill[fill=orange, opacity=0.1] (-1.8,2) -- (-1.8,-2) -- (2.8,-2) -- (2.8,2);
        \draw[dashed, orange] (-1.8,2) -- (-1.8,-2);
    \end{tikzpicture}
= \;i \SumInt_X 
    \begin{tikzpicture}[line width=1,scale=0.25,baseline={([yshift=1.0ex]current bounding box.center)}]
    	\coordinate (v0) at (-2,0);
        \coordinate (v1) at (2,0);
        \draw[] (v0) -- ++(-150:2.5);
        \draw[] (v0) -- ++(-180:2.5);
        \draw[] (v0) -- ++(-210:2.5);
        \draw[] (v1) -- ++(30:2.5);
        \draw[] (v1) -- ++(0:2.5);
        \draw[] (v1) -- ++(-30:2.5);
        \draw[] (v0) -- (v1);
        \draw[] (-2,0.5) -- (2,0.5);
        \draw[] (-2,-0.5) -- (2,-0.5);
        \draw[] (-2,1) -- (2,1);
        \draw[] (-2,-1) -- (2,-1);
        \filldraw[color=black, fill=black!5] (v0) circle[radius=1.5] node[xshift=-1] {$\mathcal{M}$};
        \filldraw[color=black, fill=black!5] (v1) circle[radius=1.5] node[xshift=1] {$\mathcal{M}^\ast$};
        \draw[dashed, orange] (0,2) -- (0,-2);
        \fill[fill=orange, opacity=0.1] (0,2) -- (0,-2) -- (5,-2) -- (5,2);
        \draw[] (0,-3) node {\footnotesize$X$};
        \draw[] (-4,-3) node {\footnotesize$\text{in}$};
        \draw[] (4,-3) node {\footnotesize$\text{out}$};
        \node[scale=0.75,yshift=-2] at (0,1.6) {$\rightarrow$};
        \node[scale=0.75,yshift=0] at (0,-1.6) {$\rightarrow$};
    \end{tikzpicture}
\ee
Only the states in $X$ can cross the dashed line, and the shading to the right of it indicates the complex conjugation. In particular, the two terms on the left-hand side can be interpreted as trivial cuts.
In our pictures, energy always flows from left to right.

The unitarity equation \eqref{eq:unitarity_amps} indicates that scattering amplitudes $\mathcal{M}$ have discontinuities whenever a new unitarity cut starts being kinematically allowed. These are called \emph{normal thresholds} and are associated with singularities of $\M$, e.g., poles, or square-root or logarithmic branch points. Once iterated (by plugging in $T^\dag$ from the left-hand side into the right-hand side), unitarity also implies existence of more complicated singularities, see, e.g., \cite{Hannesdottir:2022bmo}. They are called \emph{anomalous thresholds} or \emph{Landau singularities} \cite{Bjorken:1959fd,Landau:1959fi,10.1143/PTP.22.128} and are associated with the cuts explored in   Sec.~\ref{sec:individual} and \ref{sec:generalized}. Indeed, anomalous thresholds are central objects of those sections.

We emphasize that, since \eqref{eq:unitarity_amps} is a direct consequence of the unitarity of the $S$-matrix, \eqref{eq:unitarity_amps}  is valid fully nonperturbatively. In particular, if $|\textrm{in}\rangle = |\textrm{out}\rangle = |p\rangle$ is a single-particle state for a particle species with on-shell momentum $p$, then \eqref{eq:unitarity_amps} implies that the imaginary part of the two-point function $\mathcal{M}(p \to p)$ is proportional to the total decay width of the particle. Similarly, if $|\textrm{in}\rangle = |\textrm{out}\rangle = |p_1 p_2\rangle$ is a two-particle state, then \eqref{eq:unitarity_amps} reduces to the well-known \emph{optical theorem}, which relates the imaginary part of the forward ($t=0$) scattering amplitude $\mathcal{M}(p_1 p_2 \to p_1 p_2 )$ to the total cross section,
\begin{align}
\sigma_{p_1 p_2 \to X}^{\text{tot}}(s) & = \frac{1}{2 \sqrt{\lambda(s,p_1^2,p_2^2)}} \SumInt_{X} 
\bm{\hat{\delta}}_{p_1 p_2 \text{,X}} |\mathcal{M}(p_1 p_2{\to}X)|^2
\nonumber
\\
& =\frac{\Im\, \mathcal{M}(p_1 p_2 \to p_1 p_2)}{\sqrt{\lambda(s,p_1^2,p_2^2)}}\, ,\label{eq:optical-thm}
\end{align}
where $s=(p_1+p_2)^2$. The K\"all\'en function is
\be \label{eq:Kallen-fn}
\lambda(a,b,c) := a^2 + b^2 + c^2 - 2ab - 2bc - 2ca\, . 
\ee
We will come back to the relation between unitarity cuts and cross section computations in Sec.~\ref{sec:reverse-unitarity}.

\subsubsection{\label{sec:largest-time}Largest time equation}

Unitarity cuts are not independent of each other. One constraint they satisfy descends from the following algebraic identity:
\be\label{eq:T-Tbar-identity}
\sum_{k=0}^{n} (-1)^k \!\!\!\!\! \sum_{\sigma \in P(k,n-k)} \!\!\! \bar{\T} [\mathcal{O}_{\sigma_1} \cdots \mathcal{O}_{\sigma_k} ]\, \T [\mathcal{O}_{\sigma_{k+1}} \cdots \mathcal{O}_{\sigma_n}] = 0\, ,
\ee
where $P(k, n-k)$ is the set of partitions of $n$ labels into two sets of size $k$ and $n-k$. Here, $\T$ and $\bar{\T}$ denote the time and anti-time ordering of the operators $\mathcal{O}_i = \mathcal{O}_i(x_i)$.
After sandwiching \eqref{eq:T-Tbar-identity} between vacuum states $\langle 0|$ and $|0\rangle$, inserting a complete basis of states $\mathbbm{1} = \SumInt_{X} |X \rangle \langle X |$ between $\bar\T$ and $\T$, and applying the LSZ reduction on every term, we obtain the \emph{largest time equation}. It states that the sum over all possible unitarity cuts (including trivial ones) with appropriate $\pm$ signs, vanishes. This statement was already illustrated in \eqref{eq:unitarity-diagram}.

The largest time equation can be also derived in perturbation theory, see Sec.~\ref{sec:largest-time-again}.

\begin{mdexample}
Consider $n=4$. The identity \eqref{eq:T-Tbar-identity} expands to
\begin{gather}
\T[\O_1 \O_2 \O_3 \O_4] + \bar{\T}[\O_1 \O_2 \O_3 \O_4] \nonumber\\
+ \bar{\T}[\O_3 \O_4 ] \T[\O_1 \O_2 ] + \text{(all other partitions)} = 0\, .\label{eq:largest-time-n4}
\end{gather}
Except for the first two, all the remaining terms have support only in one kinematic channel. For example, 
specializing to $s$-channel kinematics (with $p_1^0, p_2^0 > 0$ and $p_3^0,p_4^0<0$), only the first term in the second line of \eqref{eq:largest-time-n4} gives a nonzero contribution. After applying the LSZ procedure and taking $\O_i$ to be operators that create the external states, it evaluates to
\begin{align}
\prod_{i=1}^{4} &\lim_{p_i^2 \to m_i^2} \SumInt_X
\int \prod_{j=1}^{4} \d^\D x_j\, \e^{-i p_j \cdot x_j} (\square_{x_j} {+} m_j^2) \nonumber\\
&\quad\qquad\qquad \langle 0 | \bar{\mathcal{T}}[\O_3 \O_4] | X \rangle \langle X | \mathcal{T}[\O_1 \O_2] | 0 \rangle \nonumber\\
=
&\SumInt_X \bm{\hat{\delta}}_{p_1 p_2, X} \mathcal{M}(p_1 p_2 \to X)\,
\mathcal{M}(p_3 p_4 \to X)^*\, .
\label{eq:LTE_unitarity}
\end{align}
Likewise, the first two terms in \eqref{eq:largest-time-n4} lead to the time-ordered and anti-time-ordered amplitudes $i \M$ and $-i\M^\ast$. The result is an instance of the unitarity equation \eqref{eq:unitarity_amps}.
\end{mdexample}

\subsection{Perturbative unitarity}

The computational power of the unitarity equation \eqref{eq:unitarity_amps} is that it can be applied order-by-order in perturbation theory. In fact, it also applies to individual Feynman integrals. Perturbative unitarity cuts are natural extensions of those we encountered above.
\begin{mddefinition}
\label{def:unitarity_cuts}
For any connected Feynman diagram, 
a unitarity cut is a curve (illustrated by a dashed line) through a subset $C$ of its edges that splits the diagram into two parts. Let us call the (possibly disconnected) set of edges to the left of the cut $C_L$ and likewise $C_R$ for those to the right.
The unitarity cut of the corresponding Feynman integral $\I$ is then defined as
\begin{align} \label{eq:pert-unitarity-cuts}
\UCut_{C}\, \I :=  &\int_{\ell_1, \ldots \ell_{\L}} \frac{\N_L}{\prod_{e \in C_L} (q_e^2 - m_e^2 + i\eps)} \\
&\times \prod_{e \in C} \ddelta^+ (q_e^2 - m_e^2) \frac{\N_R^\ast}{\prod_{e \in C_R} (q_e^2 - m_e^2 - i\eps)}\, , \nonumber
\end{align}
where $\N_L$ and $\N_R$ are the sets of numerators to the left and right of the cut respectively.
\end{mddefinition}
Note that unitarity cuts are in general not analytic and depend on the kinematic channel they are evaluated in. We illustrate this property and the notation on the following examples.
\begin{mdexample} \label{ex:I1m-cuts}
Consider $\I^{1m}$ from Ex.~\ref{ex:I-2m} in the $s$-channel for which $s>0$ and $t<0$.  In this example, we always have $\N=\N_L=\N_R=1$. The unitarity cut through the edges $C=\{1,3\}$, with $C_L=\{2\}$ and $C_R=\{4\}$, (with the edges labelled as in Ex.~\ref{ex:I-2m}) gives:
\begin{align}
    \Cut & ^U_{\{1,3\}} \, \I^{1m} \big\vert^{s>0}_{t<0}\, = \nonumber\begin{gathered}
    \begin{tikzpicture}[line width=1,scale=0.35]
    	\coordinate (v1) at (-1,-1);
        \coordinate (v2) at (-1,1);
        \coordinate (v3) at (1,1);
        \coordinate (v4) at (1,-1);
        \draw[] (v1) -- (v2) -- (v3) -- (v4) -- (v1);
        \draw[] (v1) -- ++(-150:0.7);
        \draw[] (v1) -- ++(-120:0.7) node[left,yshift=-2,scale=0.7] {$p_1$};
        \draw[] (v2) -- ++(135:0.7)
        node[left,scale=0.7] {$p_2$};
        \draw[] (v3) -- ++(45:0.7) node[right,yshift=0,scale=0.7] {$p_3$};
        \draw[] (v4) -- ++(-45:0.7) node[right,scale=0.7,yshift=-3] {$p_4$};
        \draw[dashed, orange] (0,1.4) -- (0,-2);
        \fill[fill=orange, opacity=0.1] (0,2) -- (0,-2) -- (2,-2) -- (2,2);
        \node[scale=0.7] at (0,1.8) {$p_{12}{-}\ell$};
        \node[scale=0.7,yshift=-13.5] at (0,1.6) {$\rightarrow$};
        \node[scale=0.7] at (-0.5,-1.6) {$\ell$};
        \node[scale=0.7,yshift=12] at (0,-1.6) {$\rightarrow$};
    \end{tikzpicture}
    \end{gathered}\\
    &= \int_\ell \frac{\ddelta^+(\ell^2) \ddelta^+ [(p_{12}{-}\ell)^2]}{[(p_1{-}\ell)^2{+}i\varepsilon] [(p_{123}{-}\ell)^2{-}i\varepsilon]}
    \nonumber
    \\
    \label{eq:cutI1m_s}
    & = \frac{4 i}{s t}
    \Big[
    \sin (\pi \epsilon) \Big(\frac{1}{\epsilon^2} - \frac{\pi^2}{12} \Big) s^{-\epsilon} \\
    & \hspace{2.5cm}
    + \pi \log\Big(\frac{p_1^2{+}i\varepsilon-s}{t}\Big)
    \Big]  + \mathcal{O}(\epsilon) \,.\nonumber
\end{align}
Note that the same cut evaluated in the $t$-channel would be zero.

Similarly, the unitarity cut through $C=\{1,2\}$ is nonzero only if $p_1^2 > 0$:
\begin{align}
    &\Cut^U_{\{1,2\}} \I^{1m} \big\vert^{s>0}_{t<0} =
    \begin{gathered}
    \begin{tikzpicture}[line width=1,scale=0.25]
    	\coordinate (v1) at (-1,-1);
        \coordinate (v2) at (1,1);
        \coordinate (v3) at (3,1);
        \coordinate (v4) at (1,-1);
        \draw[] (v1) -- (v2) -- (v3) -- (v4) -- (v1);
        \draw[] (v1) -- ++(-150:0.7);
        \draw[] (v1) -- ++(-120:0.7) node[left,yshift=-2,scale=0.7] {$p_1$};
        \draw[] (v2) -- ++(175:2.7) node[left,scale=0.7] {$p_2$};
        \draw[] (v3) -- ++(30:0.7) node[right,yshift=2,scale=0.7] {$p_3$};
        \draw[] (v4) -- ++(-5:2.7) node[right,scale=0.7] {$p_4$};
        \draw[dashed, orange] (0,2) -- (0,-2);
        \fill[fill=orange, opacity=0.1] (0,2) -- (0,-2) -- (4,-2) -- (4,2);
        \node[scale=0.7,xshift=-23,yshift=-20] at (0,1.8) {$p_{1}{-}\ell$};
        \node[scale=0.7,yshift=-5,rotate=50,xshift=-10] at (0,1.6) {$\rightarrow$};
        \node[scale=0.7,xshift=0] at (-0.5,-1.6) {$\ell$};
        \node[scale=0.7,yshift=10,xshift=3] at (0,-1.6) {$\rightarrow$};
    \end{tikzpicture}
    \end{gathered}
    \nonumber
    \\
    &= \frac{4 i}{s t}
    \Big[
    -\sin (\pi \epsilon) \Big(\frac{1}{\epsilon^2} - \frac{\pi^2}{12} \Big)
    (p_1^2)^{-\epsilon} 
    \\
    & \hspace{1.5cm}
    {+} \pi \log\Big(\frac{s t}{(p_1^2{+}i\varepsilon {-}s)(p_1^2{-}t )}\Big) + \mathcal{O}(\epsilon) 
    \Big] \theta (p_1^2)  \, .
    \nonumber
\end{align}
If $p_1^2 \leqslant 0$, this cut is not kinematically allowed because the bottom-left vertex cannot be realized as an on-shell interaction. We therefore explicitly included the step function $\theta(p_1^2)$ above.  Notice that the computation did not depend on the sign of energy of $p_2$, and therefore the same unitarity cut also contributes in the $p_1^2$-channel kinematics (the $1 \to 234$ decay process), where the energy of $p_2$ has the opposite sign.
Note that when $p_1^2 > 0$, the $i\eps$ in the argument of the logarithm is crucial to obtain the correct branches of the unitarity cut. It also illustrates that individual unitarity cuts can have nonvanishing imaginary parts.
\end{mdexample}

In order to state the perturbative version of the unitarity equation \eqref{eq:unitarity_amps} more concisely, let us introduce the notation for the sum of all kinematically-allowed unitarity cuts in a given physical region:
\be
\UCut_{s_J} \I := \sum_{C \in C_{s_J}} \!\pm\, \UCut_{C} \I\, ,
\ee
where $C_{s_J}$ denotes the set of all such admissible cuts and the sign depends on the number of disconnected components to the left and right of the cut, see, e.g., \cite{Pius:2016jsl} for details. 
Unitarity in the form of \eqref{eq:unitarity_amps} then gives\footnote{Since most of the perturbative computations are done at the level of scalar ``master integrals'', we will now assume CP invariance.}
\be
\label{eq:im-unitarity}
2i \, \Im \, \I \big|_{s_J\text{-channel}} =  \UCut_{s_J} \I\, .
\ee
Note once again that it is important to include all disconnected contributions. We illustrate this formula with the following examples.
\begin{mdexample}
Consider the running example of $\I^{1m}$ in the $s$-channel.
Once again, the value of the imaginary part depends on the sign of $p_1^2$. Using the identity $({-}x{+}i\varepsilon)^{a} - ({-}x{-}i\varepsilon)^{a} = 2 i \sin(\pi a) x^a \theta(x)$, we find
\begin{align}
    2 i \,\Im\, \I^{1m}\big\vert^{s>0}_{p_1^2,t<0}
    &
    = \frac{4 i}{s t}
    \Big[
    \sin (\pi \epsilon) \Big(\frac{1}{\epsilon^2} - \frac{\pi^2}{12} \Big) s^{-\epsilon}
    \nonumber
    \\
    & \hspace{0.8cm}
    + \pi \log\left(\frac{p_1^2-s}{t}\right)
    \Big] + \mathcal{O}(\epsilon) \nonumber\\
    & = \Cut^U_{\{1,3\}} \I^{1m} \big\vert_{t,p_1^2<0}^{s>0} \, ,
\end{align}
where the last step follows from comparing to~\eqref{eq:cutI1m_s} in Ex.~\ref{ex:I1m-cuts}.
Similarly, we have
\begin{align}
    2 i\,\Im &\, \I^{1m}\big\vert^{s,p_1^2>0}_{t<0}
     = \frac{4 i}{s t}
    \Big[
    \sin (\pi \epsilon) \Big(\frac{1}{\epsilon^2} - \frac{\pi^2}{12} \Big) \left( s^{-\epsilon} {-} (p_1^2)^{-\epsilon}\right) 
    \nonumber
    \\ & \hspace{2.5cm}
    + \pi \log\left(\frac{s}{p_1^2-t}\right)
    \Big] + \mathcal{O}(\epsilon) \nonumber\\
 &   = \big(\Cut^U_{\{1,3\}} \I^{1m} + \Cut^U_{\{2,3\}} \I^{1m} \big)\big\vert_{t<0}^{s,p_1^2>0} \,.
\end{align}

Note that in the second case the real parts of the two unitarity cuts cancelled out in such a way that $2 i \,\Im \I^{1m}$ is purely imaginary.
\end{mdexample}

\subsubsection{Proofs}
There are various proofs of \eqref{eq:im-unitarity} at different levels of rigor. The original derivation was sketched by Cutkosky based on unitarity arguments \cite[Sec.~II.C]{Cutkosky:1960sp}, which is why this method for computing imaginary parts is often referred to as the \emph{Cutkosky (or cutting) rules}. 

Another approach is that of the largest time equation \cite{tHooft:1973wag}, further reviewed in Sec.~\ref{sec:largest-time}, which proves \eqref{eq:im-unitarity} as an algebraic identity using the combinatorics of retarded, advanced, and cut propagators. Yet another strategy is to use time-ordered (or old-fashioned) perturbation theory \cite{Sterman:1993hfp}, where one splits every term in \eqref{eq:im-unitarity} into different time orderings of vertices and proves the equality term-by-term. See also \cite{Pius:2016jsl} for a proof using contour manipulations.

Let us explain why \eqref{eq:im-unitarity} is intuitively true. One starts by engineering a Lagrangian that produces a single Feynman diagram at the leading order in the couplings. For example, the bubble diagram from Ex.~\ref{ex:bubble} needs 
\be
{\cal L} = \sum_{i=1}^{6} \tfrac{1}{2}[(\partial \phi_i)^2 - m^2 \phi_i^2] + (g_1 \phi_1 \phi_2 + g_2 \phi_3 \phi_4) \phi_5 \phi_6 \, ,
\ee
such that ${\cal M}(\phi_1 \phi_2 \to \phi_3 \phi_4)$ at the leading order ${\cal O}(g_1 g_2)$ equals to the Feynman diagram from Ex.~\ref{ex:bubble}. Expanding the nonperturbative unitarity equation \eqref{eq:unitarity_amps} order-by-order, the leading term reproduces \eqref{eq:im-unitarity}.

\subsubsection{\label{sec:largest-time-again}The largest time equation in perturbation theory}

The largest time equation can also be shown to hold diagram-by-diagram in perturbation theory~\cite{Veltman:1994wz}. We sketch the derivation for scalar Feynman diagrams for simplicity. We start with writing scalar Feynman propagators in position space,
\begin{equation}
    \Delta_F (x) = \lim_{\eps \to 0^+} \int \frac{\d^\D q}{(2\pi)^\D} \frac{i  \e^{- i q \cdot x}}{q^2-m^2+ i \varepsilon} \,,
\end{equation}
as a sum of positive- and negative-frequency components,
\begin{equation}
    \Delta_F (x) = \theta(x^0) \Delta^+(x) + \theta(-x^0) \Delta^-(x) \,,
    \label{eq:DeltaF}
\end{equation}
where
\begin{equation}
    \Delta^{\pm} (x) = i \int \frac{\d^\D q}{(2\pi)^{\D}} \e^{-i q \cdot x}\, \hat{\delta}^{\pm}(q^2-m^2) \,.
\end{equation}
Similarly, the complex conjugated propagator satisfies $\Delta_F^\ast (x) = \theta(x^0) \Delta^- + \theta(-x^0) \Delta^+(x)$. The largest time equation for the simplest Feynman diagram consisting of just one propagator reads
\begin{equation}
    \Delta_F + \Delta_F^\ast = \Delta^+ + \Delta^- \,,
    \label{eq:LTE_propagator}
\end{equation}
for any $x^0$.
It easily follows from~\eqref{eq:DeltaF} and its conjugate, since if $x^0>0$, $\Delta_F=\Delta^+$ and $\Delta^\ast_F=\Delta^-$, and vice versa if $x^0<0$.

Anticipating a generalization to more complicated diagrams, we associate the different propagators to colorings of vertices,
\begin{align}
\Delta_F(x_2{-}x_1)& = \phantom{+}\!
\begin{gathered}
    \begin{tikzpicture}[baseline={([yshift=2ex]current bounding box.center)}]
    \coordinate (a) at (0,0);
    \coordinate (b) at (1,0);
    \draw[thick] (a) -- (b);
    \draw[thick,fill=white] (a) circle (2pt) node[below] {$x_1$};
    \draw[thick,fill=white] (b) circle (2pt) node[below] {$x_2$};
    \end{tikzpicture}
\end{gathered}\!,
\hspace{0.3cm}
\Delta_F^\ast(x_2{-}x_1)= \phantom{+}\!
\begin{gathered}
    \begin{tikzpicture}[baseline={([yshift=2ex]current bounding box.center)}]
    \coordinate (a) at (0,0);
    \coordinate (b) at (1,0);
    \draw[thick] (a) -- (b);
    \draw[thick,fill=black] (a) circle (2pt) node[below] {$x_1$};
    \draw[thick,fill=black] (b) circle (2pt) node[below] {$x_2$};
    \end{tikzpicture}
\end{gathered}
\nonumber
\\
\Delta^+(x_2{-}x_1) & = -\!
\begin{gathered}
    \begin{tikzpicture}[baseline={([yshift=2ex]current bounding box.center)}]
    \coordinate (a) at (0,0);
    \coordinate (b) at (1,0);
    \draw[thick] (a) -- (b);
    \draw[thick,fill=white] (a) circle (2pt) node[below] {$x_1$};
    \draw[thick,fill=black] (b) circle (2pt) node[below] {$x_2$};
    \end{tikzpicture}
\end{gathered}\!,
\hspace{0.3cm}
\Delta^-(x_2{-}x_1)= -\!
\begin{gathered}
    \begin{tikzpicture}[baseline={([yshift=2ex]current bounding box.center)}]
    \coordinate (a) at (0,0);
    \coordinate (b) at (1,0);
    \draw[thick] (a) -- (b);
    \draw[thick,fill=black] (a) circle (2pt) node[below] {$x_1$};
    \draw[thick,fill=white] (b) circle (2pt) node[below] {$x_2$};
    \end{tikzpicture}
\end{gathered}
\label{eq:colorings}
\end{align}
Following~\cite{Veltman:1994wz}, we use the convention that black vertices come with a minus sign.
Note that energy always flows from white to black vertices. With this notation, the generalization of~\eqref{eq:LTE_propagator} to a Feynman integral $\I$ with multiple propagators and $\mathrm{V}$ vertices becomes
\begin{equation}
    \sum_{\text{colorings } c} \!\!\!\! \I_c = 0 \,,
    \label{eq:LTE}
\end{equation}
where each of the $2^\mathrm{V}$ colorings $c$ corresponds to assigning black or white color to a vertex, following which $\I_c$ is computed by using the rules in \eqref{eq:colorings}.
To show that this equation holds, one first notes that one of the vertices, say $x_i$, must correspond to the one with the largest time, i.e., $x_i^0>x_j^0$ for all $j \neq i$. The diagrams in~\eqref{eq:LTE} then cancel pairwise between those with $x_i$ colored black and those with $x_i$ white. This equation is simply~\eqref{eq:T-Tbar-identity} applied to an individual Feynman diagram. Note that here we have used sign conventions for $\I_c$ consistent with~\cite{Veltman:1994wz}.

To tie~\eqref{eq:LTE} back to the unitarity equation from~\eqref{eq:unitarity_amps}, we note that the terms with all white or all black vertices correspond to the amplitude and complex-conjugated amplitude, respectively. Note that the sets of white and black vertices might be each disconnected. Propagators connecting white and black vertices lie on the unitarity cut. 

\begin{mdexample}\label{ex:bubble_cut}
    The bubble Feynman integral from Ex.~\ref{ex:bubble} admits four distinct colorings of vertices,
    \begin{subequations}
    \begin{align}
    \begin{gathered}
    \begin{tikzpicture}[line width=1,scale=0.35]
	\coordinate (v1) at (-1,0);
    \coordinate (v2) at (1,0);
    \draw[] (v1) [out=55,in=125] to (v2);
    \draw[] (v1) [out=-55,in=-125] to (v2);
    \draw[] (v1) -- ++(-135:0.7) node[left,scale=0.75] {$p_1$};
    \draw[] (v1) -- ++(135:0.7) node[left,scale=0.75] {$p_2$};
    \draw[] (v2) -- ++(45:0.7) node[right,scale=0.75] {$p_3$};
    \draw[] (v2) -- ++(-45:0.7) node[right,scale=0.75] {$p_4$};
    \draw[thick,fill=white] (v1) circle (4pt);
    \draw[thick,fill=white] (v2) circle (4pt);
    \draw[dashed, orange] (1.3,1.4) -- (1.3,-1.4);
    \node[scale=0.75,yshift=0] at (1.3,0.8) {$\rightarrow$};
    \node[scale=0.75,yshift=-1.5] at (1.3,-0.8) {$\rightarrow$};
    \fill[fill=orange, opacity=0.1] (1.3,1.4) -- (1.3,-1.4) -- (2.5,-1.4) -- (2.5,1.4);
    \end{tikzpicture}
    \end{gathered}
    \hspace{0.5cm}
    \begin{gathered}
    \begin{tikzpicture}[line width=1,scale=0.35]
	\coordinate (v1) at (-1,0);
    \coordinate (v2) at (1,0);
    \draw[] (v1) [out=55,in=125] to (v2);
    \draw[] (v1) [out=-55,in=-125] to (v2);
    \draw[] (v1) -- ++(-135:0.7) node[left,scale=0.75] {$p_1$};
    \draw[] (v1) -- ++(135:0.7) node[left,scale=0.75] {$p_2$};
    \draw[] (v2) -- ++(45:0.7) node[right,scale=0.75] {$p_3$};
    \draw[] (v2) -- ++(-45:0.7) node[right,scale=0.75] {$p_4$};
    \draw[thick,fill=black] (v1) circle (4pt);
    \draw[thick,fill=black] (v2) circle (4pt);
    \draw[dashed, orange] (-1.3,1.4) -- (-1.3,-1.4);
    \fill[fill=orange, opacity=0.1] (-1.3,1.4) -- (-1.3,-1.4) -- (2.5,-1.4) -- (2.5,1.4);
    \node[scale=0.75,yshift=0] at (-1.3,0.8) {$\rightarrow$};
    \node[scale=0.75,yshift=-1.5] at (-1.3,-0.8) {$\rightarrow$};
    \end{tikzpicture}
    \end{gathered}
    \\
        \begin{gathered}
    \begin{tikzpicture}[line width=1,scale=0.35]
	\coordinate (v1) at (-1,0);
    \coordinate (v2) at (1,0);
    \draw[] (v1) [out=55,in=125] to (v2);
    \draw[] (v1) [out=-55,in=-125] to (v2);
    \draw[] (v1) -- ++(-135:0.7) node[left,scale=0.75] {$p_1$};
    \draw[] (v1) -- ++(135:0.7) node[left,scale=0.75] {$p_2$};
    \draw[] (v2) -- ++(45:0.7) node[right,scale=0.75] {$p_3$};
    \draw[] (v2) -- ++(-45:0.7) node[right,scale=0.75] {$p_4$};
    \draw[thick,fill=white] (v1) circle (4pt);
    \draw[thick,fill=black] (v2) circle (4pt);
    \draw[dashed, orange] (0,1.4) -- (0,-1.4);
    \fill[fill=orange, opacity=0.1] (0,1.4) -- (0,-1.4) -- (2.5,-1.4) -- (2.5,1.4);
    \node[scale=0.75,yshift=0] at (0,0.8) {$\rightarrow$};
    \node[scale=0.75,yshift=-1.5] at (0,-0.8) {$\rightarrow$};
    \end{tikzpicture}
    \end{gathered}
    \hspace{0.5cm}
    \begin{gathered}
    \begin{tikzpicture}[line width=1,scale=0.35]
	\coordinate (v1) at (1,0.7);
    \coordinate (v2) at (-1,-0.7);
    \draw[] (v1) [out=200,in=70] to (v2);
    \draw[] (v1) [out=-115,in=15] to (v2);
    \draw[] (v1) -- ++(171:2.5) node[left,scale=0.75] {$p_2$};
    \draw[] (v1) -- ++(189:2.5) node[left,scale=0.75] {$p_1$};
    \draw[] (v2) -- ++(9:2.5) node[right,scale=0.75] {$p_3$};
    \draw[] (v2) -- ++(-9:2.5) node[right,scale=0.75] {$p_4$};
    \draw[thick,fill=black] (v1) circle (4pt);
    \draw[thick,fill=white] (v2) circle (4pt);
    \draw[dashed, orange] (0,1.4) -- (0,-1.4);
    \fill[fill=orange, opacity=0.1] (0,1.4) -- (0,-1.4) -- (2.5,-1.4) -- (2.5,1.4);
    \node[scale=0.75,yshift=0] at (0,1.1) {$\rightarrow$};
    \node[scale=0.75,yshift=-1.5] at (0,-1.1) {$\rightarrow$};
    \end{tikzpicture} \,.
    \end{gathered}
    \end{align}
    \end{subequations}
Assuming that $p_1$, $p_2$ are incoming, and $p_3$, $p_4$ are outgoing, and recalling that energy flows from white to black vertices, the bottom-right configuration vanishes by energy conservation. The largest-time equation~\eqref{eq:LTE} says that
\begin{equation}
    \I_{\text{bub}} {-} \I_{\text{bub}}^\ast {-} \int_\ell \hat{\delta}^+ (\ell^2{-}m^2) \hat{\delta}^+ [(p_{12}{-}\ell)^2{-}m^2] = 0\, .
    \label{eq:cut_bubble}
\end{equation}
Note that the extra minus sign between $\I_{\text{bub}}$ and $\I_{\text{bub}}^\ast$ traces its roots to the $i$ in the definition of the loop-momentum phase space in~\eqref{eq:phasespace}. We can recognize the integral as the unitarity cut $\UCut_{C}\, \I$ from Def.~\ref{def:unitarity_cuts} for the bubble Feynman integral.

The unitarity cut of the scalar bubble is simply a phase-space integral for the two cut particles. The result of the integration in $\D=2$ spacetime dimensions is
\begin{equation}\label{eq:bubble-ucut}
    \UCut_{C}\, \I^{\D=2}_{\text{bub}} = \frac{4 \pi i}{\sqrt{s (s-4m^2)}}\,, 
\end{equation}
which agrees with $ 2 i \, \Im\, \I^{\D=2}_{\text{bub}}$ for $s>4m^2$ according to~\eqref{eq:I_bub_int} and thus verifies~\eqref{eq:cut_bubble}.
\end{mdexample}

\subsubsection{Unstable particles}

If the field theory under consideration contains unstable particles (which is typically the case for phenomenologically relevant theories), the unitarity of the $S$-matrix, and consequently the notion of unitarity cuts, is more subtle. A consistent way of defining a unitary $S$-matrix in the presence of unstable particles is to only allow stable particles as asymptotic states, and to replace all propagators by their dressed version, where all 1PI contributions to the two-point function have been resummed~\cite{Veltman:1963th}. This amounts to working with complex propagator poles, which can be implemented, e.g., when working in the so-called \emph{complex mass scheme}~\cite{Denner:1999gp,Denner:2005fg,Denner:2006ic}. Importantly, it can be shown that in this scheme all contributions from unitarity cuts which involve a cut propagator of an unstable particle cancel in the final contribution to the cut of the scattering amplitude~\cite{Veltman:1963th}. For the nonperturbative perspective, see e.g., \cite{IAGOLNITZER2006475}.

\subsubsection{Reverse unitarity}
\label{sec:reverse-unitarity}

The optical theorem~\eqref{eq:optical-thm} allows one to relate total cross sections to the imaginary part of the forward scattering amplitude, which may be written as a sum of Feynman integrals. Since the imaginary parts of Feynman integrals are computable from unitarity cuts, the latter provide an effective tool to compute (total) cross sections in perturbation theory. 

Unitarity cuts, and thus the imaginary parts of Feynman integrals, satisfy the same algebraic relations as the full Feynman integrals since they can be defined using the same integrand but different integration contours.
We will come back to this point in Sec.~\ref{sec:generalized} from a different perspective. In particular, the unitarity cuts satisfy the same linear \emph{integration-by-parts} (IBP) relations~\cite{Tkachov:1981wb,Chetyrkin:1981qh} (note that a diagram vanishes if an absent propagator is cut).
IBP relations are the basis for computing Feynman integrals using the differential equations technique~\cite{Kotikov:1990kg,Kotikov:1991hm,Kotikov:1991pm,Remiddi:1997ny}. Hence, this technique can be applied to compute unitarity cuts relevant to cross section computations. This result, which is known as \emph{reverse unitarity}~\cite{Anastasiou:2002yz,Anastasiou:2003yy}, forms the basis for many analytic state-of-the-art computations for (total) cross sections for collider processes, e.g.~\cite{Anastasiou:2015vya,Mistlberger:2018etf,Duhr:2019kwi,Duhr:2020seh,Duhr:2020sdp,Duhr:2021vwj,Chen:2021isd}.

\subsection{Branch cuts and dispersion relations}

\subsubsection{Mathematical preliminaries}

A multivalued analytic function $f(z)$ is typically introduced as a function that has \emph{branch cuts} in the complex plane, i.e., a curve such that $f(z)$ behaves discontinuously when it is crossed. Multivalued functions and branch cuts in particular arise when integrating a function that has a pole in the complex plane. If the residue at the pole is nonvanishing, the result of the integration will depend on how the integration contour encircles the singularity. The notion of encircling a singularity will be made more precise in Sec.~\ref{sec:individual}.

\begin{mdexample}\label{ex:log_branch}
The prime example of a multivalued function is the (complex) logarithm,
\be
\log(z) = \int_1^z\frac{\d t}{t}\,.
\ee
As long as $z$ is nonnegative, we can integrate along the straight line $[1,z]$. This defines the \emph{principal sheet} of the (complex) logarithm function. For negative $z$, however, the integral along the straight path is ill-defined, because of the pole at $t=0$. As a consequence, the function is discontinuous as we approach the negative real axis from above or below:
\be
\lim_{\delta\to 0^+} \log(z\pm i\delta) = \log|z| \pm i\pi\,,\quad z<0\,.
\ee
We therefore say that $\log z$ has a branch cut extending from $z=0$ to $z=-\infty$ in the complex plane, and $z=0,-\infty$ are called \emph{branch points}.
\end{mdexample}

We would like to stress the following important point: while it is conventional to choose the principal sheet of the logarithm so that the branch cut is along the negative real axis, this choice is completely arbitrary and conventional, and we could have chosen the branch cut to be any other semi-infinite non-self-intersecting curve extending from the origin. In the following, we will always assume that branch cuts for real branch points are located along the real axis.

\begin{mddefinition}
\label{def:discontinuity}
The discontinuity $\Disc_z f(z)$ of a function $f(z)$ across the real axis is defined as the difference
\be\label{eq:Schwarz}
\Disc_z f(z) := \lim_{\delta \to 0^+} \left[ f(z + i\delta) - f(z - i\delta)\right]\, .
\ee
\end{mddefinition}
The discontinuity is nonzero if $f(z)$ has a branch cut on the real line. For example, for the logarithm with the branch cut as in Ex.~\ref{ex:log_branch}, we have
$\Disc_{z} \log z = 2\pi i\,\theta(-z)$.
Note that the discontinuity is in general not an analytic function, e.g.,  $\Disc_{z} \frac{1}{z} = -2\pi i\delta(z)$. For $\Disc_z\log z$ the non-analyticity is apparent from the step function, and $\Disc_z\log z$ is only piecewise analytic. More generally, a discontinuity can be thought of as the difference between the value of a function after analytic continuation to the second sheet of (possibly multiple) branch cuts and the value on the original sheet. For this reason, $\Disc_z f(z)$ may have singularities that do not exist on the first sheet of $f(z)$. Finally, we mention that there is a  connection between the discontinuity and the imaginary part. If $f(z)$ takes real values on some real interval, then Schwarz's reflection principle implies that for real $z$, $f(z-i\delta) = f(z+i\delta)^*$, and so in those cases we find
\be\label{eq:disc_Im}
\Disc_z f(z) = \lim_{\delta\to 0^+}2i\,\Im\,f(z+i\delta)\,.
\ee
For instance, $f(z) = \log(z)$ satisfies this identity because the logarithm is real for $z > 0$, but $f(z) = i \log(z)$ does not.

\begin{mdexample}\label{ex:dilog}
The dilogarithm
\be
\Li_2(z) := - \int_0^z \frac{\log(1-t)}{t} \d t
\ee
on the principal sheet has a logarithmic branch point at $z=1$ with a branch cut going to the right. On the second sheet of this branch cut, it has another logarithmic singularity at $z=0$ which is not present on the first. The analytic structure on the two sheets can be illustrated as, respectively:
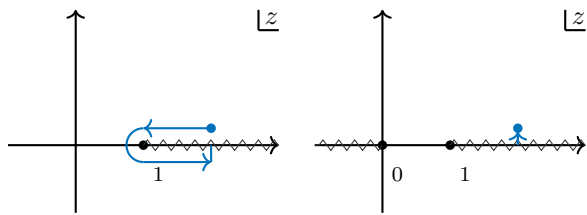
\be\nonumber
    \begin{gathered}
    \begin{tikzpicture}[scale=0.9]
    \draw[->,thick] (-1,0) -- (3,0);
    \draw[->,thick] (0,-1) -- (0,2);
    \draw[thick] (2.7,1.7) -- (3,1.7);
    \draw[thick] (2.7,1.7) -- (2.7,2);
    \draw[thick] (2.65,1.65) node[above right] {$z$};
    \fill[] (1,0) circle[radius=0.07] node[below right, yshift=-0.5em] {\footnotesize$1$};
    \draw[decorate, decoration={zigzag, segment length=6, amplitude=2}, black!80] (1,0) -- (3,0);
    \draw[RoyalBlue, thick, ->] (2,0.25) -- (1,0.25) node[above, yshift=0.5em] {};
    \draw[RoyalBlue, thick] (1,0.25) arc (90:270:0.25);
    \draw[RoyalBlue, thick, <-] (2,-0.25) -- (1,-0.25) node[above, yshift=0.5em] {};
    \draw[RoyalBlue, thick] (2,-0.25) -- (2,0) node[above, yshift=0.5em] {};
    \fill[RoyalBlue] (2,0.25) circle[radius=0.07];
    \end{tikzpicture}
    \end{gathered}
    \quad
    \begin{gathered}
    \begin{tikzpicture}[scale=0.9]
    \draw[->,thick] (-1,0) -- (3,0);
    \draw[->,thick] (0,-1) -- (0,2);
    \draw[thick] (2.7,1.7) -- (3,1.7);
    \draw[thick] (2.7,1.7) -- (2.7,2);
    \draw[thick] (2.65,1.65) node[above right] {$z$};
    \fill[] (0,0) circle[radius=0.07] node[below right, yshift=-0.5em] {\footnotesize$0$};
    \fill[] (1,0) circle[radius=0.07] node[below right, yshift=-0.5em] {\footnotesize$1$};
    \draw[decorate, decoration={zigzag, segment length=6, amplitude=2}, black!80] (1,0) -- (3,0);
    \draw[decorate, decoration={zigzag, segment length=6, amplitude=2}, black!80] (0,0) -- (-1,0);
    \draw[RoyalBlue, thick] (2,0.25) -- (2,0) node[above, yshift=0.5em] {};
    \draw[RoyalBlue, thick, ->] (2,0) -- (2,0.2);
    \fill[RoyalBlue] (2,0.25) circle[radius=0.07];
    \end{tikzpicture}
    \end{gathered}
    \ee
The discontinuity of the dilogarithm is computed as the difference between the blue dots on both sheets (in the limit as $\delta \to 0^+$):
\be\label{eq:DiscLi2}
\Disc_{z} \Li_2(z) = 2\pi i \log(z)\, \theta(z-1)\,.
\ee
 The identity theorem guarantees that its extension from $z > 1$ to the $z$-plane is unique. The result \eqref{eq:DiscLi2} has a branch point at $z=0$, which is a singularity of $\Li_2(z)$ on the second sheet.
\end{mdexample}

It is well known that for multivalued complex functions it is possible to recover the full function from its discontinuity via a \emph{dispersion relation}. To simplify the discussion, we assume that $f(z)$ vanishes at infinity and that it has a branch point at $z_0\geqslant 0$ (and the branch cut extends along the half-line $[z_0,+\infty]$), but no other singularities. Consider the following contour $\Gamma$:
   \be
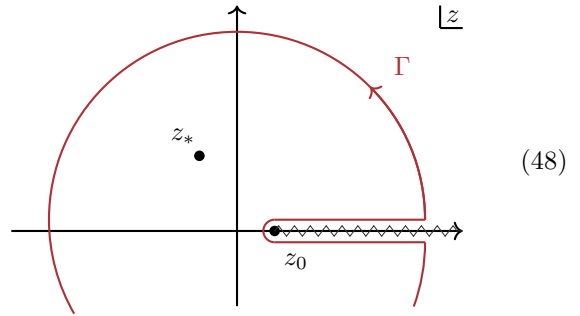
    \begin{gathered}
    \begin{tikzpicture}
    \draw[->,thick] (-3,0) -- (3,0);
    \draw[->,thick] (0,-1) -- (0,3);
    \draw[thick] (2.7,2.7) -- (3,2.7);
    \draw[thick] (2.7,2.7) -- (2.7,3);
    \draw[thick] (2.65,2.65) node[above right] {$z$};
    \coordinate (s0) at (-1,0);
    \coordinate (s1) at (0.5,0);
    \coordinate (sast) at (-0.5,1);
    \draw[] (2.2,2.2) node {$\textcolor{Maroon}{\Gamma}$};
    \fill[] (s1) circle[radius=0.07] node[below right, yshift=-0.5em] {$z_0$};
    \fill[] (sast) circle[radius=0.07] node[above left, yshift=0.1em, xshift=0.2em] {$z_\ast$};
    \draw[decorate, decoration={zigzag, segment length=6, amplitude=2}, black!80] (s1) -- (3,0);
    \draw[Maroon, thick] (2.5,0.15) -- (0.5,0.15);
    \draw[Maroon, thick] (2.5,-0.15) -- (0.5,-0.15);
    \draw[Maroon, thick] (0.5,0.15) arc (90:270:0.15);
    \draw[Maroon, thick] (2.5,0.15) arc (0:210:2.5);
    \draw[Maroon, thick, ->] (2.5,0.15) arc (0:45:2.5);
    \draw[Maroon, thick] (2.5,-0.15) arc (0:-20:2.5);
    \end{tikzpicture}
    \end{gathered}
\ee
Applying the residue theorem, we find:
\be
\frac{1}{2\pi i}\oint_{\Gamma}\frac{\d z}{z-z_*} f(z) = f(z_*)\,.
\ee
Since $f(z)$ vanishes at infinity, the contribution from the (infinitely large) circle vanishes, while the integration along the cuts gives rise to the discontinuity: 
\be
f(z_*) = \frac{1}{2\pi i}\int_{z_0}^{\infty}\frac{\d z}{z-z_*} \Disc_zf(z)\,.
\ee
It is possible to extend this result to the case where $f(z)$ has additional singularities by appropriately subtracting them before performing the contour integral.

\subsubsection{Dispersion relations for amplitudes}

We can apply the previous results to scattering amplitudes and individual Feynman integrals. A detailed discussion is given in \cite{BROS200687}. Here we only give some highlights. In general, establishing the validity of dispersion relations depends on proving analyticity in the complexified kinematics, e.g., the center-of-mass energy squared $z=s$. Further, relating discontinuities to unitarity cuts hinges on the statement \eqref{eq:disc_Im}. The latter requires the existence of a \emph{Euclidean region} in which the amplitude is real and free of branch cuts.

In the simplest case of off-shell two-point functions, dispersion relations are well established nonperturbatively. The following example illustrates that we can also use dispersion relations to compute individual Feynman integrals from their unitarity cuts, cf.~\cite{Veltman:1994wz,Remiddi:1981hn}.
\begin{mdexample}
Consider the bubble diagram from Ex.~\ref{ex:bubble} treated as a $1 \to 1$ process. In this case, the Euclidean region is $s<0$, and so the relation between $\Disc_s$ and the imaginary part from~\eqref{eq:disc_Im} applies.
We have seen in Ex.~\ref{ex:bubble_cut} that the unitarity cut $\UCut_{C}\, \I^{\D=2}_{\text{bub}}$ computes the imaginary part of the bubble integral. Consequently, we also have
\be
\Disc_{s}\I^{\D=2}_{\text{bub}} = \UCut_{C}\, \I^{\D=2}_{\text{bub}}\,.
\ee
Since we can recover a multivalued function from its discontinuity via dispersion relations, we see that we can write the bubble integral as a dispersion integral over its unitarity cut:
\be
\I^{\D=2}_{\text{bub}}(s) = \frac{1}{2\pi i}\int_{4m^2}^\infty\frac{\d s'}{s'-s}\UCut_{C}\I^{\D=2}_{\text{bub}}(s')\,.
\ee
\end{mdexample}

Beyond two-point functions, the best-studied case is that of $2\to 2$ amplitudes $\mathcal{A}(s,t)$ of stable particles in gapped theories. Provided that the momentum transfer $-t$ is sufficiently low compared to the mass gap scale, dispersion relations can be written down in the complex $s$-plane minus the $s$-channel (where $\mathcal{A}(s,t) = \lim_{\delta \to 0^+} \mathcal{A}(s+i\delta,t)$) and $u$-channel (where $\mathcal{A}(s,t) = \lim_{\delta \to 0^+} \mathcal{A}(s-i\delta,t)$) branch cuts \cite{Martin:1969ina}. For general $t<0$, the domain of analyticity established with axiomatic quantum field theory \cite{Bros:1965kbd} is not sufficiently large to write down dispersion relations in the same form nonperturbatively. For recent approaches with applications to bounds on Wilson coefficients, see, e.g., \cite{Adams:2006sv,deRham:2022hpx}. We refer to \cite{BROS200687,BROS2006465} for a more thorough discussion.

Complications arise for higher-point amplitudes or when external unstable particles are involved. For example, it might happen that $\mathcal{A}(s,t) \neq \lim_{\delta \to 0^+} \mathcal{A}(s+i\delta,t)$ in the $s$-channel or that $2i\, \Im\, \mathcal{A} \neq \Disc_s \mathcal{A}$, see \cite{Hannesdottir:2022bmo} for examples.
Dispersion relations for amplitudes with more than four external legs have not been exhaustively studied; see, e.g., \cite{Rubin:1966zz,Stapp:1982mq} for partial results.

\subsection{Double discontinuity}

One can iterate the computation of discontinuities. Let us consider the simplest case of a double discontinuity for $2\to2$ scattering. We can define the \emph{double spectral density}:
\be\label{eq:double-spectral-density}
\rho(s,t) := \lim_{\delta \to 0^+}\!\!\! \sum_{\eta_s, \eta_t \in \{\pm 1\}}\!\! \eta_s\, \eta_t\, \mathcal{A}(s + i\eta_s \delta,\, t + i\eta_t \delta)\, .
\ee
Often, the notation $\Disc_t \Disc_s \mathcal{A}(s,t)$ is used for the same quantity. This formula requires a level of care. It is predicated on the assumption of \emph{maximal analyticity}, which states that $\I(s,t)$ is analytic everywhere in $(s,t) \in \mathbb{C}^2 \setminus \{ s,t,u \geqslant \mathfrak{m}^2\}$, where $\mathfrak{m}$ is the mass of the lightest (multi-particle) exchanged state. In this case, the four terms in \eqref{eq:double-spectral-density} are to be understood as the boundary values of $\I(s,t)$ approached from different directions.

\begin{mdexample}
For the massless box $\I^{0m}$ from Ex.~\ref{ex:I-0m}, only the term proportional to $\log(-s)\log(-t)$ contributes to the double spectral density, and it is given by
\begin{align}
    \rho^{0m}(s,t) & = \frac{2}{s t} \!\lim_{\delta \to 0^+}\!\!\!\! \sum_{\eta_s, \eta_t \in \{\pm 1\}} \!\!\!\!\! \eta_s \eta_t \log({-}s {-} i\eta_s \delta) \log({-}t{-}i\eta_t \delta) \nonumber \\
    & = -\frac{8\pi^2}{s t} \theta(s) \theta(t)\,.
\end{align}
\end{mdexample}

\subsubsection{S-matrix bootstrap}

The double spectral density $\rho(s,t)$ has support in unphysical kinematics, e.g., when $s,t>0$, which can be realized by embedding the external momenta in $(2,2)$ space-time signature. In this case, $\rho(s,t)$ can be related to a generalized version of a cut, but only in the sense of a specific analytic continuation of $\Disc_s \I$ to $t>0$ \cite{Martin:1969ina}.

The \emph{Mandelstam representation} is a natural generalization of dispersion relations to two complex variables, which relates $\I(s,t)$ to two-fold integrals of $\rho(s,t)$ and possibly subtraction terms \cite{Mandelstam:1958xc}. It sets the starting point for the modern nonperturbative \emph{S-matrix bootstrap} program, as it imposes a nonlinear constraint on the scattering amplitude embodying the assumption of maximal analyticity. See \cite{Correia:2020xtr,Kruczenski:2022lot} for reviews.

\subsubsection{Steinmann relations}

For the scattering of stable states in gapped theories, it is known that $\rho(s,t) = 0$ within the physical region. This is the simplest instance of \emph{Steinmann relations}, which state that double discontinuities in partially overlapping channels vanish when evaluated in physical kinematics:
\be
\label{eq:Steinmann-rel}
\Disc_{s_J} \Disc_{s_I} \mathcal{A} \big|_{\text{physical}} = 0
\ee
if the sets $I, J$ are partially overlapping: neither disjoint nor contained in one another \cite{Steinmann1960a,Steinmann1960b}. This relation is trivial for $2 \to 2$ scattering, since only one out of $s$, $t$, $u$ can be timelike and have a nonzero discontinuity.
Steinmann relations may not hold in theories with massless particles, as illustrated in examples below.

\begin{mdexample}\label{ex:I-2m-steinmann}
    We can interpret the diagram $\mathcal{I}^{2m}$ with $s>0$ and $t>0$ from Ex.~\ref{ex:I-2m} as a $2 \to 4$ scattering process, if $p_1^2$ is large. Then, $s$ and $t$ are partially-overlapping physical kinematic channels, so we can check~\eqref{eq:Steinmann-rel}. The only terms that can potentially lead to a double discontinuity are
    \begin{equation}
        \I^{2m} \supset \frac{1}{s t - p_1^2 p_3^2} \left[ 2 \Li_2 \left(1-\frac{p_1^2 p_3^2}{st } \right) - \log^2 \left( \frac{s}{t} \right) \right] \,.
    \end{equation}
    We have to perform an analytic continuation of the expression given in~\eqref{eq:I2m_full}, making sure that the branch cuts of the resulting expression are at $s>0$ and $t>0$.
    
    After doing so, we find that in physical-kinematic regions,
    \begin{equation}\label{eq:I-2m-steinmann}
        \left. \Disc_s \Disc_t \I^{2m} \right\vert_{\text{physical}} = 0\,,
    \end{equation}
    in accordance with~\eqref{eq:Steinmann-rel}.
    Even though the Steinmann relations are not proven for this diagram since massless particles are involved, they happen to work in this case.
    
\end{mdexample}

\begin{mdexample}\label{ex:Steinmann-I1m}
The Feynman integral $\mathcal{I}^{1m}$ from Ex.~\ref{ex:I-2m} can be made into a $2 \to 3$ scalar scattering process where $p_1$ is the sum of the momenta of the two incoming particles, and we take $p_1^2>s>0$,  $p_1^2>t>0$ and $s+t>p_1^2$ for the kinematics to be physical. For such a configuration, the Steinmann relations from~\eqref{eq:Steinmann-rel} do not hold, since
\begin{equation}
    \left. \Disc_s \Disc_t \, \mathcal{I}^{1m} \right\vert_{p_1^2>s,t>0} = - \frac{8 \pi^2}{s t} \,.
\end{equation}
We computed this expression by noticing that the term proportional to $\log(s) \log(t)$ in~\eqref{eq:I2m_full} is the only one that contributes to the double discontinuity in $s$ and $t$, and after analytic continuation to $s>0$ and $t>0$, it becomes $\log(-s) \log(-t)$.
Thus, the double discontinuity in the partially overlapping channels $s$ and $t$ does \emph{not} vanish, even though $s>0,t>0$ correspond to a physical-kinematic channel. The breakdown of the Steinmann relations can be traced to the presence of massless particles.
\end{mdexample}

\subsection{Other applications}

So far we have discussed unitarity cuts primarily in the context of quantum field theory and high-energy physics. Unitarity cuts also appear in many other applications throughout physics. Here, we collect some of them.

\subsubsection{Gravitational radiation}

The expectation value $\langle \text{in} | {\cal O} | \text{in} \rangle$ of an operator ${\cal O}$ in the in-state $|\text{in}\rangle$ can be computed using quantum field theory techniques.
A notable application is that of gravitational Bremsstrahlung emitted in a scattering event of two heavy bodies in the classical limit~\cite{Kosower:2018adc}, where ${\cal O}$ (say, the Riemann tensor) is an outgoing mode sourced by a graviton field and $|\text{in}\rangle = |p_1 p_2\rangle$. After inserting a complete basis of out-states $X$ before ${\cal O}$, the computation involves unitarity cuts:
\be
\langle \text{in}' | {\cal O} | \text{in} \rangle =
\SumInt_X 
    \begin{tikzpicture}[line width=1,scale=0.25,baseline={([yshift=0.0ex]current bounding box.center)}]
    	\coordinate (v0) at (-2,0);
        \coordinate (v1) at (2,0);
        \draw[RoyalBlue] (v0) -- ++(-150:2.5);
        \draw[Maroon] (v0) -- ++(-210:2.5);
        \draw[Maroon] (v1) -- ++(30:2.5);
        \draw[RoyalBlue] (v1) -- ++(-30:2.5);
        \draw[vector] (v0) -- (v1);
        \draw[Maroon] (-2,0.5) -- (2,0.5);
        \draw[RoyalBlue] (-2,-0.5) -- (2,-0.5);
        \draw[vector] (0,1.5) -- (v0);
        \filldraw[color=black, fill=black!5] (v0) circle[radius=1.5] node[xshift=-1] {$S$};
        \filldraw[color=black, fill=black!5] (v1) circle[radius=1.5] node[xshift=1] {$S^\ast$};
        \draw[dashed, orange] (0,2) -- (0,-2);
        \fill[fill=orange, opacity=0.1] (0,2) -- (0,-2) -- (5,-2) -- (5,2);
        \draw[] (0,-3) node {\footnotesize$X$};
        \draw[] (-0.5,2.5) node {\footnotesize$\mathcal{O}$};
        \draw[] (-4,-3) node {\footnotesize$\text{in}$};
        \draw[] (4,-3) node {\footnotesize$\text{in}'$};
        \node[scale=0.75,yshift=-3] at (0,-1.2) {$\rightarrow$};
    \end{tikzpicture}\,.
\ee
Here, $S$ is the full $S$-matrix operator including the identity. The two heavy bodies are represented by solid, colored lines and the graviton by a wavy line. The sum over $X$ involves the two bodies themselves, but also gravitational radiation. Therefore, cut-based techniques are useful even in classical physics.

\subsubsection{Thermal physics}

At finite temperature or in out-of-equilibrium systems one is interested in computing thermal correlations functions, often evaluated in the Schwinger--Keldysh formalism \cite{Schwinger:1960qe,Keldysh:1964ud}, where operators are distributed among two time-folds, one running forward and the other one backwards. One encounters correlations functions involving mixed time orderings of the type $\bar{\mathcal{T}}[\cdots] \mathcal{T}[\cdots]$ encountered in Sec.~\ref{sec:largest-time} and hence also generalized versions of unitarity cuts. We refer to \cite{Chou:1984es} for details. Conversely, Schwinger--Keldysh techniques can be also used to study scattering amplitudes and other quantities at zero temperature \cite{Gelis:2015kya,Caron-Huot:2023vxl}.

\subsubsection{Cosmology and (anti-)de Sitter space}

A number of generalizations of cuts have been introduced in curved backgrounds. For example, one can extend the optical theorem and the largest-time equation to (anti-)de Sitter space correlation functions and use them practically on Witten diagrams \cite{Meltzer:2020qbr}. Different versions of cutting rules extending those from time-ordered perturbation theory \cite{Sterman:1993hfp} can be applied in cosmological backgrounds. See e.g.~\cite{Baumann:2022jpr} for a review.

\subsubsection{String theory}

Unitarity cuts in superstring theory require separate treatment, because the discussion so far relied on the existence of local operators, while strings interact nonlocally. There have been two approaches to this problem. In the context of string field theory, where amplitudes are sums of infinitely many Feynman diagrams, Cutkosky rules can be derived in $\D>4$ with a careful treatment of loop-momentum contours \cite{Pius:2016jsl}. In the worldsheet formulation, unitarity cuts are seen as a modified contour prescription on the moduli space of Riemann surfaces and computed practically at one-loop level \cite{Eberhardt:2022zay}.

\section{\label{sec:individual}Cuts and monodromies}

\subsection{Encircling singularities}

In the previous section, we related unitarity cuts, imaginary parts, and total discontinuities. The total discontinuity corresponds to simultaneously crossing multiple branch cuts in the kinematic space. Using unitarity, we related this discontinuity to a sum over all allowed unitarity cuts in physical kinematics. In this section, we will instead ask a question from a more mathematical point of view: what happens if we analytically continue around an \emph{individual} branch point in the complexified kinematic space?

\begin{mdexample}
\label{ex:log-roots}
    Let us consider the following function of two complex variables $z_1$ and $z_2$:
    \begin{align}\label{eq:log}
        g(z_1,z_2) :&= \int_0^\infty \!\!\!\!\frac{\d t}{(t-z_1)(t-z_2)}\\ &= \frac{\log(-z_2)- \log(-z_1)}{z_1-z_2}\, .\nonumber
    \end{align}
    The integrand in the complex $t$ plane can be represented as
    \be
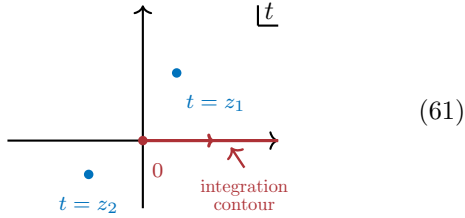
\begin{gathered}
    \begin{tikzpicture}[scale=0.9]
    \draw[->,thick] (-2,0) -- (2,0);
    \draw[->,thick] (0,-1) -- (0,2);
    \draw[thick] (1.7,1.7) -- (2,1.7);
    \draw[thick] (1.7,1.7) -- (1.7,2);
    \draw[thick] (1.65,1.65) node[above right] {$t$};
    \fill[Maroon] (0,0) circle[radius=0.07] node[below right, yshift=-0.5em] {\footnotesize$0$};
    \draw[line width=1.2,Maroon] (0,0) -- (2,0);
    \fill[RoyalBlue] (0.5,1) circle[radius=0.07] node[below right, yshift=-0.5em] {\footnotesize$t=z_1$};
    \fill[RoyalBlue] (-0.8,-0.5) circle[radius=0.07] node[below, yshift=-0.5em] {\footnotesize$t=z_2$};
    \draw[->, line width=0.9, Maroon] (1.5,-0.4) -- (1.3,-0.1) node[below, midway, yshift=-0.3em] {};
    \node[Maroon, scale=0.7, align=center] at (1.5,-0.8) {integration\\[-0.5em] contour}; 
    \draw[Maroon, line width=1.2, -{>[scale=0.6]}] ($(1,0)+(0,0)$) -- ($(1,0)+(0.05,0)$);
    \end{tikzpicture}
\end{gathered}
\ee
    where the two simple poles of the integrand are indicated with blue dots.
    The integral representation of~\eqref{eq:log} is ambiguous when the denominator goes to zero, which happens when $t=z_1$ or $t=z_2$ for $z_1\geq 0$ and $z_2\geq 0$, i.e., when either of the two roots approaches the integration contour. In such situations, Cauchy's theorem can be used to deform the contour to avoid the singularities at $t = z_1$ and $t=z_2$. Thus, the function $g(z_1,z_2)$ has \emph{branch cuts} for $z_1\geq 0$ and $z_2\geq 0$. Deforming the integration contour to avoid the singularities amounts to an analytic continuation of the integral in~\eqref{eq:log}, which effectively moves away the branch cuts.
    For example, deforming the contour $t \to t \e^{-i\eps t}$ makes the integral well-defined directly for positive $z_1$ and $z_2$ since the branch cuts moved to slightly below the positive $z_1$ and $z_2$ axes.
    More generally, the choice of the integration contour translates to the choice of how branch cuts of the resulting integral are placed.

    However, when one of the singular points of the denominator, $t=z_1$ or $t=z_2$, coincides with the endpoints of the integral at $t=0$ or $t\to \infty$, one cannot use a contour deformation to avoid the denominator becoming zero on the integration contour. Thus, these are places of potential \emph{singularities} of the integral. In this example, the values $z_1=0,\infty$ and $z_2=0,\infty$ are precisely the singular points of the functions $\log(-z_1)$ and $\log(-z_2)$.

    The denominator of the function in~\eqref{eq:log} has an additional singularity when $z_1=z_2$. However, on the principal sheets of the logarithms, the numerator cancels the pole to render the function finite at this point. This is generically not the case on other sheets. One example is when one of the roots, say $z_1$, approaches a real and positive value from the upper half-plane, while the other root $z_2$ approaches the same value for the lower half-plane. The limit in that case is $\lim_{z_1\to z_2} g(z_1,z_2) = \frac{2\pi i}{z_1-z_2}$, so with this approach, $g(z_1,z_2)$ is singular at $z_1=z_2$. What happens is that the integration contour is pinched between the two roots, so it cannot be deformed to simultaneously avoid both of them:
        \be
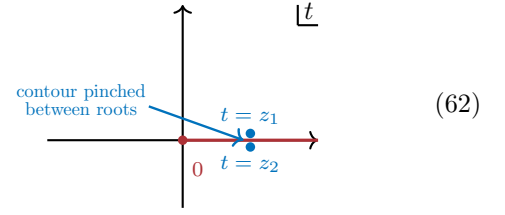
\label{fig:example-roots-pinch}
\begin{gathered}
    \begin{tikzpicture}[scale=0.9]
    \draw[->,thick] (-2,0) -- (2,0);
    \draw[->,thick] (0,-1) -- (0,2);
    \draw[thick] (1.7,1.7) -- (2,1.7);
    \draw[thick] (1.7,1.7) -- (1.7,2);
    \draw[thick] (1.65,1.65) node[above right] {$t$};
    \fill[Maroon] (0,0) circle[radius=0.07] node[below right, yshift=-0.5em] {\footnotesize$0$};
    \draw[line width=1.2,Maroon] (0,0) -- (2,0);
    \fill[RoyalBlue] (1,0.10) circle[radius=0.07] node[above] {\footnotesize$t=z_1$};
    \fill[RoyalBlue] (1,-0.10) circle[radius=0.07] node[below] {\footnotesize$t=z_2$};
    \node[RoyalBlue, scale=0.7, align=center] at (-1.5,0.6) {contour pinched\\[-0.5em] between roots}; 
    \draw[->, line width=0.9, RoyalBlue] (-0.5,0.5) -- (0.9,0) node[below, midway, yshift=-0.3em] {};
    \end{tikzpicture}
\end{gathered}
\ee
This is the simplest example of a \emph{pinch singularity} in one complex dimension.
\end{mdexample}

Let us study singularity of an analytic function at a given codimension-$1$ locus $L = 0$ in a multidimensional kinematic space. Singularity is any place where the function is ill-defined: it could be a pole or a branch point. We pick a one-dimensional linear subspace parametrized by a complex variable $z$, such that it intersects $L = 0$ transversely at $z=0$. We write the function as $f(z;w)$ where $w$ represents all other arguments.

Analytically continuing around a singularity is made precise with a notion of monodromy.
\begin{mddefinition}\label{def:monodromy}
    The \emph{monodromy} (also called \emph{monodromy variation}) $\Delta_{L=0} f(z_0; w)$ of a function $f(z; w)$ around $L = 0$, as described above,  is defined as the difference
    \be\label{eq:mon}
    \disc_{L=0} f(z_0; w) := f(z_0; w) -  f(z_0 + \gamma; w) \, ,
    \ee
    where $f(z_0 + \gamma; w)$ is the result of analytically continuing $f(z_0;w)$ along a path $\gamma$ in the complex $z$-plane, starting at the base point $z=z_0$:
    \be
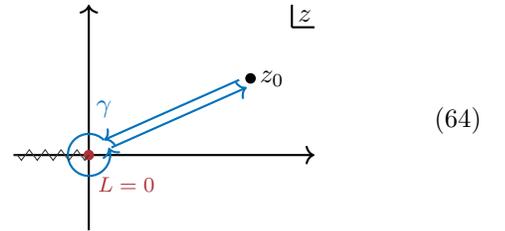
    \begin{gathered}
    \begin{tikzpicture}
    \draw[->,thick] (-1,0) -- (3,0);
    \draw[->,thick] (0,-1) -- (0,2);
    \draw[thick] (2.7,1.7) -- (3,1.7);
    \draw[thick] (2.7,1.7) -- (2.7,2);
    \draw[thick] (2.65,1.65) node[above right] {$z$};
    \fill[Maroon] (0,0) circle[radius=0.07] node[below right, yshift=-0.5em] {\footnotesize$L=0$};
    \draw[RoyalBlue, thick, ->] (2,1) -- (0.2,0.2) node[above, yshift=0.5em] {$\gamma$};
    \draw[RoyalBlue, thick, ->] (0.3,0.1) -- (2.1,0.9);
    \fill[] (2.15,1.02) circle [radius=0.07] node [right] {$z_0$};
    \draw[RoyalBlue, thick, ->] (0.2,0.2) arc (45:380:0.28);
    \draw[decorate, decoration={zigzag, segment length=6, amplitude=2}, black!80] (0,0) -- (-1,0);
    \end{tikzpicture}
    \end{gathered}
    \ee
    which goes from $z_0$ to a neighborhood of the origin, along an infinitesimal counterclockwise circle, and goes back to $z_0$, in such a way that no other singularities of $f(z;w)$ are enclosed.

    If \eqref{eq:mon} is nonzero, the function $f(z;w)$ is said to have a \emph{branch point} at $z = 0$. The value of the monodromy is independent of the choice of branch cuts.
    \label{def:cut-pm}
\end{mddefinition}

The action of $\Delta_{L=0}$ is often called \emph{variation} or simply \emph{discontinuity} around $L=0$, but we use the term \emph{monodromy} to distinguish this action from $\Disc$ specified in Def.~\ref{def:discontinuity}.

\begin{mdexample}
    We can compute the monodromy around the branch point at $z_1=0$ of the function $g(z_1,z_2)$ from Ex.~\ref{ex:log-roots} by tracking the root at $t=z_1$ as $z_1$ encircles 0 in an counterclockwise direction. As the root in the $t$ complex plane encircles $t=0$, the integration contour needs to be deformed accordingly:
    \be
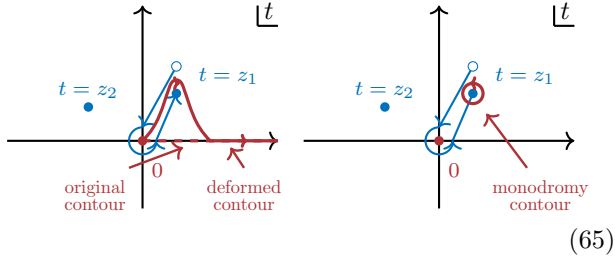
    \begin{gathered}
    \begin{tikzpicture}[scale=0.9]
    \draw[->,thick] (-2,0) -- (2,0);
    \draw[->,thick] (0,-1) -- (0,2);
    \draw[thick] (1.7,1.7) -- (2,1.7);
    \draw[thick] (1.7,1.7) -- (1.7,2);
    \draw[thick] (1.65,1.65) node[above right] {$t$};
    \fill[Maroon] (0,0) circle[radius=0.07] node[below right, yshift=-0.5em] {\footnotesize$0$};
    \draw[line width=1.2,Maroon, dashed] (0,0) -- (2,0);
    \draw[->, line width=0.9, Maroon] (1.5,-0.4) -- (1.3,-0.1) node[below, midway, yshift=-0.3em] {};
    \node[Maroon, scale=0.7, align=center] at (1.5,-0.8) {deformed\\[-0.5em] contour}; 
    \draw[->, line width=0.9, Maroon] (-0.2,-0.4) -- (0.6,-0.1) node[below, midway, yshift=-0.3em] {};
    \node[Maroon, scale=0.7, align=center] at (-0.7,-0.8) {original\\[-0.5em] contour}; 
    \coordinate (start) at (0.5,1.1); 
    \coordinate (end) at (0.5,0.7); 
    \draw[RoyalBlue, thick, ->] (start) -- (0,0.2);
    \draw[RoyalBlue, thick, ->] (0.2,0) -- (end);
    \draw[RoyalBlue, thick, ->] (0,0.2) arc (90:360:0.2);
    \fill[RoyalBlue] (-0.8,0.5) circle[radius=0.07] node[above] {\footnotesize$t=z_2$};
    \fill[RoyalBlue] (end) circle[radius=0.07] node[above right,xshift=5] {\footnotesize$t=z_1$};
    \draw[RoyalBlue,fill=white] (start) circle[radius=0.07];
    \coordinate (start2) at (0,0); 
    \coordinate (end2) at (2,0); 
    \coordinate (p1) at (0.4,0.8);      
    \coordinate (p2) at (0.5,0.9);     
    \coordinate (p3) at (0.6,0.8);     
    \coordinate (p4) at (1,0);     
    \draw[line width=1.2, Maroon] 
        (start2) to[out=45, in=-110] 
        (p1) to[out=70, in=180] 
        (p2) to[out=0, in=110] 
        (p3) to[out=-70, in=135] 
        (p4) to[out=0, in=180]
        (end2);
    \draw[Maroon, line width=1.2, -{>[scale=0.6]}] ($(end)+(0,0.15)$) -- ($(end)+(0.05,0.15)$);
    \draw[Maroon, line width=1.2, -{>[scale=0.6]}] ($(1.5,0)+(0,0)$) -- ($(1.5,0)+(0.05,0)$);
    \end{tikzpicture}
    \hspace{0.2cm}
        \begin{tikzpicture}[scale=0.9]
    \draw[->,thick] (-2,0) -- (2,0);
    \draw[->,thick] (0,-1) -- (0,2);
    \draw[thick] (1.7,1.7) -- (2,1.7);
    \draw[thick] (1.7,1.7) -- (1.7,2);
    \draw[thick] (1.65,1.65) node[above right] {$t$};
    \fill[Maroon] (0,0) circle[radius=0.07] node[below right, yshift=-0.5em] {\footnotesize$0$};
    \draw[->, line width=0.9, Maroon] (1.3,-0.4) -- (0.7,0.4) node[below, midway, yshift=-0.3em] {};
    \node[Maroon, scale=0.7, align=center] at (1.5,-0.8) {monodromy\\[-0.5em] contour}; 
    \coordinate (start) at (0.5,1.1); 
    \coordinate (end) at (0.5,0.7); 
    \draw[RoyalBlue, thick, ->] (start) -- (0,0.2);
    \draw[RoyalBlue, thick] (0.2,0) -- (end);
    \draw[RoyalBlue, thick, ->] (0,0.2) arc (90:360:0.2);
    \fill[RoyalBlue] (-0.8,0.5) circle[radius=0.07] node[above] {\footnotesize$t=z_2$};
    \fill[RoyalBlue] (end) circle[radius=0.07] node[above right,xshift=5] {\footnotesize$t=z_1$};
    \draw[RoyalBlue,fill=white] (start) circle[radius=0.07];
    \draw[Maroon, line width=1.2] ($(end)+(0,0.15)$) arc (90:450:0.15);
    \draw[Maroon, line width=1.2, -{<[scale=0.6]}] ($(end)+(0,0.15)$) -- ($(end)+(0.05,0.15)$);
    \end{tikzpicture}
    \end{gathered}
    \ee
    As a result, the difference between the original integration contour and the one obtained after analytic continuation is a contour encircling the singularity at $t=z_1$, as depicted above. As a result, the monodromy is proportional to the residue around $t=z_1$,
    \begin{equation}
        \Delta_{z_1=0} g(z_1,z_2) = 2 \pi i \, \text{Res}_{t=z_1} \frac{1}{(t-z_1)(t-z_2)} \!=\! \frac{2 \pi i}{z_1-z_2} . 
    \end{equation}
    This result can also be obtained using the functional form of the right-hand side of~\eqref{eq:log}, recognizing that a monodromy around the branch point of a logarithm is equal to $2 \pi i$.
\end{mdexample}

It is worth illustrating the distinction between the imaginary part, discontinuity, and monodromy on a simple example.

\begin{mdexample}
Consider the function
\be
f(z) = i \log(-z) \sqrt{1-z}\, .
\ee
Its analytic structure in the $z$-plane is as follows:
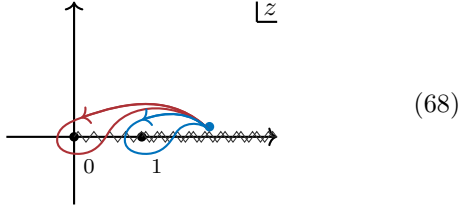
\be\label{fig:example-mon}
\begin{gathered}
    \begin{tikzpicture}[scale=0.9]
    \draw[->,thick] (-1,0) -- (3,0);
    \draw[->,thick] (0,-1) -- (0,2);
    \draw[thick] (2.7,1.7) -- (3,1.7);
    \draw[thick] (2.7,1.7) -- (2.7,2);
    \draw[thick] (2.65,1.65) node[above right] {$z$};
    \fill[] (0,0) circle[radius=0.07] node[below right, yshift=-0.5em] {\footnotesize$0$};
    \fill[] (1,0) circle[radius=0.07] node[below right, yshift=-0.5em] {\footnotesize$1$};
    \draw[decorate, decoration={zigzag, segment length=6, amplitude=2}, black!80] (0,0) -- (3,0);
    \draw[decorate, decoration={zigzag, segment length=6, amplitude=2}, black!80] (1,0) -- (3,0);
    \fill[RoyalBlue] (2,0.15) circle[radius=0.07];
    \coordinate (start) at (2,0.1); 
    \coordinate (p1) at (0.1,0.28);      
    \coordinate (p2) at (-0.25,-0.05);     
    \coordinate (pp2) at (0.1,-0.25);     
    \coordinate (p3) at (0.5,0.05);     
    \draw[thick, Maroon] 
        (start) to[out=135, in=20] 
        (p1) to[out=200, in=90] 
        (p2) to[out=270, in=-175] 
        (pp2) to[out=5, in=-120]
        (p3) to[out=60, in=-225] 
        (start);
    \draw[thick,->,Maroon] (start) to[out=135, in=20] (p1);
    \draw[thick, RoyalBlue] 
        (2,0.1) to[out=135, in=20] 
        (1,0.25) to[out=200, in=90] 
        (0.75,-0.05) to[out=270, in=-175] 
        (1.1,-0.25) to[out=5, in=-120]
        (1.5,0.05) to[out=60, in=-225] 
        (2,0.1);
    \draw[thick,->,RoyalBlue] (start) to[out=135, in=20] (1,0.25);
    \end{tikzpicture}
\end{gathered}
\ee

We have
\begin{subequations}
   \begin{align}
   \Im\, f(z) &= \begin{dcases} \log(|z|) \sqrt{1-z} &\text{if}\quad z \leqslant 1\, ,\\
   - \pi \sqrt{z-1} &\text{if}\quad z>1\,,
   \end{dcases}\\
   \Disc_z f(z) &= \begin{dcases}
   0 &\text{if}\quad z\leqslant 0\, ,\\
   2\pi \sqrt{1-z} &\text{if}\quad 0 < z \leqslant 1\, ,\\
   2  \log(z) \sqrt{z-1} &\text{if}\quad z>1\,,
   \end{dcases}\\
   \Delta_{z=0} f(z) &= 2\pi \sqrt{1-z}\, ,\\
   \Delta_{z=1} f(z) &= 2 i \log(-z) \sqrt{1-z}\, .
   \end{align}
\end{subequations}
The two monodromies are computed using the red and blue contours in \eqref{fig:example-mon}. Note that $\Disc_z f(z) = \Delta_{z=0} f(z)$ for $0 < z \leqslant 1$, because they are computed using the same contour.
Moreover, $\disc_{z=z_\ast} f(z)$ vanishes for any point $z_\ast \in \C \setminus \{0,1\}$.
\end{mdexample}

The monodromy is an analytic function, while the discontinuity is only piecewise analytic, i.e., its value jumps as the discontinuity contour encloses new singularities.

It is worth asking how to detect singularities of Feynman integrals in the first place, without direct computation. It is clear that they can arise only if something violent happens to the analytic properties of the \emph{integrand}, e.g., when some of its poles collide (see Ex.~\ref{ex:log-roots}) or when one of them degenerates from a hyperboloid to a lightcone. More generally, singularities can occur for special values of the kinematics for which the pole locus of the integrand changes topology.

    Let us apply this logic to Feynman integrals $\I$ from \eqref{eq:FeynInt-def}. In fact, the combinatorics of pole degenerations can be streamlined by introducing Schwinger (or Feynman) parameters $\alpha_e$:
    \begin{equation}\label{eq:I-mixed}
        \I = \int_{\ell_1, \ldots, \ell_\L} \int_0^\infty \frac{\d^\E \alpha}{\text{GL}(1)} \frac{\N \,  \Gamma(\E) }{\left[\sum_{e=1}^{\E} \alpha_e (q_e^2 - m_e^2 + i\varepsilon)\right]^{\E}} \,.
    \end{equation}
    The invariance under $\text{GL}(1)$ rescalings ($\alpha_e \to \lambda \alpha_e$ for any $\lambda>0$) is commonly used to fix $\sum_{e=1}^\E \alpha_e=1$.

    Were it not for the insertion of $i\eps$, the integral $\I$ would be ambiguous whenever its denominator goes to zero on the integration contour. Points in the external kinematic space for which this happens correspond to branch cuts (see Ex.~\ref{ex:log-roots}).
    At such points, the $i \varepsilon$ prescription is needed to resolve which side of the branch cut the integral is evaluated on. While branch cuts of analytic functions can be placed arbitrarily without changing the function, the integral representation \eqref{eq:I-mixed}
    selects how such cuts are placed in the kinematic space. This defines the \emph{physical sheet} of $\I$.
    
    The integral $\I$ may be genuinely singular when the denominator vanishes and, in addition, all the integration variables $\alpha_e$ and $\ell_{a}$ are either at the endpoint of the integration contour ($\alpha_e = 0$), or the contour is pinched (all derivatives $\frac{\partial}{\partial \alpha_e}$ and $\frac{\partial}{\partial \ell_a}$ of the denominator in \eqref{eq:I-mixed} vanish). These conditions can be summarized in the textbook form of \emph{Landau equations} \cite{Bjorken:1959fd,Landau:1959fi,10.1143/PTP.22.128}:
    \begin{subequations}\label{eq:Landau-eqs}
    \begin{align}
        \alpha_e (q_e^2-m_e^2)& =0 \quad && \text{for all } e = 1,2, \ldots, \E \,,
        \label{eq:Landau-eqs-1}
        \\
        \sum_{e \in \text{loop } a} \!\!\pm\, \alpha_e q_e & = 0 \quad && \text{for all } a = 1,2, \ldots, \L \,.
        \label{eq:Landau-eqs-2}
    \end{align}
    \end{subequations}
    In the last equation, the sum goes over all edges $e$ forming a given loop $a$, and 
    the sign in front of $\alpha_e q_e$ is $+$ if $q_e$ is oriented in the same direction as $\ell_a$ and $-$ otherwise.

The system \eqref{eq:Landau-eqs} splits into different families depending on which subset of edges $\mathcal{E}$ satisfies $\alpha_e = 0$ for $e \in \mathcal{E}$. Each such family corresponds to the \emph{reduced diagram} obtained from the original Feynman diagram by contracting all the edges in $\mathcal{E}$ and putting the surviving propagators on their mass shell.  

In general, the equations \eqref{eq:Landau-eqs} impose conditions on the external kinematics. The most interesting case, though not the only one, is that in which they yield a codimension-$1$ variety $L=0$ (UV and IR divergences would show up as codimension-$0$ solutions). Likewise, the solution set in the internal $(\alpha_e, \ell_a)$ variables might be isolated, but it could also form continuous manifolds.

The Landau equations capture a subset of possible singularities of massive Feynman integrals that occur at finite values of the loop momenta. Singularities at infinite loop momenta are sometimes referred to as \emph{second and mixed-type} singularities~\cite{Cutkosky:1960sp,doi:10.1063/1.1724262}. In the presence of massless particles, the Landau equations \eqref{eq:Landau-eqs} must be treated more carefully since they mix kinematic singularities with infrared divergences. In a more careful treatment, one needs to consider the cases in which $\alpha_e$ and $q_e^2 - m_e^2$ in \eqref{eq:Landau-eqs-1} vanish, but at different relative rates \cite{Fevola:2023kaw}. In the mathematical language, one uses a compactification of the loop (and/or Schwinger parameter) integration domain. For related geometric constructions see \cite{AIHPA_1967__6_2_89_0,Brown:2009ta}.

\begin{mddefinition}
    The original integration contour of $\I$ passes through the solutions to the Landau equations~\eqref{eq:Landau-eqs} for which 
    \begin{equation}
        \alpha_e \geqslant  0 \, \text{ and } \ell_a \in \R^{1,\D-1} \,.
    \end{equation}
    Singularities corresponding to such solutions are referred to as \emph{physical-sheet singularities}. Further, those obtained for physical values of the external momenta $p_i$, are called \emph{physical singularities}.
\end{mddefinition}

Physical singularities of a Feynman integral $\I$ can  be interpreted as those for which the corresponding Feynman diagram can be realized as a \emph{classical} spacetime process. To see why, one can define $\tau_e= m_e \alpha_e$ as the proper time of a particle of positive mass $m_e$. The equation \eqref{eq:Landau-eqs-2} then indicates that the distance between any two interaction points is given by the path taken by classical particles moving with momentum $q_e$. That is, physical singularities occur when all particles in the Feynman diagram are on shell and the interactions happen at fixed spacetime points $x_e$, with $x_{e+1}-x_e = \alpha_e q_e$. This perspective on the Landau equations is referred to as the \emph{Coleman--Norton interpretation}~\cite{Coleman:1965xm}. 

Most of the singularities of Feynman integrals are not physical-sheet and are instead encountered when analytically continuing $\I$ on other Riemann sheets.

\subsection{Connection to cuts}

We will also need a refined definition of cuts, also due to Cutkosky \cite[Sec.~II.A]{Cutkosky:1960sp}, which allow us to encircle any combination of poles in the integrand.
\begin{mddefinition}\label{def:cut-gen}
For a Feynman integral $\I$ as in~\eqref{eq:FeynInt-def} and a subset of edges $\{e_1^\pm, e_2^\pm, \ldots, e_k^\pm\}$ with associated signs $\pm$, the \emph{cut} is defined as
\be
\Cut_{e_1^{\pm} e_2^{\pm} \ldots e_k^{\pm}} \I := \int_{\ell_1 \ell_2 \ldots \ell_\L} \frac{\N \prod_{i=1}^{k} \ddelta^{\pm}(q_i^2 - m_i^2)}{\prod_{e = k+1}^{\E} (q_e^2 - m_e^2 + i\eps)}\, ,
\label{eq:generic-cut}
\ee
where the definitions of $q_e$, $\N$, $\ell$ and $\hat{\delta}$ were given in and around~\eqref{eq:FeynInt-def} and~\eqref{eq:delta-hat}. The choice of $\pm$ sign of each edge selects which $\ddelta^\pm$ is used.
\end{mddefinition}
The difference between this definition and the unitarity cut from Def.~\ref{def:unitarity_cuts} is that we do not complex-conjugate any part of the diagram, and energies of cut particles may have different signs.

Let us now connect Def.~\ref{def:cut-gen} and the computation of monodromies from Def.~\ref{def:monodromy}.
We consider the reduced diagram obtained imposing on-shell conditions $\delta^\pm(q_{e_i}^2 - m_{e_i}^2)$ on the set of edges $\{e_1^\pm, e_2^\pm, \ldots, e_k^\pm\}$ and $\alpha_e = 0$ on the complementary set.
Consider further the case in which the resulting singularity is physical and given by a codimension-$1$ variety $L = 0$ coming from a dimension-$0$ pinch in the loop-momentum space (i.e., where the Landau equations uniquely determine the Schwinger parameters and the loop momenta).
Provided that the $L=0$ singularity does not arise from any other pinch, we have 
\be\label{eq:monodromy-pham}
\Delta_{L=0} \I \,\stackrel{?}{=}\, \pm \,\Cut_{e_1^{\pm} e_2^{\pm} \ldots e_k^{\pm}} \I\, ,
\ee
where the $\pm$ sign has to do with the choice of orientation.
The equality is marked with a question mark because it is only known to hold in favorable cases. A proof was given by Pham for finite Feynman integrals with generic masses where the above conditions are satisfied \cite{pham2011singularities} (see also \cite{Bloch:2015efx}). Despite this restriction, there are some examples of divergent integrals or those involving massless particles for which \eqref{eq:monodromy-pham} still holds, as illustrated below in Ex.~\ref{sec:max_cut_tri}.

The mathematical theory used to study monodromies around individual branch points and which cuts they correspond to is called Picard--Lefschetz theory \cite{pham2011singularities}, which we discuss in Sec.~\ref{sec:PL}.

\subsection{Discontinuities and monodromies}

The simplest way to connect individual discontinuities and cuts is to work in a kinematic region where only one cut is allowed, and using the unitarity equations from the previous section.
\begin{mdexample}
    Going back to the example of the one-mass box from Ex.~\ref{eq:Ex-1mbox}, whose cuts
    were computed in Ex.~\ref{ex:I1m-cuts}, we consider the kinematic region $s>0$, $p_1^2<0$ and $t<0$. The only nonvanishing cut is in the $s$ channel. We can also easily connect the discontinuity and monodromy, since there we are above the a branch cut starting at $s=0$, but we are below the branch cuts in both $t$ and $p_1^2$. In this case, we get
    \begin{subequations}
    \begin{align} 
        \disc_{s=0} \I^{1m} (t,p_1^2;s) & = \Disc_s \I^{1m} (s,t,p_1^2) \\ & = \Cut_{\{1,3\}}^U \I^{1m} \\ & = \Cut_{1^+ 3^+} \I^{1m} \,,
    \end{align}
    \end{subequations}
    if we start at a kinematic point where $\Re(s)>0$ and $s \to s+i\varepsilon$.
    Note that the unitarity cut $\Cut_{\{1,3\}}^U$ and $\Cut_{1^+ 3^+}$ are the same in this case, since they impose the same energy flow for the cut propagators and the $i \varepsilon$'s of remaining propagators do not matter as they can never go on shell. Hence, in this case where only one cut is allowed in the $s$-channel, we have shown that $\disc_{s=0} \I^{1m} (t,p_1^2;s) = \Cut_{1^+ 3^+} \I^{1m} $.
\end{mdexample}
With the definition of a monodromy, we can start in different kinematic regions, and hold a subset of the variables fixed. In some cases, they can be related to the cuts from Def.~\ref{def:cut-gen}.
\begin{mdexample}
In the one-mass box example from Ex.~\ref{eq:Ex-1mbox}, the monodromies in different channels can be computed from the full expression. We can, for example, start at a physical kinematic point where $s,p_1^2>0$, $t<0$ and take $p_1^2 \to p_1^2 + i\varepsilon$, $s \to s+ i\varepsilon$. If we analytically continue counterclockwise around $s=0$ while holding $p_1^2$ fixed, we get
\begin{align}
    \disc&{}_{s=0}  \I^{1m}\big\vert^{p_1^2,s>0}_{t<0}
    \\\nonumber & =
    \frac{4 i }{s t}
    \Big[
    \frac{\sin (\pi \epsilon)}{\epsilon^2} s^{-\epsilon} 
    + \pi \log\left(\frac{p_1^2 + i\varepsilon -s}{t}\right)
    +\mathcal{O}(\epsilon)\Big],
\end{align}
where $p_1^2>s$ by the assumption that $\Delta_{s=0}$ corresponds to a small circle around $s=0$. In this case, we see from~\eqref{eq:cutI1m_s} that $\disc_{s=0} \I^{1m}\big\vert^{p_1^2,s>0}_{t<0} = \Cut_{1^+ 3^+} \I^{1m}$.
Similarly, we can compute the monodromy in $p_1^2$, $\disc_{p_1^2=0} \I^{1m}\big\vert^{p_1^2,s>0}_{t<0}$, which agrees with the following cut (using the labeling from Ex.~\ref{ex:I1m-cuts}):
\begin{align}
   &\Cut_{1^+ 2^+} \I^{1m} \big\vert^{p_1^2,s>0}_{t<0} = 
    \begin{gathered}
    \begin{tikzpicture}[line width=1,scale=0.25]
    	\coordinate (v1) at (-1,-1);
        \coordinate (v2) at (1,1);
        \coordinate (v3) at (3,1);
        \coordinate (v4) at (1,-1);
        \draw[] (v1) -- (v2) -- (v3) -- (v4) -- (v1);
        \draw[] (v1) -- ++(-150:0.7);
        \draw[] (v1) -- ++(-120:0.7); 
        \draw[] (v2) -- ++(175:2.7); 
        \draw[] (v3) -- ++(60:0.7);
        \draw[] (v3) -- ++(30:0.7);
        \draw[] (v4) -- ++(-45:0.7);
        \draw[dashed, orange] (0,2) -- (0,-2);
        \node[scale=0.7,rotate=50] at (-0.1,0.4) {$\rightarrow$};
        \node[scale=0.7,rotate=0] at (0.15,-1.5) {$\rightarrow$};
    \end{tikzpicture}
    \end{gathered}
   \\ & = \int_\ell \frac{\delta^+(\ell^2) \delta^+ [(p_{1}{-}\ell)^2]}{[(p_{12}{-}\ell)^2{+}i\varepsilon] (p_{123}{-}\ell)^2}
   \nonumber
    \\ \nonumber& = \frac{4i }{s t}
    \Big[
    \frac{-\sin (\pi \epsilon)}{\epsilon^2} (p_1^2)^{-\epsilon} 
    + \pi \log\Big(\frac{s t}{(p_1^2{-}i\varepsilon {-}s)(p_1^2{-}t )}\Big)\\
    \nonumber&\qquad \qquad +\mathcal{O}(\epsilon)
    \Big]  \,.\nonumber
\end{align}
\end{mdexample}

In this simple one-loop example, there was a way to relate the individual monodromies around $s=0$ and $p_1^2=0$ to the cuts in the $s$-channel and $p_1^2$, respectively, with the proper $i \varepsilon$ prescription for the remaining propagators. We should not be misled into generalizing from this example: it is not generally true that individual unitarity cuts agree with individual discontinuities. This one-loop example is however deceivingly simple, since the last propagator $(\ell{+}p_{123})^2$ can never go on shell, so we can fully separate out the dependence on the imaginary parts of $p_1^2$ and $s$. Beyond such examples, no such correspondence is known to hold.

\subsection{Cuts for anomalous thresholds}
The correspondence between monodromies around individual branch points and cuts in different kinematic channels is much more intricate, as previously discussed around~\eqref{eq:monodromy-pham}. As a first step, we must solve the Landau equations to find the points in the kinematic space that are potential singularities or branch points of the scattering amplitude. We sometimes find that such solutions are physical.
\begin{mdexample}
\label{ex:tri_threshold}
In a $3 \to 3$ scattering process of particles with mass $m$, the location of the anomalous threshold can be found by solving for the momenta at which the classical scattering process can occur, i.e., when it factorizes into a product of three $2\to 2$ on-shell scattering processes as follows:
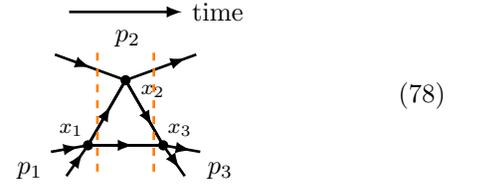
\begin{equation} \label{fig:anom_threshold}
    \begin{gathered}
    \begin{tikzpicture}[line width=1,scale=0.5]
    	\coordinate (v1) at (0,0);
        \coordinate (v2) at (1,1.73205);
        \coordinate (v3) at (2,0);
        \draw[fill=black] (v1) circle (0.1) node[left, xshift=2, yshift=6] {\footnotesize $x_1$};
        \draw[fill=black] (v2) circle (0.1) node[right, xshift=2, yshift=-4] {\footnotesize $x_2$};
        \draw[fill=black] (v3) circle (0.1) node[right, xshift=-2, yshift=6] {\footnotesize $x_3$};
        \draw[] (v1) -- (v2) -- (v3) -- (v1);
        \draw[arrow style at position={0.6}{latex}]  (v1) -- (v2);
        \draw[arrow style at position={0.7}{latex}]  (v2) -- (v3);
        \draw[arrow style at position={0.6}{latex}]  (v1) -- (v3);
        \draw[arrow style at position={0.75}{latex}]  ++(-170:1) -- (v1);
        \draw[arrow style at position={0.75}{latex}] ++(-125:1) -- (v1);
        \draw[arrow style at position={0.75}{latex}] (v3) -- ++(-10:1);
        \draw[arrow style at position={0.75}{latex}] (v3) -- ++(-55:1);
        \draw[arrow style at position={0.85}{latex}] (v2) -- ++(20:2);
        \draw[arrow style at position={0.5}{latex}] (v2)++(160:2) -- (v2);
        \draw[-latex] ($(v1)+(-0.5,3.55)$) -- ($(v3)+(0.5,3.55)$) node[right] {time};
        \draw[dashed, orange] (0.25,-0.7) -- (0.25,2.6);
        \draw[dashed, orange] (1.75,-0.7) -- (1.75,2.6);
        \node[scale=1,xshift=-22,yshift=-35] at (0,1.8) {$p_1$};
        \node[scale=1,xshift=15,yshift=15] at (0,1.8) {$p_2$};
        \node[scale=1,xshift=50,yshift=-35] at (0,1.8) {$p_3$};
    \end{tikzpicture}
    \end{gathered}
\end{equation}
Such a classical configuration occurs when the momenta $q_i$ of all the cut particles are put on shell, $q_i^2 = m^2$ with $i=1,2,3$, and interaction vertices happen at fixed space-time points $x_i$. Since $x_{i+1} - x_i = \alpha_i q_i$ (with $x_4 = x_1$) for Schwinger proper time $\alpha_i$, the latter constraint reads
\be\label{eq:sum-alpha-q}
\sum_{i=1}^{3} \alpha_i q_i = 0\, .
\ee
By dotting this constraint with $q_{j}$, we can rephrase it in terms of the $3 \times 3$ matrix $y$ with entries:
\be\label{eq:y-ij}
    y_{ij} = -\frac{q_i \cdot q_j}{m^2} = \frac{p_{ij}^2-2m^2}{2 m^2 } \quad i,j = 1,2,3\, .
\ee
Equation \eqref{eq:sum-alpha-q} implies that the determinant of $y$ vanishes: 
    \begin{equation}
    \det y =  \frac{p_1^2 p_2^2 p_3^2 + m^2 \lambda}{4 m^6}= 0 \,,
    \end{equation}
    where the $\lambda = \lambda(p_1^2, p_2^2, p_3^2)$ is the K\"all\'en function from~\eqref{eq:Kallen-fn}. This is a second-order equation in $p_2^2$, and when $p_1^2>4 m^2$ and $p_3^2>4 m^2$, one of the solutions in $p_2^2$ is found to be a physical singularity, i.e., $\alpha_i \geqslant 0$.
\end{mdexample}

We follow up with an example of a concrete formula in which such singularities are displayed.

\begin{mdexample}
\label{ex:tri_3D}
    In $\D=3$ space-time dimensions in the region $y_{ij}<-1$, the triangle diagram from Ex.~\ref{ex:tri_threshold} is given by
    \begin{align}
            \label{eq:3D_triangle}
        \I_{\text{tri}}^{\D=3} & (y_{12},y_{23},y_{13}) \Big|_{y_{ij}<-1} 
    \\
        & =
        \frac{\sqrt{\pi}}{4 \sqrt{\det y}} \log \Big( \mfrac{y_{12}{+}y_{23} y_{13} {+} \sqrt{\det y}}{y_{12}{+}y_{23} y_{13} {-} \sqrt{\det y}}
        \nonumber
        \\
        & \hspace{0.5cm} \times
        \mfrac{y_{23}{+}y_{12} y_{13} {+} \sqrt{\det y}}{y_{23}{+}y_{12} y_{13} {-} \sqrt{\det y}}
        \times
        \mfrac{y_{13}{+}y_{23} y_{12} {+} \sqrt{\det y}}{y_{13}{+}y_{23} y_{12} {-} \sqrt{\det y}}
        \Big)
        \nonumber
    \end{align}
    with $y_{ij}$ given in \eqref{eq:y-ij}. In this kinematic region, there is no singularity at $\det y =0$, since the naive square-root branch cuts cancel between the prefactor and the square roots inside the logarithm.

    After analytic continuation to other physical-kinematic regions, the logarithms can pick up imaginary parts, and the singularity at $\det y=0$ can become a physical singularity.
\end{mdexample}

In cases where anomalous thresholds can be realized in physical kinematics, one can compute cuts corresponding to putting all particles in the loop simultaneously on shell.
\begin{mdexample}\label{sec:max_cut_tri}
    For the triangle integral we can compute unitarity cuts in the $p_1^2$, $p_2^2$ and $p_3^2$ channels using the same methods as in Ex.~\ref{ex:I1m-cuts} and \ref{ex:bubble_cut}. We also have a new type of a cut in addition: the one where all particles in the triangle can simultaneously go on shell. In the kinematic channel where $p_1^2>4m^2$, $p_2^2<0$ and $p_3^2>4m^2$, it evaluates to

    \begin{align}
    & \Cut_{1^+ 2^+ 3^+} \I^{\D=3}_{\text{tri}}  
    =\begin{gathered}
    \begin{tikzpicture}[line width=1,scale=0.25]
    	\coordinate (v1) at (0,0);
        \coordinate (v2) at (1,1.73205);
        \coordinate (v3) at (2,0);
        \draw[] (v1) -- (v2) -- (v3) -- (v1);
        \draw[] (v1) -- ++(-150:0.7);
        \draw[] (v1) -- ++(-120:0.7);
        \draw[] (v3) -- ++(-30:0.7);
        \draw[] (v3) -- ++(-60:0.7);
        \draw[] (v2) -- ++(60:0.7);
        \draw[] (v2) -- ++(120:0.7);
        \draw[dashed, orange] (1,-0.7) -- (1,0.5);
        \draw[dashed, orange] (1,0.5) -- (0,1.5);
        \draw[dashed, orange] (1,0.5) -- (2,1.5);
        \node[scale=0.7,xshift=-6,yshift=-3] at (0,1.8) {$\ell$};
        \node[scale=0.7,yshift=-5,rotate=60,xshift=2] at (0,1.6) {$\rightarrow$};
        \node[scale=0.7,xshift=13] at (-0.5,-1.6) {$p_1{-}\ell$};
        \node[scale=0.7,yshift=5,xshift=10] at (0,-1.6) {$\rightarrow$};
        \node[scale=0.7,xshift=35,yshift=-3] at (0,1.8) {$p_2{+}\ell$};
        \node[scale=0.7,yshift=30,rotate=-60,xshift=40] at (0,1.6) {$\rightarrow$};
        \node[scale=0.7,xshift=-16,yshift=-25] at (0,1.8) {$p_1$};
        \node[scale=0.7,xshift=10,yshift=15] at (0,1.8) {$p_2$};
        \node[scale=0.7,xshift=35,yshift=-25] at (0,1.8) {$p_3$};
    \end{tikzpicture}
    \end{gathered}
    \\& = \int_\ell \hat{\delta}^+(\ell^2{-}m^2) \hat{\delta}^+[(p_1{-}\ell)^2{-}m^2] \hat{\delta}^+[(\ell{+}p_2)^2{-}m^2]
    \nonumber \\ & =  \frac{2 \pi^{3/2}}{\sqrt{-\det y}} \,.
    \nonumber
    \end{align}
It is natural to ask whether this cut corresponds to a monodromy in the kinematic space. In this simple example, it turns out that the cut is precisely the monodromy around the anomalous threshold at $\det y = 0$. Let us verify it.

The value of the monodromy around $\det y=0$  depends heavily on the value of the kinematics. As an example, let us assume that we fix $y_{12}$ and $y_{23}$, and analytically continue $y_{13}$ in the counterclockwise direction such that it encircles $\det y = 0$. This amounts to fixing $p_1^2$ and $p_3^2$ and analytically continuing in $p_2^2$. Since we already established in Ex.~\ref{ex:tri_3D} that $\det y = 0$ is not a singularity of $\I^{\D=3}_{\text{tri}}$ when all the invariants $p_i^2$ are negative, we get
\begin{equation}
    \Delta_{\det y = 0}\, \I^{\D=3}_{\text{tri}} \big\vert_{p_1^2, p_2^2,p_3^2<0} = 0\,.
\end{equation}
In this kinematic region, the cut is also zero. The reason why there is no singularity at $\det y =0$ when all the invariants are negative can be explained with the physical intuition mentioned before: we cannot factorize the process into on-shell processes unless at least two of the invariants are positive. 
However, if we instead fix $p_1^2,p_3^2>4m^2$, with $p_1^2 \to p_1^2+i\varepsilon$ and $p_3^2 \to p_3^2 + i\varepsilon$, and take a monodromy around the branch of $\det y=0$ which is physical (i.e., its solutions of~\eqref{eq:sum-alpha-q} have $\alpha_i \geq 0$ in physical kinematics), we get
    \begin{equation}
        \Delta_{\det y =0} \, \I_{\text{tri}}^{\D=3} \big\vert_{p_2^2<0}^{p_1^2, p_3^2>4m^2} =  \frac{2 \pi^{3/2}}{\sqrt{-\det y}} \,.
    \end{equation}
When $p_1^2>4m^2$ and $p_3^2>4m^2$, the anomalous-threshold process can be allowed in physical kinematics as in \eqref{fig:anom_threshold},
and the monodromy has support. Furthermore, the monodromy is, in this case, equal to the cut,
and thus
\begin{align}
    \Delta_{\det y = 0}\, \I^{\D=3}_{\text{tri}} & \big\vert^{p_1^2, p_3^2>4m^2}_{p_2^2 <0} \\ &  = \Cut_{1^+ 2^+ 3^+} \I^{\D=3}_{\text{tri}} \big\vert^{p_1^2, p_3^2>4m^2}_{p_2^2 <0}\,.
\end{align}
Note that this last expression does not have a hope of being obtained by only using the unitarity equation in~\eqref{eq:unitarity}. We are relying on keeping $p_1^2$ and $p_3^2$ fixed to be positive and with a small positive imaginary part, and only analytically continuing in $p_2^2$. This is a very different manipulation from the one using unitarity constraints, where we cross branch cuts of all invariants involved.
\end{mdexample}

\subsubsection{Non-Mandelstam variables}
The discussion of analytic structure depends heavily  on the choice of variables for the problem.

\begin{mdexample}
\label{ex:Itri_Mandelstams}
    Consider the triangle Feynman integral for a $3\to 3$ scattering process in a theory of scalar massless particles.
The external kinematic invariants are allowed to take any complex values.
This integral evaluates to
\begin{align}
    \I^{\textrm{D}=4}_{\text{tri}} & =
    \begin{gathered}
    \begin{tikzpicture}[line width=1,scale=0.25]
    	\coordinate (v1) at (0,0);
        \coordinate (v2) at (1,1.73205);
        \coordinate (v3) at (2,0);
        \draw[] (v1) -- (v2) -- (v3) -- (v1);
        \draw[] (v1) -- ++(-150:0.7);
        \draw[] (v1) -- ++(-120:0.7);
        \draw[] (v3) -- ++(-30:0.7);
        \draw[] (v3) -- ++(-60:0.7);
        \draw[] (v2) -- ++(60:0.7);
        \draw[] (v2) -- ++(120:0.7);
        \node[scale=0.7,xshift=-16,yshift=-25] at (0,1.8) {$p_1^2$};
        \node[scale=0.7,xshift=10,yshift=15] at (0,1.8) {$p_2^2$};
        \node[scale=0.7,xshift=35,yshift=-25] at (0,1.8) {$p_3^2$};
    \end{tikzpicture}
    \end{gathered}
    =
    \frac{2}{\sqrt{\lambda}} \Bigg[ \Li_2\left( 
    \frac{p_1^2+p_2^2-p_3^2+ \sqrt{\lambda}}{2 p_1^2}
    \right)
    \nonumber
    \\ 
    & \quad -\Li_2\left(\frac{p_1^2+p_2^2-p_3^2- \sqrt{\lambda}}{2 p_1^2} \right)
    \label{eq:Itri_Mandelstams}
    \\ 
    & \quad + \frac{1}{2} \log\left( \frac{p_2^2}{p_1^2} \right) \log\left( \frac{p_3^2+p_1^2-p_2^2 - \sqrt{\lambda}}{p_3^2+p_1^2-p_2^2 + \sqrt{\lambda}}  \right)\Bigg]
    \,,
    \nonumber
\end{align}
where $\lambda = \lambda(p_1^2, p_2^2, p_3^2)$ is the K\"all\'en function given in~\eqref{eq:Kallen-fn}. The expression is valid in the kinematic region where $p_1^2, p_2^2, p_3^2<0$ and $\lambda<0$, when the square root, $\Li_2$ and $\log$ are taken to be on their principal branches. We see from the expression that it has branch points at $p_1^2=0$, $p_2^2=0$, $p_3^2=0$ and at $\lambda=0$. The last one, $\lambda=0$, is a second-type singularity and does not correspond to the triangle anomalous threshold.
\end{mdexample}

\begin{mdexample}
    The triangle integral from Ex.~\ref{ex:Itri_Mandelstams} can also be written as 
    \begin{align}
            \I^{\textrm{D}=4}_{\text{tri}} & = \mfrac{2}{p_1^2(z{-}\zb)} \left[ \Li_2(z) {-} \Li_2 (\zb) + \frac{1}{2} \log(z\zb) \log \left( \mfrac{1{-}z}{1{-}\zb}\right) \right]\! ,
            \label{eq:Itri_zzb}
    \end{align}
    where $z$ and $\zb$ are two independent variables defined by $z \zb = p_2^2/ p_1^2$ and $(1-z)(1-\zb) = p_3^2/p_1^2$. We take $\zb<z$ without loss of generality, and then the expression above is valid in the region where $0<\zb<z<1$. In terms of the Mandelstam invariants, this region is the one where either $\{p_1^2,p_2^2,p_3^2<0, \lambda>0\}$ or $\{p_1^2,p_2^2,p_3^2>0, \lambda>0\}$. In other kinematic regions, $z$ and $\zb$ pick up $\pm i \varepsilon$'s to capture the correct analytic continuation of the logarithms and dilogarithms in~\eqref{eq:Itri_zzb}.

    The triangle anomalous threshold at $\lambda = z - \bar{z} = 0$, which is a \emph{square root} singularity of the expression in~\eqref{eq:Itri_Mandelstams} has become a \emph{pole} at $z-\zb=0$ in the $z$, $\zb$ variable representation in~\eqref{eq:Itri_zzb}. Thus, the change of variables (which itself is branched) has washed out the square root branch cut, and changed the  singularity into a pole instead. This phenomenon is not surprising; it is a common feature of analytic functions that changes of variables generally modifies the analytic structure.

    Let us briefly comment on the analytic structure of the one-loop triangle in~\eqref{eq:Itri_zzb} seen as a function of $z$ and $\bar{z}$. First, we note that in the region where $\lambda<0$, $z$ and $\bar{z}$ are complex conjugates of each other. In that region we may see~\eqref{eq:Itri_zzb} as a function of the single complex variable $z$. Although the individual logarithms and dilogarithms in~\eqref{eq:Itri_zzb} have branch cuts, \eqref{eq:Itri_zzb} is a single-valued function of the complex variable $z$, cf., e.g.,~\cite{Chavez:2012kn,Brown:2015ztw}. There is no contradiction with the fact that $\I^{\textrm{D}=4}_{\text{tri}}$ must have branch cuts dictated by unitarity. In fact, the branch cut structure of $\I^{\textrm{D}=4}_{\text{tri}}$ seen as a function of the invariants $p_i^2$ translates directly into the fact that~\eqref{eq:Itri_zzb} is a single-valued function of the complex variable $z$ in the region where $\lambda<0$.
\end{mdexample}

\subsubsection{Loop-tree theorems}

There are multiple theorems relating loop-level scattering amplitudes to on-shell phase integrals over tree amplitudes. Let us explain the simplest of them: the Feynman tree theorem \cite{Feynman:1963ax}. Consider a one-loop diagram computed with retarded (instead of Feynman) propagators $G_R(q,m) = 1/(q^2 - m^2 + i\eps q^0)$:
\be\label{eq:retarded-loop}
\int_\ell \N \prod_{i=0}^{n-1} G_R(\ell + p_{1\cdots i}, m_i) = 0\, .
\ee
Causality dictates that such an amplitude has to vanish, as otherwise we would create a closed time-like loop. One way to see it practically is by deforming the loop-energy $\ell^0$ contour into the upper half-plane, which vanishes since all the poles are located in the lower half-plane.

Plugging in the relation between retarded and Feynman propagators $G_F(q,m) = 1/(q^2 - m^2 + i\eps)$,
\be
G_R(q,m) = G_F(q,m) - \ddelta^+(q^2 - m^2)
\ee
into \eqref{eq:retarded-loop}, one obtains a relation between the time-ordered amplitude, all of its single cuts, double cuts, etc. Note here that we require real external kinematics. For example, we have
\begin{align}
\begin{gathered}
\begin{tikzpicture}[line width=1,scale=0.35,baseline={([yshift=2.0ex]current bounding box.center)}]
    	\coordinate (v1) at (-1,-1);
        \coordinate (v2) at (-1,1);
        \coordinate (v3) at (1,1);
        \coordinate (v4) at (1,-1);
        \draw[] (v1) -- (v2) -- (v3) -- (v4) -- (v1);
        \draw[] (v1) -- ++(-150:0.7);
        \draw[] (v1) -- ++(-120:0.7);
        \draw[] (v2) -- ++(135:0.7);
        \draw[] (v3) -- ++(45:0.7);
        \draw[] (v4) -- ++(-45:0.7);
\end{tikzpicture}
\end{gathered}
=&
\sum_{1\text{-cuts}}\;
\begin{gathered}
\begin{tikzpicture}[line width=1,scale=0.35,baseline={([yshift=2.0ex]current bounding box.center)}]
    	\coordinate (v1) at (-1,-1);
        \coordinate (v2) at (-1,1);
        \coordinate (v3) at (1,1);
        \coordinate (v4) at (1,-1);
        \draw[] (v1) -- (v2) -- (v3) -- (v4) -- (v1);
        \draw[] (v1) -- ++(-150:0.7);
        \draw[] (v1) -- ++(-120:0.7);
        \draw[] (v2) -- ++(135:0.7);
        \draw[] (v3) -- ++(45:0.7);
        \draw[] (v4) -- ++(-45:0.7);
        \draw[dashed,orange] (0,0.2) -- (0,1.8);
        \node[scale=0.7,rotate=0,] at (0,1.25) {$\rightarrow$};
\end{tikzpicture}
\end{gathered}
-
\sum_{2\text{-cuts}}\;
\begin{gathered}
\begin{tikzpicture}[line width=1,scale=0.35,baseline={([yshift=2.0ex]current bounding box.center)}]
    	\coordinate (v1) at (-1,-1);
        \coordinate (v2) at (-1,1);
        \coordinate (v3) at (1,1);
        \coordinate (v4) at (1,-1);
        \draw[] (v1) -- (v2) -- (v3) -- (v4) -- (v1);
        \draw[] (v1) -- ++(-150:0.7);
        \draw[] (v1) -- ++(-120:0.7);
        \draw[] (v2) -- ++(135:0.7);
        \draw[] (v3) -- ++(45:0.7);
        \draw[] (v4) -- ++(-45:0.7);
        \draw[dashed,orange] (0,0.2) -- (0,1.8);
        \draw[dashed,orange] (0.2,0) -- (1.8,0);
        \node[scale=0.7,rotate=0] at (0,1.25) {$\rightarrow$};
        \node[scale=0.7,rotate=-90] at (1.25,0) {$\rightarrow$};
\end{tikzpicture}
\end{gathered}
\\
&+
\sum_{3\text{-cuts}}\;
\begin{gathered}
\begin{tikzpicture}[line width=1,scale=0.35,baseline={([yshift=2.0ex]current bounding box.center)}]
    	\coordinate (v1) at (-1,-1);
        \coordinate (v2) at (-1,1);
        \coordinate (v3) at (1,1);
        \coordinate (v4) at (1,-1);
        \draw[] (v1) -- (v2) -- (v3) -- (v4) -- (v1);
        \draw[] (v1) -- ++(-150:0.7);
        \draw[] (v1) -- ++(-120:0.7);
        \draw[] (v2) -- ++(135:0.7);
        \draw[] (v3) -- ++(45:0.7);
        \draw[] (v4) -- ++(-45:0.7);
        \draw[dashed,orange] (0,0.2) -- (0,1.8);
        \draw[dashed,orange] (0.2,0) -- (1.8,0);
        \draw[dashed,orange] (0,-0.2) -- (0,-1.8);
        \node[scale=0.7,rotate=0] at (0,1.25) {$\rightarrow$};
        \node[scale=0.7,rotate=-90] at (1.25,0) {$\rightarrow$};
        \node[scale=0.7,rotate=-180] at (0,-1.25) {$\rightarrow$};
\end{tikzpicture}
\end{gathered}
\;-\;
\begin{gathered}
\begin{tikzpicture}[line width=1,scale=0.35,baseline={([yshift=2.0ex]current bounding box.center)}]
    	\coordinate (v1) at (-1,-1);
        \coordinate (v2) at (-1,1);
        \coordinate (v3) at (1,1);
        \coordinate (v4) at (1,-1);
        \draw[] (v1) -- (v2) -- (v3) -- (v4) -- (v1);
        \draw[] (v1) -- ++(-150:0.7);
        \draw[] (v1) -- ++(-120:0.7);
        \draw[] (v2) -- ++(135:0.7);
        \draw[] (v3) -- ++(45:0.7);
        \draw[] (v4) -- ++(-45:0.7);
        \draw[dashed,orange] (0,0.2) -- (0,1.8);
        \draw[dashed,orange] (0.2,0) -- (1.8,0);
        \draw[dashed,orange] (0,-0.2) -- (0,-1.8);
        \draw[dashed,orange] (-0.2,0) -- (-1.8,0);
        \node[scale=0.7,rotate=0] at (0,1.25) {$\rightarrow$};
        \node[scale=0.7,rotate=-90] at (1.25,0) {$\rightarrow$};
        \node[scale=0.7,rotate=-180] at (0,-1.25) {$\rightarrow$};
        \node[scale=0.7,rotate=-270] at (-1.25,0) {$\rightarrow$};
\end{tikzpicture}
\end{gathered}
\nonumber
\end{align}
Many of the cuts might vanish due to kinematical reasons, e.g., the final term always vanishes since it involves positive energy circulating in a loop. The same result can of course be obtained by carefully deforming the $\ell^0$-contour in the original Feynman integral. 
This strategy can be extended to higher loops and used in numerical evaluations of amplitudes \cite{Runkel:2019yrs,Capatti:2019ypt,deJesusAguilera-Verdugo:2021mvg}. 

The distinction between the Feynman tree theorem and the optical theorem is that the former involves the full integral (not just its imaginary part) and the cuts do not have a one-directional flow of energy.
\begin{mdexample}
Consider the bubble diagram in Ex.~\ref{ex:bubble} in the $s$-channel kinematics. In this case, the Feynman tree theorem reads
\be\label{eq:FTF-bub}
\I_{\text{bub}}^{\D=2} =
\begin{gathered}
    \begin{tikzpicture}[line width=1,scale=0.35,baseline={([yshift=-2ex]current bounding box.center)}]
	\coordinate (v1) at (-1,0);
    \coordinate (v2) at (1,0);
    \draw[] (v1) [out=55,in=125] to (v2);
    \draw[] (v1) [out=-55,in=-125] to (v2);
    \draw[] (v1) -- ++(-135:0.7);
    \draw[] (v1) -- ++(135:0.7);
    \draw[] (v2) -- ++(45:0.7);
    \draw[] (v2) -- ++(-45:0.7);
    \draw[dashed,orange] (0,0) -- (0,1);
    \node[scale=0.7,rotate=0] at (0,0.75) {$\rightarrow$};
    \end{tikzpicture}
    \end{gathered}
\,+\,
\begin{gathered}
    \begin{tikzpicture}[line width=1,scale=0.35,baseline={([yshift=1ex]current bounding box.center)}]
	\coordinate (v1) at (-1,0);
    \coordinate (v2) at (1,0);
    \draw[] (v1) [out=55,in=125] to (v2);
    \draw[] (v1) [out=-55,in=-125] to (v2);
    \draw[] (v1) -- ++(-135:0.7);
    \draw[] (v1) -- ++(135:0.7);
    \draw[] (v2) -- ++(45:0.7);
    \draw[] (v2) -- ++(-45:0.7);
    \draw[dashed,orange] (0,0) -- (0,-1);
    \node[scale=0.7,rotate=-180] at (0,-0.75) {$\rightarrow$};
    \end{tikzpicture}
    \end{gathered}
    \ee
As above, the maximal cut does not contribute. In $\D=2$ dimensions, the first cut evaluates to
\begin{align}
\begin{gathered}
    \begin{tikzpicture}[line width=1,scale=0.35,baseline={([yshift=0.0ex]current bounding box.center)}]
	\coordinate (v1) at (-1,0);
    \coordinate (v2) at (1,0);
    \draw[] (v1) [out=55,in=125] to (v2);
    \draw[] (v1) [out=-55,in=-125] to (v2);
    \draw[] (v1) -- ++(-135:0.7);
    \draw[] (v1) -- ++(135:0.7);
    \draw[] (v2) -- ++(45:0.7);
    \draw[] (v2) -- ++(-45:0.7);
    \draw[dashed,orange] (0,0) -- (0,1);
    \node[scale=0.7,rotate=0] at (0,0.75) {$\rightarrow$};
    \end{tikzpicture}
    \end{gathered}
    &= 
    \int_\ell \frac{\hat{\delta}^+ (\ell^2{-}m^2)}{(p_{12}{-}\ell)^2{-}m^2 + i\eps} \\
    &= \frac{i}{\sqrt{s(4m^2-s)}} \bigg[ \log \left( \mfrac{\sqrt{4m^2 {-}s } {-} i \sqrt{s}}{\sqrt{4m^2 {-}s } {+} i \sqrt{s}} \right) 
     - i \pi \bigg],\nonumber
\end{align}
where we take $s \to s + i\varepsilon$.
The remaining cut doubles the contribution from the first term in the square brackets and cancels the $i \pi$ term, such that the identity \eqref{eq:FTF-bub} is satisfied.
\end{mdexample}

\subsection{Symbols and monodromies}

\subsubsection{Definitions}

A convenient way of probing the analytic structure of (multiple) polylogarithms (cf. Ex.~\ref{ex:dilog}) is through the \emph{symbol}~\cite{Goncharov.A.B.:2009tja,Chen:1977oja,Brown:2009qja,Goncharov:2010jf,Duhr:2011zq}. 
We refer to \cite{Duhr:2019tlz} for a review and notation. Let us say that a symbol of a function $f$ is given by
\be\label{eq:symbol}
\mathcal{S}[f] = \!\!\!\sum_{i_1, i_2, \ldots, i_k}\!\! c_{i_1, i_2, \ldots, i_k} a_{i_1} \otimes a_{i_2} \otimes \cdots \otimes a_{i_k}\, ,
\ee
where $a_i$ are the symbol letters and $c_{i_1, i_2, \ldots, i_k}$ are the coefficients, which are typically rational numbers. Then, the symbol of the monodromy around a given singularity at $a_j=0$ is given by 
\be\label{eq:symbol-monodromy}
\mathcal{S}[\disc_{a_j=0} f] = 2\pi i \!\!\!\sum_{ i_2, \ldots, i_k}\!\!  c_{j, i_2, \ldots, i_k} a_{i_2} \otimes \cdots \otimes a_{i_k}\,.
\ee
In other words, only terms in the symbol  which have $a_j$ in the first entry survive. A similar statement can be made at the level of the coaction \cite{Duhr:2012fh,Brown:coaction}. More generally, we can iterate this procedure, and we see that the sequence of $a_j$'s in a symbol captures the effect of computing iterated monodromies, e.g.,
\be\begin{split}
\label{eq:symbol-monodromy-iterated}
\mathcal{S}[\disc_{a_{l}=0}&\disc_{a_j=0} f] =\\
&\,(2\pi i )^2\!\!\!\sum_{ i_3, \ldots, i_k}\!\!c_{j,l, i_3, \ldots, i_k} a_{i_3} \otimes \cdots \otimes a_{i_k}\,.
\end{split}
\ee
Note that while the symbol captures the (iterated) monodromies of a polylogarithmic function, its value is independent of the Riemann sheet on which the function is evaluated.

\begin{mdexample}
Consider the dilogarithm $\Li_2(z)$ from Ex.~\ref{ex:dilog}. Its symbol is
\be\label{eq:dilog_symbol}
\mathcal{S}[\Li_2(z)] = - (1-z) \otimes z\, .
\ee
On the other hand, computing the monodromy around $z=1$ gives
\be
\disc_{z=1} \Li_2(z) = 2\pi i \log(z) \, ,
\ee
which results in 
\be
\mathcal{S}[\disc_{z=1} \Li_2(z)] = 2\pi i\, z\, .
\ee
This expression is an instance of \eqref{eq:symbol-monodromy}. 

Note that we can read off the analytic structure of the dilogarithm directly from the symbol in~\eqref{eq:dilog_symbol}. The first entry of the symbol vanishes or becomes infinite when $z=1$ or $z=\infty$, which reflects the fact that on its principal sheet (where the function is real),  the dilogarithm only has a branch cut starting at $z=1$ along the positive real axis, cf. Ex.~\ref{ex:dilog}. The second entry vanishes for $z=0$, and encodes the fact that on the second sheet, the dilogarithm also has a branch point at $z=0$.
\end{mdexample}

\subsubsection{Symbols of Feynman integrals}

If a scattering amplitude or an individual Feynman integral can be expressed in terms of polylogarithms, we can compute its symbol. Since the symbol captures the (iterated) monodromies, the possible symbol letters that can arise are constrained by the position of the branch points. 

In particular, the first entries in the symbol vanish only at physical-sheet singularities.
In massless theories it is common to use the \emph{first entry condition}~\cite{Gaiotto:2011dt}, which says that the first entries in a symbol can only be Mandelstam invariants. 
Since the symbol also encodes iterated monodromies, the second entries are constrained by relations inspired by those of Steinmann \eqref{eq:Steinmann-rel} extrapolated to massless theories and with the monodromy $\disc$ instead of the discontinuity $\Disc$~\cite{Caron-Huot:2019bsq}:
\be\label{eq:steinmann-Delta}
\disc_{s_J} \disc_{s_I} \mathcal{A} = 0\, .
\ee
Generalizations to multiple iterated monodromies exist in some cases \cite{Hannesdottir:2022xki,Berghoff:2022mqu}.

The constraints on the symbol 
allow one to put constraints on the coefficients $c$ in \eqref{eq:symbol}.
For example, this strategy becomes useful in the perturbative bootstrap of $\mathcal{N}=4$ SYM amplitudes in the planar limit, see, e.g., \cite{Caron-Huot:2019bsq}.

\begin{mdexample}
We illustrate the structure of the symbol on the example of the ``two-mass easy box'' function from Ex.~\ref{ex:I-2m}. If we define $\mathcal{J}^{2m} = \frac{1}{2}(st-p_1^2p_3^2)\mathcal{I}^{2m}$, then the symbol is given by
\be\begin{split}
\label{eq:I-2m-symb}
\mathcal{S}&\big[\mathcal{J}^{2m}\big] =
-\frac{1}{\epsilon}\big(s+t-p_1^2-p_3^2\big)\\
&+\big[s\otimes s -s\otimes (s-p_1^2)-s\otimes (s-p_3^2)\\
&\quad+s\otimes (st-p_1^2p_3^2) +(s\leftrightarrow t)\big]+\\
&\,-\big[p_1^2\otimes p_1^2 - p_1^2\otimes (s-p_1^2)- p_1^2\otimes (t-p_1^2)\\
&\quad+p_1^2\otimes (st-p_1^2p_3^2)+(p_1^2\leftrightarrow p_3^2)\big] + \mathcal{O}(\epsilon)\,.
\end{split}
\ee
We see that the first entries are all drawn from the set $\{s,t,p_1^2,p_3^2\}$, in agreement with the first entry condition. We also know from 
Ex.~\ref{ex:I-2m-steinmann} that the Steinmann relations (in the from of~\eqref{eq:steinmann-Delta}) apply. Equation~\eqref{eq:symbol-monodromy-iterated} then implies 
\be
\mathcal{S}\big[\disc_s\disc_t \mathcal{J}^{2m}] = 0\,.
\ee
This equation is indeed true because the symbol in~\eqref{eq:I-2m-symb} does not contain any terms of the form $s\otimes t$ or $t\otimes s$.
\end{mdexample}

\begin{mdexample}
For the one-mass box function $\mathcal{I}^{1m}$, we define $\mathcal{I}^{1m}=\frac{1}{2}st\mathcal{J}^{2m}$. The symbol is given by
\begin{align}\nonumber
\mathcal{S}\big[\mathcal{J}^{1m}\big]&\, = \frac{1}{\epsilon^2}+\frac{1}{\epsilon}\big(p_1^2-s-t\big)+\Big[p_1^2\otimes(p_1^2-s)\\
\nonumber&-p_1^2\otimes s-s\otimes(p_1^2-s)-\frac{1}{2}p_1^2\otimes p_1^2 + t\otimes s \\
&+ s\otimes s+(s\leftrightarrow t)\Big]+\mathcal{O}(\epsilon)\,.
\end{align}
The first entry condition is clearly satisfied. This time, however, the symbol contains terms of the form $s\otimes t$ and $t\otimes s$, which implies 
\be
\mathcal{S}\big[\disc_s\disc_t \mathcal{J}^{1m}] \neq 0\,.
\ee
This is a manifestation of the Steinmann relations not being satisfied for the one-mass box function, cf. Ex.~\ref{ex:Steinmann-I1m}.
\end{mdexample}

\subsection{\label{sec:PL}Picard--Lefschetz theory for Feynman integrals}

There have been attempts to systematize computations of monodromies using tools from algebraic topology, with a goal of putting \eqref{eq:monodromy-pham} on a firm mathematical footing. Consider the favorable cases discussed above \eqref{eq:monodromy-pham}, in which a pinch associated with $L=0$ is an isolated point.
Notice that a Feynman integral $\I$, as written in \eqref{eq:FeynInt-def}, is an integral over a specific contour $\Gamma$ implementing the $i\eps$ prescription.
The Picard--Lefschetz theorem prescribes how the monodromy $\Delta_{L=0}\, \I$ around a branch point $L=0$ can be rewritten in terms of a new integration contour $c$:
\begin{equation}
    \Delta_{L=0}\, \I = N_{\Gamma,c} \int_{c} \frac{\N}{\prod_{e=1}^{\E} (q_e^2 - m_e^2)}\, ,
\end{equation}
where $N_{\Gamma,c} \in \mathbb{Z}$ is called the \emph{intersection number} or \emph{Kronecker index} of $\Gamma$ and $c$. Here, $c$ is the \emph{vanishing cycle} homologous to a sphere around the pinch point \cite{pham2011singularities}.

In practice,  one starts with identifying which denominators pinch the integration contour at a physical singularity corresponding to $L=0$. Labeling the denominators which take part in this pinch as $1,2, \ldots, k$, the result is
\begin{equation}
    \Delta_{L=0}\, \I = N_{\Gamma,c} \int_{\ell_1, \ell_2 \ldots, \ell_\L} \frac{\N \prod_{e=1}^k \hat{\delta}^\pm (q_e^2-m_e^2)}{\prod_{e'=k+1}^{\E} (q_{e'}^2 - m_{e'}^2 + i\varepsilon)}\,,
    \label{eq:individualmonodromy}
\end{equation}
where the signs $\pm$ are determined by which of the two roots of the polynomial $q_e^2-m_e^2$ takes part in the pinch. The monodromy is therefore proportional to the individual cuts $\Cut_{e_1^{\pm} e_2^{\pm} \ldots e_k^{\pm}}\mathcal{I}$ defined in~\eqref{eq:generic-cut}. For the cases under consideration, one can show that the intersection number equals either $N_{\Gamma,c} = \pm 1$ or $0$, depending on whether the branch point corresponding to the physical singularity is actually present or not, see, e.g., \cite{10.1063/1.1704822,10.1063/1.1665233} for examples.

Some extensions to cases with non-generic masses, where pinches could become non-isolated or degenerate, exist for finite integrals \cite{Berghoff:2022mqu}. 

\section{\label{sec:generalized}Generalized cuts}

\subsection{Cuts of arbitrary sets of edges}

Cutting a diagram
can be considered as an operation per se, liberated from any interpretation as a specific imaginary part or discontinuity. The operation is the restriction of some subset of propagators to their mass shells, using delta functions or residues, with or without energy-flow conditions. It is an operation on an integral (integrand and/or contour). The idea of performing such an operation is to apply a specific generalized cut on two sides of an equation to obtain a set of constraints.

\begin{mddefinition}\label{def:gencut}
The generalized unitarity cut through the set of edges $C$ of the Feynman integral $\I$ is defined as
\be
\mathrm{Cut}^G_{C} \I =  \int_{\ell_1, \ldots, \ell_\L}  
\frac{\N}{\prod_{e \notin C} (q_e^2 - m_e^2 
)
}
 \prod_{c \in C} \ddelta (q_c^2 - m_c^2)\, ,
\ee
where the integration contour 
is deformed so that it 
captures the solutions of the on-shell conditions implied by the delta functions. 
\end{mddefinition}

\begin{mdexample}\label{ex:quadcut}
    All four edges of a one-loop box in $\D=4-2\epsilon$ dimensions can be cut simultaneously, by an operation known as a quadruple cut. Although the four delta functions have no common solution for real momenta in Euclidean or Minkowski signature, a pair of solutions can be identified in (2,2) signature or with complex-valued momenta \cite{Britto:2004nc}. In the case of the one-mass box, 
    \begin{align}
&\mathrm{Cut}^G_{\{1,2,3,4\}}     \I^{1m} (s,t,p_1^2)  \;=
\begin{tikzpicture}[line width=1,scale=0.20,baseline={([yshift=-0.8ex]current bounding box.center)}]
	\coordinate (v1) at (-1,-1);
    \coordinate (v2) at (-1,1);
    \coordinate (v3) at (1,1);
    \coordinate (v4) at (1,-1);
    \draw[] (v1) -- (v2) -- (v3) -- (v4) -- (v1);
    \draw[] (v1) -- ++(-135:0.7);
    \draw[] (v2) -- ++(135:0.7) node[left] {};
    \draw[] (v3) -- ++(60:0.7) node[right,yshift=2] {};
    \draw[] (v3) -- ++(30:0.7);
    \draw[] (v4) -- ++(-45:0.7) node[right] {};
    \draw[orange,dashed] (-2,0) -- (2,0);
    \draw[orange,dashed] (0,-2) -- (0,2);
\end{tikzpicture}
\\ 
& =\int_{\ell}
\hat{\delta}(\ell^2)\, \hat{\delta}[(\ell+p_1)^2]\, \hat{\delta}[(\ell+p_{12})^2]\, \hat{\delta}[(\ell-p_{4})^2] \nonumber \\
& = \frac{8\pi^2}{i  st} +
    \mathcal{O} (\epsilon) \,.
    \nonumber
\end{align}
\end{mdexample}
Although we have written delta functions in Def.~\ref{def:gencut}, the on-shell conditions can be freely interpreted in any of the other ways discussed above, such as by residues. Parametric representations are also possible (see for example \cite{Boyling1968,Arkani-Hamed:2017ahv,Britto:2023rig}) and remain to be more fully developed.

\subsection{Unitarity methods for amplitudes}

Generalized unitarity is commonly applied to identify coefficients of a basis of master integrals for a given scattering amplitude, at a fixed order in perturbation theory. Since the coefficients are rational or algebraic by construction, the branch cuts and discontinuities of the type probed by the cutting operation operate only on the master integrals. Therefore, the cut amplitude can be expanded in a basis of cut master integrals, with the same coefficients; 
\begin{align}
 {\mathcal A} &=  \sum_i c_i\, \I_i \,,
\\
\mathrm{Cut}^G_{C} {\mathcal A} &=  \sum_i c_i \,\mathrm{Cut}^G_{C} \I_i\, .
\label{eq:cutcut}
\end{align}
In the unitarity method presented in \cite{Bern:1994zx,Bern:1994cg}, this idea was used with $\mathrm{Cut}^G_{C} \to \mathrm{Cut}^U_{C}$, collecting information from various unitarity cuts. One advantage of applying cuts is that $\mathrm{Cut}^G_{C} \I_i=0$ whenever $C$ contains an edge that is not present in $\I_i$.  Another is that information can be collected from a variety of cuts.
A third advantage is that the left-hand side of \eqref{eq:cutcut} can be given the interpretation of a pair of  amplitudes, integrated over a momentum space restricted to the on-shell conditions of the cut propagators, as in Sec.~\ref{sec:non-pert-unitarity}.

In Ex.~\ref{ex:quadcut}, the same result can be obtained by taking successive unitarity cuts in the $s$ and $t$ channels after suitable analytic continuation, or by taking the residues at the singular points, which can be given the interpretation of a monodromy. The latter formalism was developed in \cite{Abreu:2017ptx}. We stress that the idea of generalized unitarity is to choose any possible interpretation of integrating with the on-shell delta functions. However, one must be precise and consistent within a given calculation, in order to restrict an ansatz for a scattering amplitude based on its functional form.

\begin{mdexample}\label{ex:boxcoeffs}
One-loop reduction uses a well known set of master integrals. 
The leading-color partial amplitude for 5-particle scattering in ${\mathcal N}=4$ supersymmetric Yang--Mills theory can be expressed in terms of scalar 1-mass boxes. 
\begin{align}
{\mathcal A}=
&\; c_1
\begin{tikzpicture}[line width=1,scale=0.20,baseline={([yshift=-0.4ex]current bounding box.center)}]
	\coordinate (v1) at (-1,-1);
    \coordinate (v2) at (-1,1);
    \coordinate (v3) at (1,1);
    \coordinate (v4) at (1,-1);
    \draw[] (v1) -- (v2) -- (v3) -- (v4) -- (v1);
    \draw[] (v1) -- ++(-135:0.7);
    \draw[] (v2) -- ++(135:0.7) node[left] {};
    \draw[] (v3) -- ++(60:0.7) node[right,yshift=2] {};
    \draw[] (v3) -- ++(30:0.7);
    \draw[] (v4) -- ++(-45:0.7) node[right] {};
    \node[yshift=-6,RoyalBlue] at (0,-1) {\footnotesize$3$};
    \node[xshift=-6,RoyalBlue] at (-1,0) {\footnotesize$4$};
    \node[yshift=6,RoyalBlue] at (0,1) {\footnotesize$5$};
    \node[xshift=6,RoyalBlue] at (1,0) {\footnotesize$2$};
\end{tikzpicture}
+
c_2
\begin{tikzpicture}[line width=1,scale=0.20,baseline={([yshift=-0.4ex]current bounding box.center)}]
	\coordinate (v1) at (-1,-1);
    \coordinate (v2) at (-1,1);
    \coordinate (v3) at (1,1);
    \coordinate (v4) at (1,-1);
    \draw[] (v1) -- (v2) -- (v3) -- (v4) -- (v1);
    \draw[] (v1) -- ++(-135:0.7);
    \draw[] (v2) -- ++(135:0.7) node[left] {};
    \draw[] (v3) -- ++(60:0.7) node[right,yshift=2] {};
    \draw[] (v3) -- ++(30:0.7);
    \draw[] (v4) -- ++(-45:0.7) node[right] {};
    \node[yshift=-6,RoyalBlue] at (0,-1) {\footnotesize $4$};
    \node[xshift=-6,RoyalBlue] at (-1,0) {\footnotesize $5$};
    \node[yshift=6,RoyalBlue] at (0,1) {\footnotesize $1$};
    \node[xshift=6,RoyalBlue] at (1,0) {\footnotesize $3$};
\end{tikzpicture}
+
c_3
\begin{tikzpicture}[line width=1,scale=0.20,baseline={([yshift=-0.4ex]current bounding box.center)}]
	\coordinate (v1) at (-1,-1);
    \coordinate (v2) at (-1,1);
    \coordinate (v3) at (1,1);
    \coordinate (v4) at (1,-1);
    \draw[] (v1) -- (v2) -- (v3) -- (v4) -- (v1);
    \draw[] (v1) -- ++(-135:0.7);
    \draw[] (v2) -- ++(135:0.7) node[left] {};
    \draw[] (v3) -- ++(60:0.7) node[right,yshift=2] {};
    \draw[] (v3) -- ++(30:0.7);
    \draw[] (v4) -- ++(-45:0.7) node[right] {};
    \node[yshift=-6,RoyalBlue] at (0,-1) {\footnotesize $5$};
    \node[xshift=-6,RoyalBlue] at (-1,0) {\footnotesize $1$};
    \node[yshift=6,RoyalBlue] at (0,1) {\footnotesize $2$};
    \node[xshift=6,RoyalBlue] at (1,0) {\footnotesize $4$};
\end{tikzpicture}
\\
& +
c_4
\begin{tikzpicture}[line width=1,scale=0.20,baseline={([yshift=-0.4ex]current bounding box.center)}]
	\coordinate (v1) at (-1,-1);
    \coordinate (v2) at (-1,1);
    \coordinate (v3) at (1,1);
    \coordinate (v4) at (1,-1);
    \draw[] (v1) -- (v2) -- (v3) -- (v4) -- (v1);
    \draw[] (v1) -- ++(-135:0.7);
    \draw[] (v2) -- ++(135:0.7) node[left] {};
    \draw[] (v3) -- ++(60:0.7) node[right,yshift=2] {};
    \draw[] (v3) -- ++(30:0.7);
    \draw[] (v4) -- ++(-45:0.7) node[right] {};
    \node[yshift=-6,RoyalBlue] at (0,-1) {\footnotesize $1$};
    \node[xshift=-6,RoyalBlue] at (-1,0) {\footnotesize $2$};
    \node[yshift=6,RoyalBlue] at (0,1) {\footnotesize $3$};
    \node[xshift=6,RoyalBlue] at (1,0) {\footnotesize $5$};
\end{tikzpicture} +
c_5
\begin{tikzpicture}[line width=1,scale=0.20,baseline={([yshift=-0.4ex]current bounding box.center)}]
	\coordinate (v1) at (-1,-1);
    \coordinate (v2) at (-1,1);
    \coordinate (v3) at (1,1);
    \coordinate (v4) at (1,-1);
    \draw[] (v1) -- (v2) -- (v3) -- (v4) -- (v1);
    \draw[] (v1) -- ++(-135:0.7);
    \draw[] (v2) -- ++(135:0.7) node[left] {};
    \draw[] (v3) -- ++(60:0.7) node[right,yshift=2] {};
    \draw[] (v3) -- ++(30:0.7);
    \draw[] (v4) -- ++(-45:0.7) node[right] {};
    \node[yshift=-6,RoyalBlue] at (0,-1) {\footnotesize $2$};
    \node[xshift=-6,RoyalBlue] at (-1,0) {\footnotesize $3$};
    \node[yshift=6,RoyalBlue] at (0,1) {\footnotesize $4$};
    \node[xshift=6,RoyalBlue] at (1,0) {\footnotesize $1$};
\end{tikzpicture}
\nonumber
\end{align}
The quadruple cut of Ex.~\ref{ex:quadcut} annihilates all master integrals except the last, leaving a linear equation for the single coefficient $c_5$:
\be
\mathrm{Cut}^G_{\{1,2,3,4\}} {\mathcal A} =
c_5\,\mathrm{Cut}^G_{\{1,2,3,4\}} \begin{tikzpicture}[line width=1,scale=0.20,baseline={([yshift=-0.4ex]current bounding box.center)}]
	\coordinate (v1) at (-1,-1);
    \coordinate (v2) at (-1,1);
    \coordinate (v3) at (1,1);
    \coordinate (v4) at (1,-1);
    \draw[] (v1) -- (v2) -- (v3) -- (v4) -- (v1);
    \draw[] (v1) -- ++(-135:0.7);
    \draw[] (v2) -- ++(135:0.7) node[left] {};
    \draw[] (v3) -- ++(60:0.7) node[right,yshift=2] {};
    \draw[] (v3) -- ++(30:0.7);
    \draw[] (v4) -- ++(-45:0.7) node[right] {};
    \node[yshift=-6,RoyalBlue] at (0,-1) {\footnotesize $2$};
    \node[xshift=-6,RoyalBlue] at (-1,0) {\footnotesize $3$};
    \node[yshift=6,RoyalBlue] at (0,1) {\footnotesize $4$};
    \node[xshift=6,RoyalBlue] at (1,0) {\footnotesize $1$};
\end{tikzpicture} 
= \frac{8 \pi^2 c_5}{i st}\, ,
\ee
where
\be
c_5 = \frac{st}{2} {\mathcal A}^{\rm tree}_1 {\mathcal A}^{\rm tree}_2 {\mathcal A}^{\rm tree}_3 {\mathcal A}^{\rm tree}_4\, .
\ee
Here ${\mathcal A}_i^{\rm tree}$ are the $3$- and $4$-particle amplitudes glued together by the cuts.
Likewise, each of the remaining coefficients $c_i$ can be isolated by a single quadruple cut.
\end{mdexample}

In Ex. \ref{ex:boxcoeffs}, we used a set of generalized cuts that isolated individual box integrals. In general, for example in a one-loop amplitude in a theory involving triangle and bubble master integrals, their coefficients could only be obtained from cuts of fewer propagators, which also act nontrivially on the boxes.
Generalized unitarity methods are most complete in cases where the master integrals are fully polylogarithmic, giving good control of the basis of functions and their associated branch cuts.
See \cite{Weinzierl:2022eaz,Brandhuber:2022qbk,Badger:2023eqz} for some recent reviews and further discussion.

\subsection{Solution space of generalized cuts}

The {\it maximal cut} of a Feynman integral  is a cut of all propagators, such as in the Ex.~\ref{ex:bubble_cut}, \ref{sec:max_cut_tri} and \ref{ex:quadcut}. Maximal cuts are closely related to the original integral but typically far simpler.

Beyond one loop, cuts of a given set of propagators are not always unique. We illustrate this fact on the maximal cut of a sunrise integral.

\begin{mdexample}\label{ex:sunrise-max-cuts}
Consider the two-loop sunrise integral with a single massive propagator in  $D=2-2\epsilon$ dimensions. The first loop integration gives the maximal cut of a scalar bubble:
\begin{align}
&\mathrm{Cut}^G_{\{1,2,3\}}     {\mathcal{I}}_{\rm sunrise} (p^2,m^2) = \begin{gathered}
    \begin{tikzpicture}[line width=1,scale=0.35,baseline={([yshift=-0.8ex]current bounding box.center)}]
	\coordinate (v1) at (-1,0);
    \coordinate (v2) at (1,0);
    \draw[] (v1) [out=55,in=125] to (v2);
    \draw[] (v1) [out=-55,in=-125] to (v2);
    \draw[] (v1) -- (v2);
    \draw[] (v1) -- ++(-135:0.7) node[left,scale=0.75] {$p_1$};
    \draw[] (v1) -- ++(135:0.7) node[left,scale=0.75] {$p_2$};
    \draw[] (v2) -- ++(45:0.7) node[right,scale=0.75] {$p_3$};
    \draw[] (v2) -- ++(-45:0.7) node[right,scale=0.75] {$p_4$};
    \draw[dashed, orange] (0,1.2) -- (0,-1.2);
    \end{tikzpicture}
    \end{gathered}
    \\ 
& =\int_{\ell_1, \ell_2}
\ddelta(\ell_1^2)\, \ddelta(\ell_2^2)\, \ddelta[(\ell_1+\ell_2+p)^2-m^2] \nonumber  \\
& = \alpha_1 \int_{\ell}
 \ddelta(\ell^2) \left[(\ell+p)^2-m^2\right]^{-1-2\epsilon} \left[(\ell+p)^2\right]^{\epsilon} 
    \nonumber \\
& = \alpha_2 \int {\d\ell^0}\,
\frac{(\ell^0)^{-1-2\epsilon} \left(p^2+2\sqrt{p^2}\ell^0\right)^{2\epsilon}}{\left(p^2-m^2+2\sqrt{p^2}\ell^0\right)^{1+2\epsilon}}	\, .
    \nonumber 
\end{align}
See \cite{Abreu:2021vhb} for  the normalization factors $\alpha_1$ and $\alpha_2$, and further details.

The contour of the last integral is unspecified, and we can identify two independent choices of the maximal cut.
The natural choices of contour extend between two endpoints that are branch points of the integrand. 
In this case, we might recognize this integral as taking the form of the Gauss hypergeometric function.
There are several related ways to understand that we expect two solutions to the maximal cut of this integral. 
One is by observing the two independent integration contours to complete the maximal cut calculation as above. 

Another is by considering the basis of master integrals related to the original one by IBP and dimension-shift identities, and observing that there is a linearly independent integral in this basis with the same set of propagators.
For example, it could be taken to contain an additional factor of $(\ell_1+\ell_2+p)^2$ in the numerator of the integrand.

A third is to consider the system of first-order linear differential equations satisfied by the set of master integrals. Generalized cuts of the master integrals satisfy the same system of differential equations~\cite{Primo:2016ebd,Frellesvig:2017aai,Bosma:2017hrk}, with possibly different boundary conditions, and possibly spanning a smaller subspace than the master integrals.
In this example of the one-mass sunrise, there are no master integrals with fewer propagators, and so the maximal cuts span the full solution space. It is possible to express both the uncut integral and the unitarity cut in $p^2$ in terms of any two independent maximal cuts.
\end{mdexample}

Especially in cases such as Ex. \ref{ex:sunrise-max-cuts}, with multiple solutions for a given set of cut propagators, it is helpful to formulate a generalized cut as a choice of a new integration contour along with the original, uncut, integrand. The number of cut solutions is the dimension of the homology group associated to a space with singularities identified with the chosen propagators. It also becomes apparent that generalized cut integrals satisfy the same linear (IBP) relations and differential equations as the original integral.\footnote{Cuts of propagators raised to nonpositive powers are zero. Therefore maximal cuts satisfy a highly simplified set of differential equations, explaining their relative simplicity as functions.}

This observation also gives a new perspective on why computational techniques for loop computations can be applied to cross section computations, cf. the discussion of reverse-unitarity in Sec.~\ref{sec:reverse-unitarity}. Thus, associated to a given Feynman integral, it is instructive to generate the so-called period matrix, whose entries are pairings of all homology and cohomology classes\footnote{In dimensional regularization, we require twisted homology and cohomology groups, due to the multivaluedness of the integrand in noninteger dimension \cite{Mastrolia:2018uzb,Matsubara-Heo:2023ylc}.} obtained from the singular loci of the Feynman integral. The entries of the period matrix are then understood as the related master integrals, and all of their independent generalized cuts. See \cite{Abreu:2022mfk} for a review of these topics with additional references. 

\section{Summary}

This article summarized different notions of cuts used throughout the literature on scattering amplitudes and related topics. We grouped them into three categories, depending on which energies are allowed to flow through a cut and whether the right-hand of the cut is conjugated or not. Such choices can make an enormous difference for the result of the computations, as highlighted in the explicit examples above. We moreover emphasized that cuts, and relations derived from them, strongly depend on the kinematic domain in which they are evaluated. Applications, ranging from dispersion relations, through symbol bootstrap, to generalized unitarity, serve as evidence that cuts are comprehensive and versatile computational tools.

\section*{Acknowledgments}
The work of C.D. is funded by the European Union (ERC Consolidator Grant LoCoMotive 101043686). Views and opinions expressed are however those of the author(s) only and do not necessarily reflect those of the European Union or the European Research Council. Neither the European Union nor the granting authority can be held responsible for them.
H.S.H. gratefully acknowledges funding provided by the William D. Loughlin Membership, an endowed fund of the Institute for Advanced Study.
S.M. gratefully acknowledges funding provided by the Sivian Fund and the Roger Dashen Member Fund at the Institute for Advanced Study. 
This material is based upon work supported by the U.S. Department of Energy, Office of Science, Office of High Energy Physics under Award Number DE-SC0009988.

\bibliographystyle{JHEP}
\bibliography{references}

\providecommand{\href}[2]{#2}\begingroup\raggedright\begin{thebibliography}{10}

\bibitem{Abreu:2022mfk}
S.~Abreu, R.~Britto and C.~Duhr, \emph{{The SAGEX review on scattering
  amplitudes Chapter 3: Mathematical structures in Feynman integrals}},
  \href{https://doi.org/10.1088/1751-8121/ac87de}{\emph{J. Phys. A} {\bfseries
  55} (2022) 443004} [\href{https://arxiv.org/abs/2203.13014}{{\ttfamily
  2203.13014}}].

\bibitem{Abreu:2017ptx}
S.~Abreu, R.~Britto, C.~Duhr and E.~Gardi, \emph{{Cuts from residues: the
  one-loop case}}, \href{https://doi.org/10.1007/JHEP06(2017)114}{\emph{JHEP}
  {\bfseries 06} (2017) 114}
  [\href{https://arxiv.org/abs/1702.03163}{{\ttfamily 1702.03163}}].

\bibitem{Abreu:2021vhb}
S.~Abreu, R.~Britto, C.~Duhr, E.~Gardi and J.~Matthew, \emph{{The diagrammatic
  coaction beyond one loop}},
  \href{https://doi.org/10.1007/JHEP10(2021)131}{\emph{JHEP} {\bfseries 10}
  (2021) 131} [\href{https://arxiv.org/abs/2106.01280}{{\ttfamily
  2106.01280}}].

\bibitem{Adams:2006sv}
A.~Adams, N.~Arkani-Hamed, S.~Dubovsky, A.~Nicolis and R.~Rattazzi,
  \emph{{Causality, analyticity and an IR obstruction to UV completion}},
  \href{https://doi.org/10.1088/1126-6708/2006/10/014}{\emph{JHEP} {\bfseries
  10} (2006) 014} [\href{https://arxiv.org/abs/hep-th/0602178}{{\ttfamily
  hep-th/0602178}}].

\bibitem{Anastasiou:2003yy}
C.~Anastasiou, L.J.~Dixon, K.~Melnikov and F.~Petriello, \emph{{Dilepton
  rapidity distribution in the Drell-Yan process at NNLO in QCD}},
  \href{https://doi.org/10.1103/PhysRevLett.91.182002}{\emph{Phys. Rev. Lett.}
  {\bfseries 91} (2003) 182002}
  [\href{https://arxiv.org/abs/hep-ph/0306192}{{\ttfamily hep-ph/0306192}}].

\bibitem{Anastasiou:2015vya}
C.~Anastasiou, C.~Duhr, F.~Dulat, F.~Herzog and B.~Mistlberger, \emph{{Higgs
  Boson Gluon-Fusion Production in QCD at Three Loops}},
  \href{https://doi.org/10.1103/PhysRevLett.114.212001}{\emph{Phys. Rev. Lett.}
  {\bfseries 114} (2015) 212001}
  [\href{https://arxiv.org/abs/1503.06056}{{\ttfamily 1503.06056}}].

\bibitem{Anastasiou:2002yz}
C.~Anastasiou and K.~Melnikov, \emph{{Higgs boson production at hadron
  colliders in NNLO QCD}},
  \href{https://doi.org/10.1016/S0550-3213(02)00837-4}{\emph{Nucl. Phys. B}
  {\bfseries 646} (2002) 220}
  [\href{https://arxiv.org/abs/hep-ph/0207004}{{\ttfamily hep-ph/0207004}}].

\bibitem{Arkani-Hamed:2017ahv}
N.~Arkani-Hamed and E.Y.~Yuan, \emph{{One-Loop Integrals from Spherical
  Projections of Planes and Quadrics}},
  \href{https://arxiv.org/abs/1712.09991}{{\ttfamily 1712.09991}}.

\bibitem{Badger:2023eqz}
S.~Badger, J.~Henn, J.C.~Plefka and S.~Zoia, \emph{{Scattering Amplitudes in
  Quantum Field Theory}},
  \href{https://doi.org/10.1007/978-3-031-46987-9}{\emph{Lect. Notes Phys.}
  {\bfseries 1021} (2024) pp.}
  [\href{https://arxiv.org/abs/2306.05976}{{\ttfamily 2306.05976}}].

\bibitem{Baumann:2022jpr}
D.~Baumann, D.~Green, A.~Joyce, E.~Pajer, G.L.~Pimentel, C.~Sleight et~al.,
  \emph{{Snowmass White Paper: The Cosmological Bootstrap}},  in
  \emph{{Snowmass 2021}}, 3, 2022
  [\href{https://arxiv.org/abs/2203.08121}{{\ttfamily 2203.08121}}].

\bibitem{Berghoff:2022mqu}
M.~Berghoff and E.~Panzer, \emph{{Hierarchies in relative Picard-Lefschetz
  theory}},  \href{https://arxiv.org/abs/2212.06661}{{\ttfamily 2212.06661}}.

\bibitem{Bern:1994zx}
Z.~Bern, L.J.~Dixon, D.C.~Dunbar and D.A.~Kosower, \emph{{One-Loop n-Point
  Gauge Theory Amplitudes, Unitarity and Collinear Limits}},
  \href{https://doi.org/10.1016/0550-3213(94)90179-1}{\emph{Nucl. Phys.}
  {\bfseries B425} (1994) 217}
  [\href{https://arxiv.org/abs/hep-ph/9403226}{{\ttfamily hep-ph/9403226}}].

\bibitem{Bern:1994cg}
Z.~Bern, L.J.~Dixon, D.C.~Dunbar and D.A.~Kosower, \emph{{Fusing gauge theory
  tree amplitudes into loop amplitudes}},
  \href{https://doi.org/10.1016/0550-3213(94)00488-Z}{\emph{Nucl. Phys.}
  {\bfseries B435} (1995) 59}
  [\href{https://arxiv.org/abs/hep-ph/9409265}{{\ttfamily hep-ph/9409265}}].

\bibitem{Bjorken:1959fd}
J.D.~Bjorken, \emph{{Experimental tests of Quantum electrodynamics and spectral
  representations of Green's functions in perturbation theory}}, Ph.D. thesis,
  Stanford U., 1959.

\bibitem{Bloch:2015efx}
S.~Bloch and D.~Kreimer, \emph{{Cutkosky Rules and Outer Space}},
  \href{https://arxiv.org/abs/1512.01705}{{\ttfamily 1512.01705}}.

\bibitem{Bosma:2017hrk}
J.~Bosma, K.J.~Larsen and Y.~Zhang, \emph{{Differential equations for loop
  integrals in Baikov representation}},
  \href{https://doi.org/10.1103/PhysRevD.97.105014}{\emph{Phys. Rev. D}
  {\bfseries 97} (2018) 105014}
  [\href{https://arxiv.org/abs/1712.03760}{{\ttfamily 1712.03760}}].

\bibitem{Boyling1968}
J.B.~Boyling, \emph{A homological approach to parametric feynman integrals},
  \href{https://doi.org/10.1007/BF02800115}{\emph{Il Nuovo Cimento A
  (1965-1970)} {\bfseries 53} (1968) 351}.

\bibitem{10.1063/1.1704822}
J.B.~Boyling, \emph{{Construction of Vanishing Cycles for Integrals over
  Hyperspheres}}, \href{https://doi.org/10.1063/1.1704822}{\emph{Journal of
  Mathematical Physics} {\bfseries 7} (2004) 1749}.

\bibitem{Brandhuber:2022qbk}
A.~Brandhuber, J.~Plefka and G.~Travaglini, \emph{{The SAGEX Review on
  Scattering Amplitudes Chapter 1: Modern Fundamentals of Amplitudes}},
  \href{https://doi.org/10.1088/1751-8121/ac8254}{\emph{J. Phys. A} {\bfseries
  55} (2022) 443002} [\href{https://arxiv.org/abs/2203.13012}{{\ttfamily
  2203.13012}}].

\bibitem{Britto:2023rig}
R.~Britto, \emph{{Generalized Cuts of Feynman Integrals in Parameter Space}},
  \href{https://doi.org/10.1103/PhysRevLett.131.091601}{\emph{Phys. Rev. Lett.}
  {\bfseries 131} (2023) 091601}
  [\href{https://arxiv.org/abs/2305.15369}{{\ttfamily 2305.15369}}].

\bibitem{Britto:2004nc}
R.~Britto, F.~Cachazo and B.~Feng, \emph{{Generalized unitarity and one-loop
  amplitudes in N=4 super-Yang-Mills}},
  \href{https://doi.org/10.1016/j.nuclphysb.2005.07.014}{\emph{Nucl.Phys.}
  {\bfseries B725} (2005) 275}
  [\href{https://arxiv.org/abs/hep-th/0412103}{{\ttfamily hep-th/0412103}}].

\bibitem{BROS200687}
J.~Bros, \emph{Dispersion relations},  in \emph{Encyclopedia of Mathematical
  Physics}, J.-P.~Françoise, G.L.~Naber and T.S.~Tsun, eds., (Oxford),
  pp.~87--101, Academic Press (2006),
  \href{https://doi.org/https://doi.org/10.1016/B0-12-512666-2/00306-0}{DOI}.

\bibitem{BROS2006465}
J.~Bros, \emph{Scattering in relativistic quantum field theory: The analytic
  program},  in \emph{Encyclopedia of Mathematical Physics}, J.-P.~Françoise,
  G.L.~Naber and T.S.~Tsun, eds., (Oxford), pp.~465--475, Academic Press
  (2006),
  \href{https://doi.org/https://doi.org/10.1016/B0-12-512666-2/00517-4}{DOI}.

\bibitem{Bros:1965kbd}
J.~Bros, H.~Epstein and V.~Glaser, \emph{{A proof of the crossing property for
  two-particle amplitudes in general quantum field theory}},
  \href{https://doi.org/10.1007/BF01646307}{\emph{Commun. Math. Phys.}
  {\bfseries 1} (1965) 240}.

\bibitem{Brown:coaction}
F.~Brown, \emph{{Notes on Motivic Periods}}, {\emph{Commun. Num. Theor Phys.}
  {\bfseries 11} (2017) 557}
  [\href{https://arxiv.org/abs/1512.06410}{{\ttfamily 1512.06410}}].

\bibitem{Brown:2015ztw}
F.~Brown and O.~Schnetz, \emph{{Single-valued multiple polylogarithms and a
  proof of the zig\textendash{}zag conjecture}},
  \href{https://doi.org/10.1016/j.jnt.2014.09.007}{\emph{J. Number Theor.}
  {\bfseries 148} (2015) 478}.

\bibitem{Brown:2009ta}
F.C.S.~Brown, \emph{{On the periods of some Feynman integrals}},
  \href{https://arxiv.org/abs/0910.0114}{{\ttfamily 0910.0114}}.

\bibitem{Brown:2009qja}
F.C.~Brown, \emph{{Multiple zeta values and periods of moduli spaces
  $\mathfrak{M}_{0,n}$}}, {\emph{Annales Sci.Ecole Norm.Sup.} {\bfseries 42}
  (2009) 371} [\href{https://arxiv.org/abs/math/0606419}{{\ttfamily
  math/0606419}}].

\bibitem{BUCHHOLZ2006456}
D.~Buchholz and S.~Summers, \emph{Scattering in relativistic quantum field
  theory: Fundamental concepts and tools},  in \emph{Encyclopedia of
  Mathematical Physics}, J.-P.~Françoise, G.L.~Naber and T.S.~Tsun, eds.,
  (Oxford), pp.~456--465, Academic Press (2006),
  \href{https://doi.org/https://doi.org/10.1016/B0-12-512666-2/00018-3}{DOI}.

\bibitem{Capatti:2019ypt}
Z.~Capatti, V.~Hirschi, D.~Kermanschah and B.~Ruijl, \emph{{Loop-Tree Duality
  for Multiloop Numerical Integration}},
  \href{https://doi.org/10.1103/PhysRevLett.123.151602}{\emph{Phys. Rev. Lett.}
  {\bfseries 123} (2019) 151602}
  [\href{https://arxiv.org/abs/1906.06138}{{\ttfamily 1906.06138}}].

\bibitem{Caron-Huot:2019bsq}
S.~Caron-Huot, L.J.~Dixon, F.~Dulat, M.~Von~Hippel, A.J.~McLeod and
  G.~Papathanasiou, \emph{{The Cosmic Galois Group and Extended Steinmann
  Relations for Planar $\mathcal{N} = 4$ SYM Amplitudes}},
  \href{https://doi.org/10.1007/JHEP09(2019)061}{\emph{JHEP} {\bfseries 09}
  (2019) 061} [\href{https://arxiv.org/abs/1906.07116}{{\ttfamily
  1906.07116}}].

\bibitem{Caron-Huot:2023vxl}
S.~Caron-Huot, M.~Giroux, H.S.~Hannesdottir and S.~Mizera, \emph{{What can be
  measured asymptotically?}},
  \href{https://doi.org/10.1007/JHEP01(2024)139}{\emph{JHEP} {\bfseries 01}
  (2024) 139} [\href{https://arxiv.org/abs/2308.02125}{{\ttfamily
  2308.02125}}].

\bibitem{Chavez:2012kn}
F.~Chavez and C.~Duhr, \emph{{Three-mass triangle integrals and single-valued
  polylogarithms}}, \href{https://doi.org/10.1007/JHEP11(2012)114}{\emph{JHEP}
  {\bfseries 11} (2012) 114} [\href{https://arxiv.org/abs/1209.2722}{{\ttfamily
  1209.2722}}].

\bibitem{Chen:1977oja}
K.-T.~Chen, \emph{{Iterated path integrals}},
  \href{https://doi.org/10.1090/S0002-9904-1977-14320-6}{\emph{Bull.Am.Math.Soc.}
  {\bfseries 83} (1977) 831}.

\bibitem{Chen:2021isd}
X.~Chen, T.~Gehrmann, E.W.N.~Glover, A.~Huss, B.~Mistlberger and A.~Pelloni,
  \emph{{Fully Differential Higgs Boson Production to Third Order in QCD}},
  \href{https://doi.org/10.1103/PhysRevLett.127.072002}{\emph{Phys. Rev. Lett.}
  {\bfseries 127} (2021) 072002}
  [\href{https://arxiv.org/abs/2102.07607}{{\ttfamily 2102.07607}}].

\bibitem{Chetyrkin:1981qh}
K.~Chetyrkin and F.~Tkachov, \emph{{Integration by Parts: The Algorithm to
  Calculate beta Functions in 4 Loops}},
  \href{https://doi.org/10.1016/0550-3213(81)90199-1}{\emph{Nucl. Phys. B}
  {\bfseries 192} (1981) 159}.

\bibitem{Chou:1984es}
K.-c.~Chou, Z.-b.~Su, B.-l.~Hao and L.~Yu, \emph{{Equilibrium and
  Nonequilibrium Formalisms Made Unified}},
  \href{https://doi.org/10.1016/0370-1573(85)90136-X}{\emph{Phys. Rept.}
  {\bfseries 118} (1985) 1}.

\bibitem{Coleman:1965xm}
S.~Coleman and R.~Norton, \emph{{Singularities in the physical region}},
  \href{https://doi.org/10.1007/BF02750472}{\emph{Nuovo Cim.} {\bfseries 38}
  (1965) 438}.

\bibitem{Correia:2020xtr}
M.~Correia, A.~Sever and A.~Zhiboedov, \emph{{An analytical toolkit for the
  S-matrix bootstrap}},
  \href{https://doi.org/10.1007/JHEP03(2021)013}{\emph{JHEP} {\bfseries 03}
  (2021) 013} [\href{https://arxiv.org/abs/2006.08221}{{\ttfamily
  2006.08221}}].

\bibitem{Cutkosky:1960sp}
R.E.~Cutkosky, \emph{{Singularities and discontinuities of Feynman
  amplitudes}}, \href{https://doi.org/10.1063/1.1703676}{\emph{J. Math. Phys.}
  {\bfseries 1} (1960) 429}.

\bibitem{deJesusAguilera-Verdugo:2021mvg}
J.~de~Jes\'us Aguilera-Verdugo et~al., \emph{{A Stroll through the Loop-Tree
  Duality}}, \href{https://doi.org/10.3390/sym13061029}{\emph{Symmetry}
  {\bfseries 13} (2021) 1029}
  [\href{https://arxiv.org/abs/2104.14621}{{\ttfamily 2104.14621}}].

\bibitem{deRham:2022hpx}
C.~de~Rham, S.~Kundu, M.~Reece, A.J.~Tolley and S.-Y.~Zhou, \emph{{Snowmass
  White Paper: UV Constraints on IR Physics}},  in \emph{{Snowmass 2021}}, 3,
  2022 [\href{https://arxiv.org/abs/2203.06805}{{\ttfamily 2203.06805}}].

\bibitem{Denner:2006ic}
A.~Denner and S.~Dittmaier, \emph{{The Complex-mass scheme for perturbative
  calculations with unstable particles}},
  \href{https://doi.org/10.1016/j.nuclphysbps.2006.09.025}{\emph{Nucl. Phys. B
  Proc. Suppl.} {\bfseries 160} (2006) 22}
  [\href{https://arxiv.org/abs/hep-ph/0605312}{{\ttfamily hep-ph/0605312}}].

\bibitem{Denner:1999gp}
A.~Denner, S.~Dittmaier, M.~Roth and D.~Wackeroth, \emph{{Predictions for all
  processes e+ e- ---\ensuremath{>} 4 fermions + gamma}},
  \href{https://doi.org/10.1016/S0550-3213(99)00437-X}{\emph{Nucl. Phys. B}
  {\bfseries 560} (1999) 33}
  [\href{https://arxiv.org/abs/hep-ph/9904472}{{\ttfamily hep-ph/9904472}}].

\bibitem{Denner:2005fg}
A.~Denner, S.~Dittmaier, M.~Roth and L.H.~Wieders, \emph{{Electroweak
  corrections to charged-current e+ e- ---\ensuremath{>} 4 fermion processes:
  Technical details and further results}},
  \href{https://doi.org/10.1016/j.nuclphysb.2011.09.001}{\emph{Nucl. Phys. B}
  {\bfseries 724} (2005) 247}
  [\href{https://arxiv.org/abs/hep-ph/0505042}{{\ttfamily hep-ph/0505042}}].

\bibitem{Duhr:2012fh}
C.~Duhr, \emph{{Hopf algebras, coproducts and symbols: an application to Higgs
  boson amplitudes}},
  \href{https://doi.org/10.1007/JHEP08(2012)043}{\emph{JHEP} {\bfseries 1208}
  (2012) 043} [\href{https://arxiv.org/abs/1203.0454}{{\ttfamily 1203.0454}}].

\bibitem{Duhr:2019tlz}
C.~Duhr and F.~Dulat, \emph{{PolyLogTools \textemdash{} polylogs for the
  masses}}, \href{https://doi.org/10.1007/JHEP08(2019)135}{\emph{JHEP}
  {\bfseries 08} (2019) 135}
  [\href{https://arxiv.org/abs/1904.07279}{{\ttfamily 1904.07279}}].

\bibitem{Duhr:2020sdp}
C.~Duhr, F.~Dulat and B.~Mistlberger, \emph{{Charged current Drell-Yan
  production at N$^{3}$LO}},
  \href{https://doi.org/10.1007/JHEP11(2020)143}{\emph{JHEP} {\bfseries 11}
  (2020) 143} [\href{https://arxiv.org/abs/2007.13313}{{\ttfamily
  2007.13313}}].

\bibitem{Duhr:2020seh}
C.~Duhr, F.~Dulat and B.~Mistlberger, \emph{{Drell-Yan Cross Section to Third
  Order in the Strong Coupling Constant}},
  \href{https://doi.org/10.1103/PhysRevLett.125.172001}{\emph{Phys. Rev. Lett.}
  {\bfseries 125} (2020) 172001}
  [\href{https://arxiv.org/abs/2001.07717}{{\ttfamily 2001.07717}}].

\bibitem{Duhr:2019kwi}
C.~Duhr, F.~Dulat and B.~Mistlberger, \emph{{Higgs Boson Production in
  Bottom-Quark Fusion to Third Order in the Strong Coupling}},
  \href{https://doi.org/10.1103/PhysRevLett.125.051804}{\emph{Phys. Rev. Lett.}
  {\bfseries 125} (2020) 051804}
  [\href{https://arxiv.org/abs/1904.09990}{{\ttfamily 1904.09990}}].

\bibitem{Duhr:2011zq}
C.~Duhr, H.~Gangl and J.R.~Rhodes, \emph{{From polygons and symbols to
  polylogarithmic functions}},
  \href{https://doi.org/10.1007/JHEP10(2012)075}{\emph{JHEP} {\bfseries 1210}
  (2012) 075} [\href{https://arxiv.org/abs/1110.0458}{{\ttfamily 1110.0458}}].

\bibitem{Duhr:2021vwj}
C.~Duhr and B.~Mistlberger, \emph{{Lepton-pair production at hadron colliders
  at N$^{3}$LO in QCD}},
  \href{https://doi.org/10.1007/JHEP03(2022)116}{\emph{JHEP} {\bfseries 03}
  (2022) 116} [\href{https://arxiv.org/abs/2111.10379}{{\ttfamily
  2111.10379}}].

\bibitem{Eberhardt:2022zay}
L.~Eberhardt and S.~Mizera, \emph{{Unitarity cuts of the worldsheet}},
  \href{https://doi.org/10.21468/SciPostPhys.14.2.015}{\emph{SciPost Phys.}
  {\bfseries 14} (2023) 015}
  [\href{https://arxiv.org/abs/2208.12233}{{\ttfamily 2208.12233}}].

\bibitem{doi:10.1063/1.1724262}
D.B.~Fairlie, P.V.~Landshoff, J.~Nuttall and J.C.~Polkinghorne,
  \emph{{Singularities of the Second Type}},
  \href{https://doi.org/10.1063/1.1724262}{\emph{Journal of Mathematical
  Physics} {\bfseries 3} (1962) 594}.

\bibitem{Fevola:2023kaw}
C.~Fevola, S.~Mizera and S.~Telen, \emph{{Landau Singularities Revisited}},
  \href{https://arxiv.org/abs/2311.14669}{{\ttfamily 2311.14669}}.

\bibitem{Feynman:1963ax}
R.P.~Feynman, \emph{{Quantum theory of gravitation}}, {\emph{Acta Phys. Polon.}
  {\bfseries 24} (1963) 697}.

\bibitem{Frellesvig:2017aai}
H.~Frellesvig and C.G.~Papadopoulos, \emph{{Cuts of Feynman Integrals in Baikov
  representation}}, \href{https://doi.org/10.1007/JHEP04(2017)083}{\emph{JHEP}
  {\bfseries 04} (2017) 083}
  [\href{https://arxiv.org/abs/1701.07356}{{\ttfamily 1701.07356}}].

\bibitem{Gaiotto:2011dt}
D.~Gaiotto, J.~Maldacena, A.~Sever and P.~Vieira, \emph{{Pulling the straps of
  polygons}}, \href{https://doi.org/10.1007/JHEP12(2011)011}{\emph{JHEP}
  {\bfseries 12} (2011) 011} [\href{https://arxiv.org/abs/1102.0062}{{\ttfamily
  1102.0062}}].

\bibitem{Gelis:2015kya}
F.~Gelis and N.~Tanji, \emph{{Schwinger mechanism revisited}},
  \href{https://doi.org/10.1016/j.ppnp.2015.11.001}{\emph{Prog. Part. Nucl.
  Phys.} {\bfseries 87} (2016) 1}
  [\href{https://arxiv.org/abs/1510.05451}{{\ttfamily 1510.05451}}].

\bibitem{10.1063/1.1665233}
P.~Goddard, \emph{{Nonphysical Region Singularities of the S Matrix}},
  \href{https://doi.org/10.1063/1.1665233}{\emph{Journal of Mathematical
  Physics} {\bfseries 11} (2003) 960}.

\bibitem{Goncharov.A.B.:2009tja}
A.~Goncharov, \emph{{A simple construction of Grassmannian polylogarithms}},
  \href{https://arxiv.org/abs/0908.2238}{{\ttfamily 0908.2238}}.

\bibitem{Goncharov:2010jf}
A.B.~Goncharov, M.~Spradlin, C.~Vergu and A.~Volovich, \emph{{Classical
  Polylogarithms for Amplitudes and Wilson Loops}},
  \href{https://doi.org/10.1103/PhysRevLett.105.151605}{\emph{Phys. Rev. Lett.}
  {\bfseries 105} (2010) 151605}
  [\href{https://arxiv.org/abs/1006.5703}{{\ttfamily 1006.5703}}].

\bibitem{Hannesdottir:2022xki}
H.S.~Hannesdottir, A.J.~McLeod, M.D.~Schwartz and C.~Vergu, \emph{{Constraints
  on sequential discontinuities from the geometry of on-shell spaces}},
  \href{https://doi.org/10.1007/JHEP07(2023)236}{\emph{JHEP} {\bfseries 07}
  (2023) 236} [\href{https://arxiv.org/abs/2211.07633}{{\ttfamily
  2211.07633}}].

\bibitem{Hannesdottir:2022bmo}
H.S.~Hannesdottir and S.~Mizera, \emph{{What is the i\ensuremath{\varepsilon}
  for the S-matrix?}}, SpringerBriefs in Physics, Springer (1, 2023),
  \href{https://doi.org/10.1007/978-3-031-18258-7}{10.1007/978-3-031-18258-7},
  [\href{https://arxiv.org/abs/2204.02988}{{\ttfamily 2204.02988}}].

\bibitem{IAGOLNITZER2006475}
D.~Iagolnitzer and J.~Magnen, \emph{Scattering, asymptotic completeness and
  bound states},  in \emph{Encyclopedia of Mathematical Physics},
  J.-P.~Françoise, G.L.~Naber and T.S.~Tsun, eds., (Oxford), pp.~475--487,
  Academic Press (2006),
  \href{https://doi.org/https://doi.org/10.1016/B0-12-512666-2/00084-5}{DOI}.

\bibitem{Keldysh:1964ud}
L.V.~Keldysh, \emph{{Diagram technique for nonequilibrium processes}},
  {\emph{Zh. Eksp. Teor. Fiz.} {\bfseries 47} (1964) 1515}.

\bibitem{Kosower:2018adc}
D.A.~Kosower, B.~Maybee and D.~O'Connell, \emph{{Amplitudes, Observables, and
  Classical Scattering}},
  \href{https://doi.org/10.1007/JHEP02(2019)137}{\emph{JHEP} {\bfseries 02}
  (2019) 137} [\href{https://arxiv.org/abs/1811.10950}{{\ttfamily
  1811.10950}}].

\bibitem{Kotikov:1991pm}
A.V.~Kotikov, \emph{{Differential equation method: The Calculation of N point
  Feynman diagrams}},
  \href{https://doi.org/10.1016/0370-2693(91)90536-Y}{\emph{Phys. Lett. B}
  {\bfseries 267} (1991) 123}.

\bibitem{Kotikov:1990kg}
A.V.~Kotikov, \emph{{Differential equations method: New technique for massive
  Feynman diagrams calculation}},
  \href{https://doi.org/10.1016/0370-2693(91)90413-K}{\emph{Phys. Lett. B}
  {\bfseries 254} (1991) 158}.

\bibitem{Kotikov:1991hm}
A.V.~Kotikov, \emph{{Differential equations method: The Calculation of vertex
  type Feynman diagrams}},
  \href{https://doi.org/10.1016/0370-2693(91)90834-D}{\emph{Phys. Lett. B}
  {\bfseries 259} (1991) 314}.

\bibitem{Kruczenski:2022lot}
M.~Kruczenski, J.~Penedones and B.C.~van Rees, \emph{{Snowmass White Paper:
  S-matrix Bootstrap}},  \href{https://arxiv.org/abs/2203.02421}{{\ttfamily
  2203.02421}}.

\bibitem{Landau:1959fi}
L.D.~Landau, \emph{{On analytic properties of vertex parts in quantum field
  theory}},
  \href{https://doi.org/10.1016/B978-0-08-010586-4.50103-6}{\emph{Nucl. Phys.}
  {\bfseries 13} (1959) 181}.

\bibitem{Mandelstam:1958xc}
S.~Mandelstam, \emph{{Determination of the pion - nucleon scattering amplitude
  from dispersion relations and unitarity. General theory}},
  \href{https://doi.org/10.1103/PhysRev.112.1344}{\emph{Phys. Rev.} {\bfseries
  112} (1958) 1344}.

\bibitem{Martin:1969ina}
A.~Martin, \emph{{Scattering Theory: Unitarity, Analyticity and Crossing}},
  vol.~3 (1969), \href{https://doi.org/10.1007/BFb0101043}{10.1007/BFb0101043}.

\bibitem{Mastrolia:2018uzb}
P.~Mastrolia and S.~Mizera, \emph{{Feynman Integrals and Intersection Theory}},
  \href{https://doi.org/10.1007/JHEP02(2019)139}{\emph{JHEP} {\bfseries 02}
  (2019) 139} [\href{https://arxiv.org/abs/1810.03818}{{\ttfamily
  1810.03818}}].

\bibitem{Matsubara-Heo:2023ylc}
S.-J.~Matsubara-Heo, S.~Mizera and S.~Telen, \emph{{Four lectures on Euler
  integrals}},
  \href{https://doi.org/10.21468/SciPostPhysLectNotes.75}{\emph{SciPost Phys.
  Lect. Notes} {\bfseries 75} (2023) 1}
  [\href{https://arxiv.org/abs/2306.13578}{{\ttfamily 2306.13578}}].

\bibitem{Meltzer:2020qbr}
D.~Meltzer and A.~Sivaramakrishnan, \emph{{CFT unitarity and the AdS Cutkosky
  rules}}, \href{https://doi.org/10.1007/JHEP11(2020)073}{\emph{JHEP}
  {\bfseries 11} (2020) 073}
  [\href{https://arxiv.org/abs/2008.11730}{{\ttfamily 2008.11730}}].

\bibitem{Mistlberger:2018etf}
B.~Mistlberger, \emph{{Higgs boson production at hadron colliders at N$^{3}$LO
  in QCD}}, \href{https://doi.org/10.1007/JHEP05(2018)028}{\emph{JHEP}
  {\bfseries 05} (2018) 028}
  [\href{https://arxiv.org/abs/1802.00833}{{\ttfamily 1802.00833}}].

\bibitem{10.1143/PTP.22.128}
N.~Nakanishi, \emph{{Ordinary and Anomalous Thresholds in Perturbation
  Theory}}, \href{https://doi.org/10.1143/PTP.22.128}{\emph{Prog. Theor. Phys.}
  {\bfseries 22} (1959) 128}.

\bibitem{AIHPA_1967__6_2_89_0}
F.~Pham, \emph{Singularit\'es des processus de diffusion multiple},
  {\emph{Annales de l'institut Henri Poincar\'e. Section A, Physique
  Th\'eorique} {\bfseries 6} (1967) 89}.

\bibitem{pham2011singularities}
F.~Pham, \emph{Singularities of Integrals: Homology, hyperfunctions and
  microlocal Analysis}, Springer Science \& Business Media (2011).

\bibitem{Pius:2016jsl}
R.~Pius and A.~Sen, \emph{{Cutkosky rules for superstring field theory}},
  \href{https://doi.org/10.1007/JHEP10(2016)024}{\emph{JHEP} {\bfseries 10}
  (2016) 024} [\href{https://arxiv.org/abs/1604.01783}{{\ttfamily
  1604.01783}}].

\bibitem{Primo:2016ebd}
A.~Primo and L.~Tancredi, \emph{{On the maximal cut of Feynman integrals and
  the solution of their differential equations}},
  \href{https://doi.org/10.1016/j.nuclphysb.2016.12.021}{\emph{Nucl. Phys. B}
  {\bfseries 916} (2017) 94}
  [\href{https://arxiv.org/abs/1610.08397}{{\ttfamily 1610.08397}}].

\bibitem{Remiddi:1981hn}
E.~Remiddi, \emph{{Dispersion Relations for Feynman Graphs}}, {\emph{Helv.
  Phys. Acta} {\bfseries 54} (1982) 364}.

\bibitem{Remiddi:1997ny}
E.~Remiddi, \emph{{Differential equations for Feynman graph amplitudes}},
  \href{https://doi.org/10.1007/BF03185566}{\emph{Nuovo Cim. A} {\bfseries 110}
  (1997) 1435} [\href{https://arxiv.org/abs/hep-th/9711188}{{\ttfamily
  hep-th/9711188}}].

\bibitem{Rubin:1966zz}
M.~Rubin, R.~Sugar and G.~Tiktopoulos, \emph{{Dispersion Relations for
  Three-Particle Scattering Amplitudes. I}},
  \href{https://doi.org/10.1103/PhysRev.146.1130}{\emph{Phys. Rev.} {\bfseries
  146} (1966) 1130}.

\bibitem{Runkel:2019yrs}
R.~Runkel, Z.~Szor, J.P.~Vesga and S.~Weinzierl, \emph{{Causality and loop-tree
  duality at higher loops}},
  \href{https://doi.org/10.1103/PhysRevLett.122.111603}{\emph{Phys. Rev. Lett.}
  {\bfseries 122} (2019) 111603}
  [\href{https://arxiv.org/abs/1902.02135}{{\ttfamily 1902.02135}}].

\bibitem{Schwinger:1960qe}
J.S.~Schwinger, \emph{{Brownian motion of a quantum oscillator}},
  \href{https://doi.org/10.1063/1.1703727}{\emph{J. Math. Phys.} {\bfseries 2}
  (1961) 407}.

\bibitem{Stapp:1982mq}
H.P.~Stapp and A.R.~White, \emph{{An Asymptotic Dispersion Relation for the Six
  Particle Amplitude}},
  \href{https://doi.org/10.1103/PhysRevD.26.2145}{\emph{Phys. Rev. D}
  {\bfseries 26} (1982) 2145}.

\bibitem{Steinmann1960a}
O.~Steinmann, \emph{{{\"U}ber den Zusammenhang zwischen den Wightmanfunktionen
  und den retardierten Kommutatoren}},
  \href{https://doi.org/10.5169/seals-113076}{\emph{Helv. Phys. Acta}
  {\bfseries 33} (1960) 257}.

\bibitem{Steinmann1960b}
O.~Steinmann, \emph{{Wightman-Funktionen und retardierte Kommutatoren. II}},
  \href{https://doi.org/10.5169/seals-113079}{\emph{Helv. Phys. Acta}
  {\bfseries 33} (1960) 347}.

\bibitem{Sterman:1993hfp}
G.F.~Sterman, \emph{{An Introduction to quantum field theory}}, Cambridge
  University Press (8, 1993).

\bibitem{SZABO200628}
R.~Szabo, \emph{Perturbation theory and its techniques},  in \emph{Encyclopedia
  of Mathematical Physics}, J.-P.~Françoise, G.L.~Naber and T.S.~Tsun, eds.,
  (Oxford), pp.~28--41, Academic Press (2006),
  \href{https://doi.org/https://doi.org/10.1016/B0-12-512666-2/00311-4}{DOI}.

\bibitem{tHooft:1973wag}
G.~'t~Hooft and M.J.G.~Veltman, \emph{{Diagrammar}},
  \href{https://doi.org/10.1007/978-1-4684-2826-1_5}{\emph{NATO Sci. Ser. B}
  {\bfseries 4} (1974) 177}.

\bibitem{Tkachov:1981wb}
F.~Tkachov, \emph{{A Theorem on Analytical Calculability of Four Loop
  Renormalization Group Functions}},
  \href{https://doi.org/10.1016/0370-2693(81)90288-4}{\emph{Phys. Lett. B}
  {\bfseries 100} (1981) 65}.

\bibitem{Veltman:1963th}
M.J.G.~Veltman, \emph{{Unitarity and causality in a renormalizable field theory
  with unstable particles}},
  \href{https://doi.org/10.1016/S0031-8914(63)80277-3}{\emph{Physica}
  {\bfseries 29} (1963) 186}.

\bibitem{Veltman:1994wz}
M.J.G.~Veltman, \emph{{Diagrammatica: The Path to Feynman rules}}, vol.~4,
  Cambridge University Press (5, 2012).

\bibitem{Weinzierl:2022eaz}
S.~Weinzierl, \emph{{Feynman Integrals}} (1, 2022),
  \href{https://doi.org/10.1007/978-3-030-99558-4}{10.1007/978-3-030-99558-4},
  [\href{https://arxiv.org/abs/2201.03593}{{\ttfamily 2201.03593}}].

\end{thebibliography}\endgroup

\end{document}